\documentclass[]{article}
\usepackage[a4paper]{geometry}
\usepackage[T1]{fontenc}
\usepackage[ansinew,utf8]{inputenc}
\usepackage{xcolor}
\usepackage{booktabs}
\usepackage{longtable}
\usepackage{hyperref}
\usepackage{graphics}
\usepackage{graphicx}
\usepackage{epsfig}
\usepackage{amssymb}
\usepackage{bm}
\usepackage[]{natbib}
\usepackage[centertags]{amsmath}
\usepackage[spanish,english]{babel}
\usepackage{amsfonts}
\usepackage{latexsym}
\usepackage{array}
\usepackage{color,soul}
\usepackage{ulem}
\usepackage{multirow}
\usepackage{times}
\usepackage[textwidth=8em,textsize=small]{todonotes}
\usepackage{xr}
\usepackage{ulem}

\newcommand{\beq}{\begin{equation}}
	\newcommand{\eeq}{\end{equation}}
\newcommand{\bed}{\begin{displaymath}}
	\newcommand{\eed}{\end{displaymath}}

\newcommand{\eps}{\varepsilon}

\newcommand{\bbeta}{\boldsymbol{\beta}}
\newcommand{\bphi}{\boldsymbol{\phi}}
\newcommand{\bPhi}{\boldsymbol{\Phi}}

\newcommand{\bdelta}{\boldsymbol\delta}

\newcommand{\beps}{\boldsymbol{\varepsilon}}
\newcommand{\btau}{\boldsymbol{\tau}}

\def \bA {{\mathbf A}}
\def \buno{{\mathbf 1}}
\def \bB {{\mathbf B}}

\def \bD {{\mathbf D}}

\def \bG {{\mathbf G}}
\def \bH {{\mathbf H}}

\def \bK {{\mathbf K}}

\def \br {{\mathbf r}}
\def \bP {{\mathbf P}}
\def \bR {{\mathbf R}}

\def \bS {{\mathbf S}}

\def \bX {{\mathbf X}}
\def \bx {{\mathbf x}}
\def \by {{\mathbf y}}

\def \bU {{\mathbf U}}

\def \bV {{\mathbf V}}
\def \bZ {{\mathbf Z}}

\def \bzero {{\mathbf 0}}

\newtheorem{mydef}{Preposition}

\title{A Nested Error Regression Model with High Dimensional Parameter for Small Area Estimation}
\author{P. Lahiri \\
	\multicolumn{1}{p{.9\textwidth}}{\centering\textit{Joint Program in Survey Methodology \& Department of Mathematics, University of Maryland College Park, USA.}\\E-mail: plahiri@umd.edu}
		\and N. Salvati \\
		\multicolumn{1}{p{.9\textwidth}}{\centering\textit{Dipartimento di Economia e Management, Universit\`a di Pisa, Italy.} \\E-mail: nicola.salvati@unipi.it}}
\date{}
\begin{document}

\maketitle

\begin{abstract}
In this paper we propose a flexible {\color{black}{nested error regression small area model}} \textcolor{black}{with high dimensional parameter}  that incorporates heterogeneity in regression coefficients and variance components. 
% Such modeling allows pooling information from a large number of areas and thus makes the notion of consistency more relevant.  
We develop a new robust small area specific estimating equations method 
% for producing consistent estimators of regression coefficients and variance components.  Moreover, since our theoretical framework
that allows appropriate pooling of a large number of areas in estimating small area specific model parameters.
% , the use of the first-order unbiasedness of the mean squared prediction error estimators is adequate and thus avoiding the cumbersome task of bias correction.  
We propose a parametric bootstrap and jackknife method to estimate not only the mean squared  errors but also other commonly used uncertainty measures such as standard errors and coefficients of variation. We conduct both model-based and design-based simulation experiments and real-life data analysis to evaluate the proposed methodology. 

\vspace{0.3cm}

\end{abstract}

\section{Introduction}\label{sec:intro}
Planning and evaluation of government programs require access to a wide range of national and sub-national socio-economic, environment, health, and other statistics. For this reason, there is a growing need for reliable statistics relating to much smaller geographical areas where data are too sparse to support the sort of standard estimation methods typically employed at the national level. These small area official statistics are routinely used for a variety of purposes, including assessing economic well-being of a nation, making public policies, and allocating funds at the federal, state and local levels. 

The main idea behind small area estimation is to borrow strength from related sources through statistical models that connect different alternative databases.  In much of the small area literature, mixed models are generally employed because such models can account for uncertainties from different sources and thereby can produce accurate estimates and associated uncertainty measures at granular levels compared to the corresponding fixed effects models. M-quantile regression can be considered as an alternative class of models for the same purpose relaxing some of the conventional modelling assumptions such as the normality of the random components and obtaining estimators that are robust against outlying values.  We refer to the well-cited Wiley book by \citet{RaoMol15} and papers by \citet{Jia06}, \cite{Pfe13}, \cite{Cha14}, \cite{Ghosh20}, and \cite{sal20} for a detailed account of different small area models and methods. 

\cite{Bat88} proposed an empirical best linear unbiased prediction (EBLUP) method, using a nested error regression (NER) model, in order to estimate acreage under corn and soybeans for 12 counties in north-central Iowa, USA.
The nested error regression model, a special case of linear mixed model,  can be viewed as an extension of a regression model where the intercept term is allowed to vary across counties (or small areas), but, in order to make the method efficient, the county specific intercepts are assumed to be generated from the same underlying distribution.  This  random area specific intercept term is introduced in order to capture a part of the leftover between county variation that is not explained by the area specific auxiliary variables included in the model.  The associated EBLUP method borrows strength from the known auxiliary variable means for the area population.

The NER model has played an important role in 
small area estimation since the publication of \cite{Bat88}. However, such a model is likely to fail when the number of small areas to be combined is large.  This is because the assumption of the same regression coefficients and/or variance components in the nested error regression model may not be tenable for all the small areas. Random area specific regression coefficients models, which extend the nested error model by treating regression coefficients as random effects,  have been suggested in the literature; see \cite{Pra90}, \cite{hobza2013}, \cite{RaoMol15} for empirical best prediction (EBP) approach and \textcolor{black}{\citet{Hoff:2009} for the Bayesian approach. Such modeling, though useful in some applications, needs more nontrivial assumptions on the joint distribution on the random regression coefficients. \cite{Jiang2012}  considered a heteroscedastic nested error regression model by allowing different fixed sampling variances.  They showed that all the parameters, except the area specific sampling variances, of their model can be consistently estimated. Interestingly, their EBP does not involve area specific sampling variances because of the assumption that the variances of the random effects are proportional to the corresponding sampling variances.  Thus, their EBP well approximates the corresponding best predictor (BP) when the number of areas is large.  However, the model does not allow second-order unbiased MSE estimator since MSE involves sampling variances, which cannot be consistently estimated.} Moreover, random area specific sampling variance models have been also proposed in order to incorporate the leftover between area heteroscedasticity of the sampling variances across areas not captured by the available auxiliary variables.  Early examples of such models can be found in \cite{otto1995sampling} and  \cite{Arora97}. For more recent research in modeling sampling variances as a way to incorporate heteroscedastic variances, see \cite{Liu2014}, \cite{Kubokawa2016}, \cite{sugasawa2017bayesian},  \cite{Neves20}, among others. Though not used, one can envision a random area specific regression coefficients model in conjunction with random area specific sampling variance model to capture variations in both regression coefficients and sampling variances. 

% \textcolor{blue}{Considering the following comment by referee 2 I think we need to add something about the methodology used in these papers: 1.	Closely related models:  These papers are relevant in terms of flexible modeling of variances, so at least these papers should be mentioned. Moreover considering comment 2 about the bayesian method we need to quote the book Hoff (2009) and write something the reason we don't use Bayesian approach. This is for you Partha. }

The random area specific regression coefficients models and/or random area specific sampling variances models involve specifications of distributions of a large number of random effects. On the other hand, fixed effects assumptions on the area specific regression coefficients and sampling variances generally lead to unstable estimates of these fixed effects due to small area specific sample sizes \citep{Jiang2012}.  In this paper, we introduce a new approach that is not considered in the literature.  Specifically, we assume fixed effects for both regression coefficients and sampling variances, but use area specific estimating equations applied to data from all areas in estimating these area specific regression coefficients and then use appropriately constructed residuals for estimation of variance components. {\color{black}The proposed model can be called a nested error regression with high dimensional parameter.}

{\color{black}{When area specific tuning parameters of the system of estimating equations are known, we have shown that parameters of our proposed nested error regression model with high dimensional parameter can be consistently estimated.  This is because we use data from all areas in estimating any area specific model parameter. When the tuning parameters are unknown, we have suggested two different estimators of these parameters. However, for obtaining consistent estimators of the tuning parameters, a basic requirement would be large area specific sample sizes. However, like most papers in small area estimation, our emphasize here is small area specific sample sizes. Our extensive model-based and design-based simulation results demonstrate that our proposed estimation method performs well under a variety of simulation conditions.}}

Estimation of mean squared error (MSE) of EBLUP has been a topic of extensive research for more than the last three decades.  The pioneering work of \citet{Pra90} inspired many in considering various extensions;  see \citet{Dat00}, \citet{Das:2004}, \citet{Jiang2002} \citet{Hal06}. In all these papers, the focus has been to derive second-order unbiased MSE estimators.
{\color{black}{In this paper, we are not requiring the second-order unbiasedness of MSE estimators for a couple of reasons.  First, our goal here is to cover a wide range of uncertainty measures (e.g., Root MSE, CV, Relative RMSE) -- not just MSE.  What is second-order unbiased is not necessarily second-order unbiased for a nonlinear function of MSE. Second, existing second-order unbiased MSE estimates do not necessarily ensure strictly positive MSE estimates.  As pointed out in \citet{Jiang2018} there is no paper in small area estimation that proves simultaneous properties of positivity and second-order unbiasedness of parametric bootstrap MSE estimates.  The McJack method ensures both properties but only in estimating a known monotone function of MSE (e.g., logarithm of MSE) -- not MSE.  For these reasons, we have not attempted to develop second-order unbiased MSE estimators in this paper. We propose a simple general parametric bootstrap and a jackknife estimators  of a wide range of uncertainty measures.  We evaluate different uncertainty measures under different situations through extensive simulations.  For known area specific tuning parameters, our estimators of uncertainty measures (not necessarily MSE) tend to the corresponding true uncertainty measures in probability. }}

% though they do not necessarily produce second-order unbiased estimator of other commonly used uncertainty measures such as standard error and coefficient of variation (CV). To achieve second-order unbiasedness, complex bias correction factors are usually applied, which may result in undesirable negative MSE estimates. \cite{Jiang2018} pointed out that there is no concrete theory of second-order unbiasedness of such strictly positive adjusted MSE estimators.

% Models considered in many small area estimation papers generally work well when the number of areas to be combined is not too large, which justifies the preference of the second-order unbiasedness over the first-order unbiasedness.   \citet{Morris2006}, in discussing \citet{Jia06}, alluded to a practical issue on the use of asymptotics in applied data analysis because of possible breakdown of the commonly used  exchangeable models when the number of areas is large.  Since our flexible high dimensional model allows more heterogeneity than the exisiting small area models, the number of small areas to be combined can be made sufficiently large to justify the use of  first-order unbiasedness criterion.  

The outline of the paper is as follows. \textcolor{black}{In Section \ref{sec:motexample} we motivate the proposed nested error regression model with high dimensional parameter by analysing data from the US Environmental Protection Agency's Environmental Monitoring and Assessment Program.} After presenting notation and EBLUP estimators for small areas in Section \ref{sec:notation}, we propose \textcolor{black}{ a nested error regression models with high dimensional parameter} in Section \ref{sec:model}. In Section \ref{alg_pred} we show its estimation algorithm, consistency property of the estimator and application to the SAE situation. In Section \ref{sec:mcjk} we discuss two different estimators of the root mean squared errors of the proposed small area estimator. In particular, the first proposal is a parametric bootstrap estimator. The second MSE estimator is based on the Monte-Carlo jackknife method proposed by \citet{Jiang2018}. In Section \ref{sim:sec} we empirically evaluate the performance of the proposed approach and its associated MSE estimators using both model-based and  design-based simulation studies, with the latter based on a real dataset: the 1995-96 Australian Agricultural Grazing Industry Survey (AAGIS) data \citep{Cha12a}. In Section \ref{emap:data}, we use the proposed method for estimating average levels of Acid Neutralising Capacity at 8-digit Hydrologic Unit Code (HUC) level using data collected in an environmental survey of lakes in the Northeast of the USA \citep{Ops08}. Finally, in Section \ref{sec:conclusion}, we summarise our main findings and provide directions for future research.

\section{A motivating example}\label{sec:motexample}
\begin{figure}[h]
	\centering    
	\includegraphics[scale = 0.40]{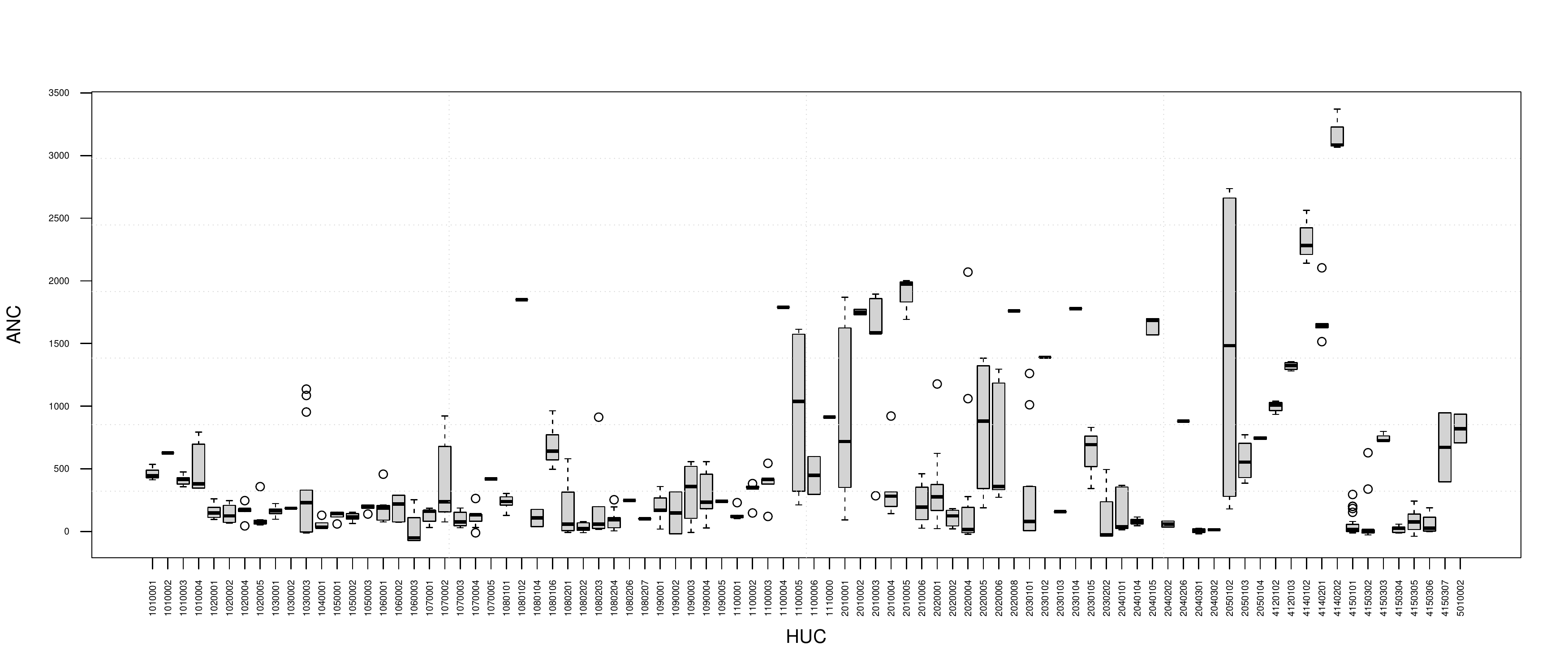}\\
	\includegraphics[scale = 0.40]{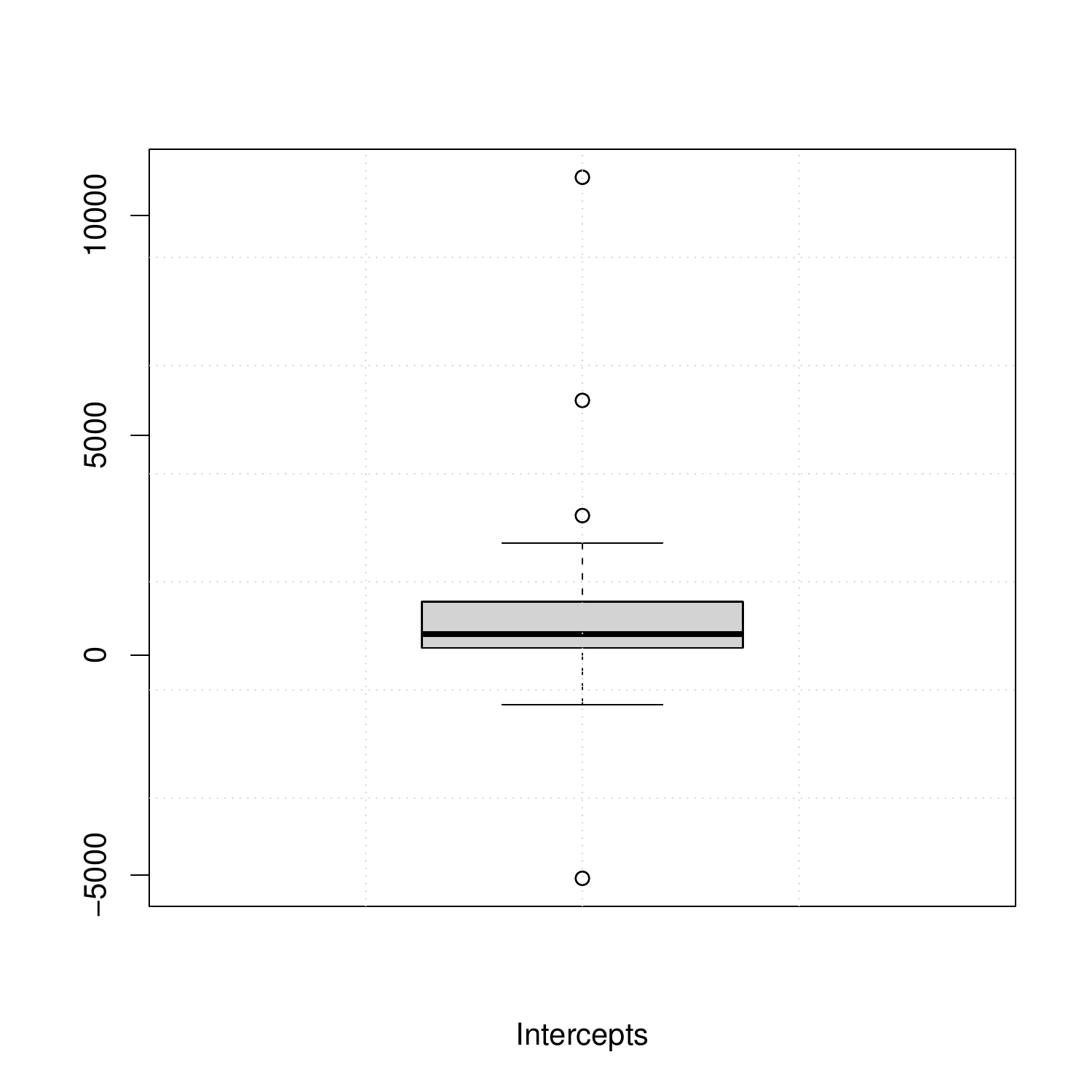} 
	\includegraphics[scale = 0.40]{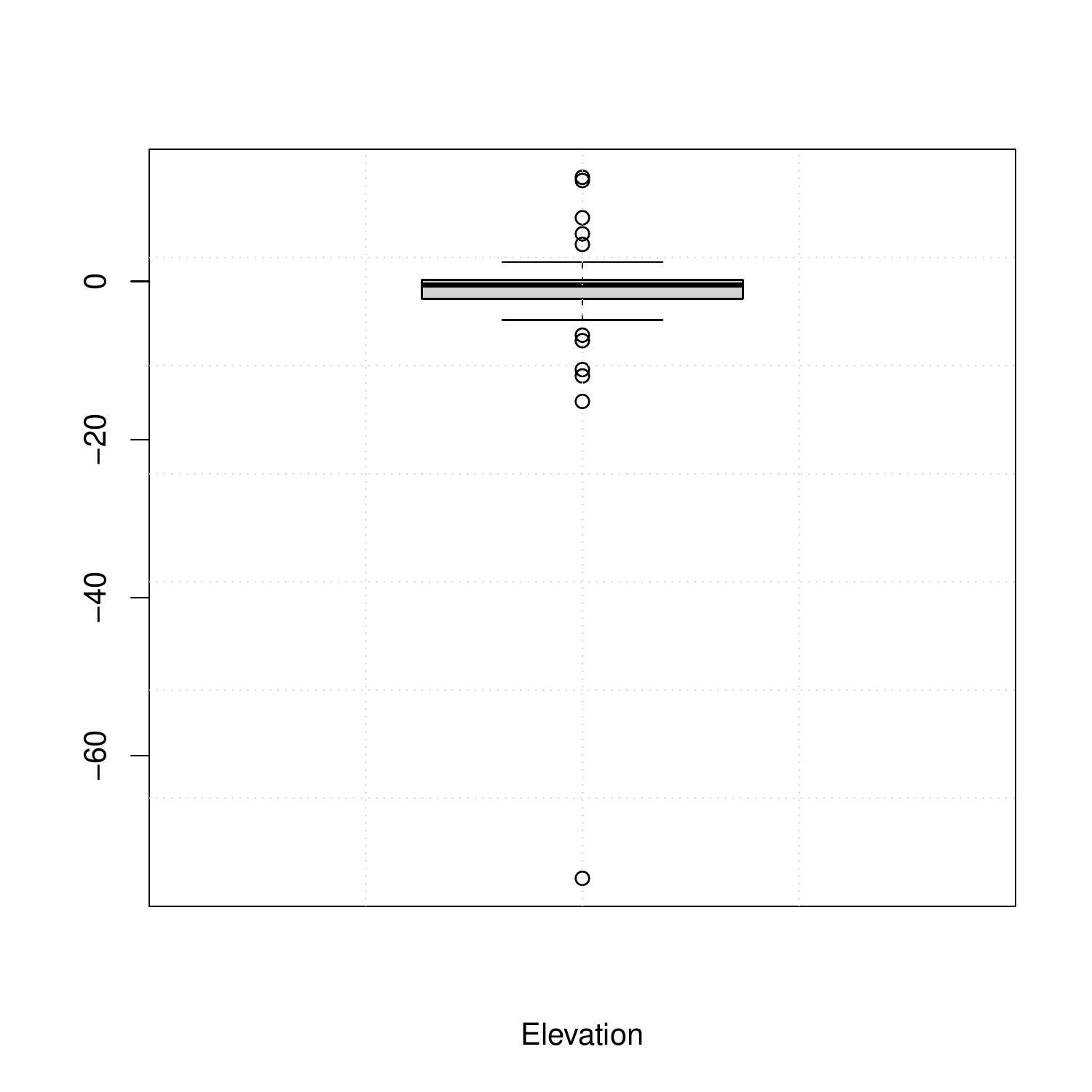} \\
	\caption{\label{EMAP_Analysis} Boxplots displaying the distribution of ANC values by area (top panel) and the distribution of the estimated values of the intercept and the slope (bottom panel) obtained fitting a regression model area by area where the ANC is the response variable and the elevation is the covariate.}
\end{figure}

\textcolor{black}{To motivate the proposed nested error regression model with high dimensional parameter, we consider data from the US Environmental Protection Agency's Environmental Monitoring and Assessment Program (EMAP) Northeast lakes survey \citep{larsen2001, Ops08, salvati2012}. Between 1991 and 1995, researchers from the US Environmental Protection Agency conducted an environmental health study of lakes in the north-eastern states of the USA. For this study, a sample of 334 lakes (or more accurately, lake locations) was selected from the population of 21,026 lakes in these states using a systematic random sample design. The lakes making up this population are grouped into 113 8-digit Hydrologic Unit Codes (HUCs), defined as small areas, of which 64 contain less than 5 observations and 27 did not have any observation. The study variable is the Acid Neutralising Capacity (ANC), an indicator of the acidification risk of water bodies. Factors affecting the ANC such as acid deposition and soil characteristics cut across HUCs, so overall spatial trends are also likely to be useful in predicting the ANC. The EMAP data set contains the elevation and geographical coordinates of the centroid of each lake in the target area. For each small area we have estimated the sample variance and  fitted a regression model where the response variable is the ANC and the covariate is the elevation. Initial exploration of the data suggests that the within-area variation and regression coefficients change dramatically across small areas. Figure \ref{EMAP_Analysis} presents box-plots of the distribution of ANC values by area (top panel) and distributions of intercept and slope estimates (bottom panel). Figure \ref{EMAP_Analysis} suggests that the assumption of identical regression coefficients and/or variance components in the nested error regression model for this real data example is unreasonable.}

%%%%%%%%%%%%%%%%%%%%%%%%%%%%%%%%%%%%%%%%%%%%%%%%%%%%%%%%%%%%%%%%%%%%%%%%%
%%%%%%%%%%%%%%%%%%%%%%%%%%%%%% Notation    %%%%%%%%%%%%%%%%%%%%%%%%%%%%%%
%%%%%%%%%%%%%%%%%%%%%%%%%%%%%%%%%%%%%%%%%%%%%%%%%%%%%%%%%%%%%%%%%%%%%%%%%

\section{Notation and background}\label{sec:notation}
Consider $m$ small areas with the $i$th small area population consisting of $N_i$ units.  Let $y_{ij}$ and $\bx_{ij}$ denote the values of the study variable and a $p \times 1$ vector of known auxiliary variables for the $j$th unit of the $i$th small area, respectively, with $i=1,\dots,m,\,j=1,\dots,N_i.$  We are interested in estimating the small area means $\bar Y_i=N_i^{-1}\sum_{j=1}^{N_i}y_{ij}$ using a simple random sample  $s$ of size $n$ drawn from the finite population covering all $m$ areas and $\bar{\bX}_i=N_i^{-1}\sum_{j=1}^{N_i}\bx_{ij}$, the vector $p \times 1$ of finite population means of the auxiliary variables for area $i$.  In a typical small area estimation situation, $n_i$, sample size for area $i$,  is not large enough to support the use of a direct estimator, the sample mean $\bar y_i=n_i^{-1}\sum_{j\in s_i}y_{ij},$ where $s_i$ denotes the part of the sample from the $i$th small area.  Here $n=\sum_{i=1}^mn_i$.
%   The vector $\by$ of dimension $N \times 1$ represents for the target variable in the population $U$.  The matrix $\bX$, $N \times p$, contains $p$ unit-level auxiliary variables.

\cite{Bat88} considered the following nested error regression model for the finite population:
\begin{equation}\label{eq:mix}
	y_{ij}=\beta_0+\bx_{ij}^{\prime}\bbeta+\gamma_i+\epsilon_{ij},\;i=1,\dots,m;\;j=1,\dots,N_i,
\end{equation}
where $\beta_0$ and $\bbeta$ are unknown fixed intercept and regression coefficients, respectively; $\gamma_i$ is a random effect for area $i$ that attempts to incorporate the leftover between area variations not captured by the auxiliary variables $\bx_{ij}$; $\epsilon_{ij}$ is the sampling error for the $j$th observation in the $i$th area, which captures the leftover variations not accounted for the other components of the model. The area specific random effects $\gamma_i$ and the sampling errors $\epsilon_{ij}$ are all assumed to be independent with $\gamma_i \sim N(0, \sigma_{\gamma}^2)$ and $\epsilon_{ij} \sim N(0, \sigma_\eps^2), \;i=1,\dots,m;\;j=1,\dots,N_i$.  The parameters $\bdelta=(\sigma_\gamma^2,\sigma_\eps^2 )$ are referred to as the variance components of model \eqref{eq:mix}.

\cite{Bat88} argued that, under the assumed nested error model \eqref{eq:mix}, the finite population mean $\bar Y_i$ can be well approximated by $\theta_i=\beta_0+\bar\bX_i'\bbeta+\gamma_i$, for large $N_i$.
\cite{Bat88} proposed an empirical best linear unbiased predictor (EBLUP) for estimating $\theta_i$ given by
\beq\label{saemix}
\hat{\theta}_{i}^{BHF}=\hat{\bbeta}_0+\bar\bX'_{i}\hat{\bbeta}+\hat{\gamma}_i,
\eeq
where $\hat{\bbeta}_0$ and $\hat{\bbeta}$ are weighted least square estimators of $\beta_0$ and $\bbeta,$ respectively; $\hat{\gamma}_i=(1-\hat B_i)(\bar y_i-\hat{\bbeta}_0-\bar\bx_{i}^{\prime}\hat{\bbeta})$  is the EBLUP of $\gamma_i$ with $\bar{\bx}_i=n_i^{-1}\sum_{j\in s_i}\bx_{ij},\;\hat B_i=\frac{\hat{\sigma}_\epsilon^2/n_i}{\hat{\sigma}_\epsilon^2/n_i+\hat\sigma_\gamma^2}$ and $\hat{\bdelta}=(\hat\sigma_{\gamma}^2,\hat\sigma_{\epsilon}^2)$ is a consistent estimator of $\bdelta=(\sigma_{\gamma}^2,\sigma_{\epsilon}^2)$ under model \eqref{eq:mix}; see \cite{Bat88} for details.  The authors implicitly assumed non-informative sampling so that the same nested error model \eqref{eq:mix} holds for the sample.  

The synthetic assumption of identical regression coefficients $\bbeta$ and sampling variance $\sigma_\epsilon^2$ across all small areas to be combined may be unrealistic when the number of small areas $m$ is large. The synthetic assumption on $\bbeta$ can be relaxed if we replace $\bbeta$ in (\ref{eq:mix}) by random area specific regression $\bbeta_i$, $i=1,\dots,m$, generated from a common model, e.g., a common $p$-dimensional multivariate normal model with common mean vector and variance-covariance matrix.  This additional assumption  for random effects is necessary to reduce the number of unknown parameters; see \citet{Pra90}. \citet[][p. 173]{RaoMol15} relaxed the homogeneity assumption of the individual errors by replacing $\sigma_\epsilon^2$ by $k_{ij}\sigma_\epsilon^2$, where $k_{ij}>0$ is a known auxiliary variable. However, identifying $k_{ij}$ in a real-life data analysis may be hard and between area variability may not be fully explained by $k_{ij}$.  
In the next section, we propose an alternative solution. 
\section{Model and method}\label{sec:model}
We propose the following extension of the nested error regression model:
\begin{equation}\label{eq:mix:new}
	y_{ij}=\beta_0+\bx_{ij}^{\prime}\bbeta_i+\gamma_i+\epsilon_{ij},\;i=1,\dots,m;\;j=1,\dots,N_i,
\end{equation}
where $\bbeta_i$ is a $p \times 1$ vector of fixed unknown  regression coefficients for area $i$; $\gamma_i$ and $\epsilon_{ij}$ are all independent with $\gamma_i\sim N(0,h_i \sigma_{\gamma}^2)$ and 
$\eps_{ij}\sim N(0,k_{ij}\sigma_{\eps i}^2)$, with $h_i$ and $k_{ij}$ are known auxiliary variables at area and individual levels, respectively. \textcolor{black}{This model can be called a nested error regression with high dimensional parameter.} \citet{Bat88} considered a special case of model \eqref{eq:mix:new} with $h_i=1$, $k_{ij}=1$, $\bbeta_i=\bbeta$ and $\sigma_{\eps i}^2=\sigma_{\eps }^2$; $i=1,\dots,m; j=1,\dots,N_i$.

% \citet{kleffe92} proposed a linear model with random effects and random error variances and a second order approximation to mean square error of the EBLUP of the mean and its approximately unbiased estimator were derived. \citet{Arora97} studied the regression case with $\bx_{ij} = \bX_i$ and known parameters $\bbeta$ and $\sigma_\gamma^2$. Under normality of the random effects and $\eps_{ij}|\sigma_{\eps i}^2$ and a specified distribution on $\sigma_{\eps i}^2$, the authors showed that the Bayes (best) estimator of the mean is more efficient than the BLUP estimator of \citet{kleffe92}. \citet{Jiang2012} proposed the heteroscedastic unit level model. The BLUP estimator of the area mean under this model is identical to the BLUP estimator under the basic unit level model \eqref{eq:mix} when the model parameters are known.

% Our approach differs from the that of \citet{kleffe92}, \citet{Arora97} and \citet{Jiang2012} in two different ways. First, the regression coefficients vary across areas as in \citet{Sal11} and \citet{Cha12a}. This could allows for incorporating the spatial information in the model. Second, our approach assume that also $\sigma_{\gamma}^2$ and $\sigma_{\eps}^2$ can vary across areas and it allows for incorporating in the model many auxiliary variables available at area level and clustering the small areas. Given a random intercept model \eqref{eq:mix:new} and the posterior linearity or the condition mean of $\bgamma$ given data, model parameter linear in observations, the Best Predictor (BP) of the mean of area $i$ is
Under model  \eqref{eq:mix:new} and non-informative sample design, the best predictor (BP) of $\theta_i=\beta_0+\bar\bX_i'\bbeta_i+\gamma_i$ is given by
\begin{eqnarray}\label{BP_prop}
	\nonumber\hat\theta_i^{BP}\equiv \hat\theta_i(\bphi_i)&=&\beta_0+\bar{\bX}'_i \bbeta_i+(1-B_i)(\bar y_i-\beta_0-{\bar{\bx}}_i ^\prime \bbeta_i)\\
	&= & (\bar{\bX}_i-\bar{\bx}_i)'\bbeta_i+\left \{B_i(\beta_0+\bar{\bx}_i^\prime \bbeta_i)+(1-B_i)\bar y_i\right \},\end{eqnarray}
where $B_i=\frac{{\sigma}_{\epsilon i}^2/n_i}{{\sigma}_{\epsilon i}^2/n_i+\sigma_\gamma^2}$ and $\bphi_i=(\beta_0,\bbeta_i,\sigma_{\gamma}^2, \sigma_{\eps i}^2)'$.

% and $\tilde\bbeta_i$ is the Best Linear Unbiased estimator of $\bbeta_i$ assuming that $\bdelta_i=(\sigma_{\gamma i}^2,\sigma_{\eps i}^2)$ is known. Here $(1-B_i) \in (0,1)$ measures the unexplained between-area variability, $\sigma_{\gamma i}^2$, relative to the total variability $\sigma_{\gamma i}^2+\sigma_{\eps i}^2/n_i$. The BP can also be expressed as a weighted average of the `survey regression' estimator $(1-B_i)(\hat\theta_i^{DIR}+(\bar{\bX}_i-\hat{\bar{\bx}}_i) ^\prime \tilde\bbeta_i)$ and the regression-synthetic estimator $\bar{\bX}_i \tilde\bbeta_i$:
% \begin{equation}\label{BP_prop}
% \tilde\theta_i^{BP}=(1-B_i)(\hat\theta_i^{DIR}+(\bar{\bX}_i-\hat{\bar{\bx}}_i) ^\prime \tilde\bbeta_i)+B_i \bar{\bX}_i \tilde\bbeta_i.
% \end{equation}
% Note that the survey regression estimator is approximately design-unbiased for the mean $\theta_i$ under simple random sampling (srs), given that the total sample size $n$ is large. 

Under simple random sampling, $\hat\theta_i^{BP}$ is design-consistent for $\theta_i$ because, as $n_i$ increases,  $\hat\theta_i^{BP}$ approaches to the design-consistent  estimator $\bar y_i$ since $B_i \rightarrow 0$   and $\bar{\bx}_i$ is design-consistent for $\bar{\bX}_i$. An empirical best predictor (EBP) of $\theta_i$ can be written as $\hat\theta_i^{EBP}\equiv \hat\theta_i(\hat\bphi_i),$ where $\hat\bphi_i$ is a consistent estimator of $\bphi_i$ under the assumed model \eqref{eq:mix:new} as $m$ tends to $\infty$.  A BP of $\bar{Y}_i$ that uses the sampling fraction $f_i=n_i/N_i$ is given by $\hat{\bar Y}_i^{BP}=f_i\bar{y}_i+(1-f_i)
\hat{\theta}_{i}^{BP}$.
%see \cite{GhoshLahiri87}, \cite{Pra90}.  
We note that  $\hat{\theta}_{i}^{BP}$ and $\hat{\bar Y}_i^{BP}$ are both design-consistent even when $f_i$ is a fixed negligible constant.

% \begin{equation}\label{EBP_prop}
% \hat\theta_i^{EBP}=\bar{\bX}_i \hat\bbeta_i+(1-\hat{B}_i)(\hat\theta_i^{DIR}-\hat{\bar{\bx}}_i ^\prime \hat\bbeta_i),
% \end{equation}
% where $\hat{B}_i=\frac{\hat\sigma_{\eps i}^2/n_i}{\hat\sigma_{\eps i}^2/n_i+\hat\sigma_{\gamma i}^2}$. 

% The estimation of parameters $\bphi_i=\{\bbeta_i^\prime, \sigma_{\gamma i}^2, \sigma_{\eps i}^2  \}^\prime$ can be problematic for this model. The varying regression coefficients can be estimated by a random slopes model but it remains the problem of the estimation of the varying variance components. Moreover, random slopes model requests the assumption of normality of the random slopes. As a consequence the model can not to mitigate the effects of outlier values on the small area estimates. 

\section{Estimation of the vector of the parameters $\bphi_i=(\beta_0,\bbeta_i,\sigma_{\gamma}^2, \sigma_{\eps i}^2)'$}\label{alg_pred}

\textcolor{black}{We begin the section by first describing the maximum likelihood method for estimating $\bphi_i$, $i=1,\dots,m$. For model \eqref{eq:mix:new}, the log-likelihood function has the expression
	\begin{eqnarray}\label{loglikeML}
		\nonumber \ell(\bphi)&=&constant-\frac{1}{2} \sum_{i=1}^{m}\left[ n_i \log{\sigma_{\epsilon i}^2}+ \log{\left[\frac{\sigma_{\epsilon i}^2+n_i \sigma_\gamma^2}{\sigma_{\epsilon i}^2}\right]} \right.\\
		& & \left.+\frac{1}{\sigma_{\epsilon i}^2} \left\{\sum_{j=1}^{n_i}(y_{ij}-\beta_0- \bx_{ij}^{\prime}\bbeta_i )^2-\frac{\sigma_\gamma^2}{\sigma_{\epsilon i}^2+n_i \sigma_\gamma^2}( n_i\bar{y}_i -n_i\beta_0-n_i\bar{\bx}_i^{\prime}\bbeta_i)^2  \right\}  \right],
	\end{eqnarray}
}
\textcolor{black}{where the constant does not depend on the parameters, $\bphi=(\bphi_1,\dots,\bphi_m)$. The maximum likelihood estimators (MLEs) of the parameters are obtained by differentiating the $\ell(\bphi)$ with respect to $\beta_0,\bbeta_i,\sigma_{\gamma}^2, \sigma_{\eps i}^2$ and solving the system of estimating equations sets out equal to zero. In particular, given the variance components, the MLE of the regression coefficients for area $i$ are:
	\begin{equation}
		\hat{\beta}_0=\left( \sum_{i=1}^m \sigma_{\epsilon i}^{-2}n_i B_i \right)^{-1}\left( \sum_{i=1}^m \sigma_{\epsilon i}^{-2}n_i B_i(\bar{y}_i -\bar{\bx}_i^{\prime}\bbeta_i) \right),
	\end{equation}
	\begin{equation}
		\hat{\bbeta}_i=\left( \sum_{j=1}^{n_i}  \bx_{ij}\bx_{ij}^{\prime}-n_i(1-B_i) \bar{\bx}_i \bar{\bx}_i^{\prime}  \right)^{-1}\left( \sum_{j=1}^{n_i}  \bx_{ij}(y_{ij}-\beta_0)-n_i(1-B_i) \bar{\bx}_i ( \bar{y}_i-\beta_0) \right),
	\end{equation}
	and, given the regression coefficients, the MLE of the sampling variance for area $i$ is:
	\begin{equation}
		\hat{\sigma}_{\epsilon i}^2=\frac{1}{n_i}\left\{\sum_{j=1}^{n_i}(y_{ij}-\beta_0- \bx_{ij}^{\prime}\bbeta_i )^2-n_i(1-B_i)( \bar{y}_i -\beta_0-\bar{\bx}_i^{\prime}\bbeta_i)^2  \right\}.
	\end{equation}
	\citet{neyman48} gave an example which shows that, when the number of nuisance parameters increases with the at the same rate as the sample size, the MLEs may not be consistent. Another example was given by \citet{Jiang2012}. The latter authors considered a heteroscedastic NER model with area-specific error variance, and noted that the MLE of the area-specific error variance is inconsistent.} 

\textcolor{black}{As in the Neyman-Scott  and Jiang and Nguyen problems, in our proposed model the number of unknown parameters is proportional to the sample size, if the area sample  sizes ($n_i$) are bounded, which is typically the case in small area estimation problems \citep{Jiang2012}. Note that consistent estimators of regression coefficients ($\beta_0, \bbeta_i$) and variance components ($\sigma_\gamma^2, \sigma_{\eps i}^2$) are all that is needed to justify the use of the proposed EBP as a point predictor.} 

\textcolor{black}{In this paper, a generalized estimating equation (GEE) approach is used to estimate the parameters of model \eqref{eq:mix:new}; this method allows to borrow strength across areas when estimating each area specific vector of parameters obtaining consistent estimators of the area specific slope parameters ($\bbeta_i$) and the area specific sampling variance ($\sigma_{\eps i}^2$).}
%In particular, the MLE of the variance components are, in general, biased. Such a bias may not vanish as the sample size increases, if the number of the fixed effects is proportional to the sample size. In fact, in the latter case the MLE will be inconsistent.
\textcolor{black}{Because of more flexibility in estimating the model parameters than that of the maximum likelihood, our estimating equation method is likely to have an edge over the maximum likelihood in terms of the predictive power, which is important in the small area estimation context.  For known area specific tuning parameter $\tau_i$, our estimating equation method yields consistent estimators of the model parameters, unlike the maximum likelihood method.}

We now describe the algorithm based on GEE approach for estimating $\bphi_i$, $i=1,\dots,m$, in the following steps.  
\begin{itemize}
	%\item[Step 1] Start with initial values, say $\{\beta_{0i}^{(0)}, \bbeta_{i}^{(0)}, \sigma_{\gamma }^{2(0)},\sigma_{\epsilon i}^{2(0)},\;i=1,\dots,m\}$, for $\{\beta_{0i}, \bbeta_{i},\sigma_{\gamma }^2,\sigma_{\epsilon i}^2,\;i=1,\dots,m\}$. Compute the following $n_i \times n_i$ matrix: 
	%$$\bV_i^{(0)}=h_i\sigma_{\gamma}^{2(0)}\buno_{n_i}\buno_{n_i}^{\prime} +\sigma_{\epsilon i}^{2(0)} \bK_i,$$ 
	%where  $\bK_i=\mbox{diag}(k_{i1},\dots,k_{in_i})$ is a $n_i \times n_i$ diagonal matrix, $i=1,\dots,m$. Let $\bU_i^{(0)}$ be a $n_i\times n_i$ diagonal matrix with diagonal elements equal to those of $\bV_i^{(0)},i=1,\dots,m.$
	\item[Step 1] \textcolor{black}{At each iteration $t=1,2,\dots$, define the following $n_l \times n_l$ matrix: 
		$$\bV_{l;i}^{(t-1)}=h_l\sigma_{\gamma}^{2(t-1)}\buno_{n_l}\buno_{n_l}^{\prime} +\sigma_{\epsilon i}^{2(t-1)} \bK_l,$$ 
		where $\buno_{n_l}$ denotes a vector of ones of length $n_l$ and $\bK_l=\mbox{diag}(k_{l1},\dots,k_{l n_l})$ is a $n_l \times n_l$ diagonal matrix, $l=1,\dots,m$. Let $\bU_{l;i}^{(t-1)}$ be a $n_l\times n_l$ diagonal matrix with diagonal elements equal to those of $\bV_{l;i}^{(t-1)},l=1,\dots,m.$ Set  ${\beta}_{0}=\sum_{i=1}^m {\alpha}_{0i}/m$ and start with initial values, say $\{\alpha_{0i}^{(0)}, \bbeta_{i}^{(0)}, \sigma_{\gamma }^{2(0)},\sigma_{\epsilon i}^{2(0)},\;i=1,\dots,m\}$, for $\{\alpha_{0i}, \bbeta_{i},\sigma_{\gamma }^2,\sigma_{\epsilon i}^2,\;i=1,\dots,m\}$.} %Compute the following $n_l \times n_l$ matrix: 

	\item[Step 2] 
	% Let  $\bU_i^{(0)}= \mbox{diag}(v_{i;11}^{(0)},\dots,v_{i;n_1n_1}^{(0)},\dots,v_{i;n-n_mn-n_m}^{(0)},\dots,v_{i;nn}^{(0)})$ be a $n \times n$ diagonal matrix, where $v_{i;jj}^{(0)}$ denotes the $j$th diagonal element of  $\bV_i^{(0)},\;j=1,\dots,n$.  
	\textcolor{black}{For $t=1,2,\dots$ define $\br_{l;i}^{(t)}=(\bU_{l;i}^{(t-1)})^{-1/2}(\by_{l}-{\color{black}\alpha_{0i}^{(t)}\buno_{n_l}-\bX_{l} \bbeta_{i}^{(t)}})$, where $\by_{l}$ is a $n_l \times 1$ vector of the response variable and $\bX_{l}$ denotes a matrix $n_l \times p$ of individual level covariates of the sampling units in area $l$. Obtain  $(\alpha_{0i}^{(t)},\bbeta_{i}^{(t)})$ by solving the following system of estimating equations for $(\alpha_{0i},\bbeta_i)$:
		\begin{equation} \label{estequbeta} \sum_{l=1}^m \left[ \bX_{l(p+1)}(\bV_{l;i}^{(t-1)})^{-1} (\bU_{l;i}^{(t-1)})^{1/2}\psi_{i}(\br_{l;i}^{(t)})\right]=\bzero, ~~i=1,\dots,m, \end{equation}
		where $\psi_i(\br_{l;i}^{(t)})$ is a $n_{l}\times 1$ vector obtained from the vector of residuals $\br_{l;i}^{(t)}$ with its $j$th component, say $r_{lj;i}^{(t)}$, replaced by $\psi_i(r_{lj;i}^{(t)}),$ a chosen known function of $r_{lj;i}^{(t)}.$  {\color{black}{Here $\bX_{l(p+1)}$ denotes a matrix of dimension $n_{l}\times (p+1)$ containing the covariates of the sampling units of area $l$ including the intercept.}}  The solution $(\alpha_{0i}^{(t)},\bbeta_{i}^{(t)})$ for $i=1,\dots,m$ can be obtained using an iteratively re-weighted least squares algorithm or the Newton-Raphson algorithm.}
	
	\vskip .2in
	\noindent{\bf Remark 1:} In this section, we assume that the function $\psi_i(r)$ is completely specified.
	% except possibly for an unknown area specific parameter $\tau_i\in \Omega$.  
	For example, 
	$$\psi_i(r)=2\psi(r)\left [\tau_iI(r>0)+(1-\tau_i)I(r\le 0)\right ],\;-\infty<r<\infty,$$ \textcolor{black}{where $\psi(r)$ is a known monotone non-decreasing function with $\psi(-\infty)<\psi(0)<\psi(\infty)$, $\tau_i\in \Omega=(0,1)$ known, and $r$ is a re-scaled residual. Note that the choice $\tau_i=0.5$ would lead to the standard weighted least square estimator of the regression coefficient vector.  The case for unknown $\tau_i$ will be discussed in the next section.}  Two popular choices of $\psi(r)$ are $\psi(r)=\mbox{sign}(r)$ and $\psi(r)=r$. In small area estimation context, \cite{Cha06} assume that $\psi(r)$ is the Huber influence function with tuning constant $c>0$ \citep{Hub81}. The authors used $c=1.345$ in producing their M-quantile-based (MQ) estimators of the small area means.  An alternative to the Huber function could be the Tukey's bisquare function. In the analysis of repeated measures, \citet{Hig93} noted that if one uses Tukey's bisquare function in robustifying log-likelihood, this function appears in the corresponding estimating equations. 
	
	% Even with this choice, we emphasize that our proposed empirical best predictor of $\theta_i$ is fundamentally different from the MQ estimator of \cite{Cha06} and its various extensions.  

	\vskip .2in
	\noindent{\bf Remark 2:}
	Note that estimates of the components $\bbeta_i$ and $\sigma_{\eps i}^2$ in $\bphi_i$ are subject to high variability if  we use data only from area $i$ to estimate these area specific fixed parameters because of small area specific sample sizes. 
	% The regression coefficients $\bbeta_{i}$ and the function $\psi_i$ rely on area specific parameters $\tau_i$. 
	Here, for known $\tau_i$, we overcome this problem using an algorithm that uses data from all areas.  

	\item[Step 3] \textcolor{black}{Define $\tilde{\br}_{l;i}^{(t)}=(\tilde{\bU}_{l;i}^{(t-1)})^{-1/2}(\by_{l}-\alpha_{0i}^{(t)}\buno_{n_l}-\bX_{l} \bbeta_{i}^{(t)})$ with $\tilde{\bU}_{l;i}^{(t-1)}$ represents a $n_l \times n_l$ diagonal matrix with diagonal elements equal to those of $\tilde{\bV}_{l;i}^{(t-1)}=h_l\sigma_{\gamma}^{2(t-1)}\buno_{n_l}\buno_{n_l}^{\prime} +\sigma_{\epsilon i}^{2(t)} \bK_l $. Obtain $\{ \sigma_{\epsilon i}^{2(t)},\;i=1,\dots,m \}$ as a solution of the following system of estimating equations:
		\begin{eqnarray}\label{score3}
			&& \sum_{l=1}^m\left [\psi_{i}(\tilde{\br}_{l;i}^{(t)})^\prime (\tilde{\bU}_{l;i}^{(t-1)})^{1/2} (\tilde{\bV}_{l;i}^{(t-1)})^{-1} 
			\frac{\partial \bV_{i;i}}{\partial \sigma_{\epsilon i}^2}|_{\sigma_\gamma^2= \sigma_\gamma^{2(t-1)} }
			(\tilde{\bV}_{l;i}^{(t-1)})^{-1}  (\tilde{\bU}_{l;i}^{(t-1)})^{1/2} \psi_i(\tilde{\br}_{l;i}^{(t)}) \right.\nonumber\\
			&&-\left.  w_i\; \mbox{tr} \left\{  (\tilde{\bV}_{l;i}^{(t-1)})^{-1} \frac{\partial \bV_{i;i}}{\partial \sigma_{\epsilon i}^2}|_{\sigma_\gamma^2=\sigma_\gamma^{2(t-1)} }    (\tilde{\bV}_{l;i}^{(t-1)})^{-1}  \frac{\partial \bV_{i;i}}{\partial \sigma_{\epsilon i}^2}|_{\sigma_\gamma^2=\sigma_\gamma^{2(t-1)} }   \right\} \right] = 0 
		\end{eqnarray}
		where $i=1,\dots,m$, ${\bV}_{i;i}=h_i\sigma_{\gamma}^{2}\buno_{n_i}\buno_{n_i}^{\prime} +\sigma_{\epsilon i}^{2} \bK_i $ and $w_i=E[\psi_i^2(u)],$ with $u\sim N(0,1)$.}

	\item[Step 4] \textcolor{black}{After Steps 2 and 3 the estimates of $\bbeta_i$ and $\sigma_{\epsilon_i}^{2}$ are obtained for each small area. Then an estimate of $\beta_0$ at iteration $t$ is obtained as ${\beta}_{0}^{(t)}=\sum_{i=1}^m {\alpha}_{0i}^{(t)}/m$. An alternative method for the estimation of $\beta_0$ can constrain the intercept value to be equal in the estimating equations \eqref{estequbeta} for all small areas. We have tested both the estimation procedures in the simulation experiments and we have not found difference in the estimates. For this reason we propose to adopt the first and much simpler method to estimate $\beta_0$. For the estimation of $\sigma_\gamma^{2}$ compute $\br^{\star (t)}=(\br_1^{\star (t) \prime},\dots,\br_i^{\star (t) \prime},\dots,\br_m^{\star (t) \prime})^{\prime}$, where $\br_i^{\star (t)}=\by_i-{\beta}_{0}^{(t)}\buno_{n_i}-\bX_{i}{\bbeta}_{i}^{(t)}$.  Obtain $\sigma_\gamma^{2(t)}$ as a 
		solution of  the following estimating equation:
		\begin{eqnarray}\label{score3a}
			\nonumber \psi((\bA^{(t)})^{-1/2}\br^{\star (t) })^\prime (\bA^{(t)})^{1/2}(\bG^{(t)})^{-1}\bZ\bZ^{\prime}(\bG^{(t)})^{-1}(\bA^{(t)})^{1/2} \psi((\bA^{(t)})^{-1/2}\br^{\star (t)})\\
			-w^{\star}\; \mbox{tr} \left ((\bG^{(t)})^{-1}\bZ\bZ^{\prime} (\bG^{(t)})^{-1}\bZ \bZ^{\prime}  \right )= 0,
		\end{eqnarray}
		where $\bZ=\mbox{diag}(\buno_{n_i},~i=1,\cdots,m)$ is the incidence matrix of dimension $n \times m$; $w^{\star}=E[\psi^2(u)],$ with $u\sim N(0,1)$; $\bG^{(t)}=\sigma_{\gamma}^{2(t)}  \bZ \bH\bZ^{\prime}+\bR^{(t)}$ of order $n \times n$ with $\bH$ is a diagonal matrix $m \times m$ with diagonal elements equal to $(h_1,\dots,h_i,\dots,h_m)$ and $\bR^{(t)}$ is a diagonal matrix $n \times n$ with diagonal elements equal to 
		\linebreak $(\underbrace{k_{11}\sigma_{\epsilon 1}^{2(t)},\dots, k_{1 n_{1}}\sigma_{\epsilon 1}^{2(t)}}_{n_1},\dots,
		\underbrace{k_{i1}\sigma_{\epsilon i}^{2(t)},\dots,k_{i n_i} \sigma_{\epsilon i}^{2(t)}}_{n_i},\dots,
		\underbrace{k_{m 1}\sigma_{\epsilon m}^{2(t)},\dots, k_{m n_m}\sigma_{\epsilon m}^{2(t)}}_{n_m})$;
		$\bA^{(t)}$ is a diagonal
		matrix with diagonal elements $a_{ij}$ equal to the diagonal elements of the covariance matrix $\bG^{(t)}$.  This estimating equation is proposed following the lines of maximum likelihood proposal II due to \citet{Ric95}.}
	
	% If the tuning constant of the influence function $\psi_i$ goes to infinity equation \eqref{score3a} becomes the traditional ML estimating equation.
	
	%\footnotesize
	%\begin{eqnarray}\label{score3}
	%(\by-\bX \bbeta_{\hat{q}_i^{ELB}})^\prime \bV_i^{-1}\bZ\bZ^{\prime}\bV_i^{-1}(\by-\bX \bbeta_{\hat{q}_i^{ELB}})-tr \left [\bV_i^{-1}\bZ\bZ^{\prime} \bV_i^{-1}\bZ \bZ^{\prime}  \right ] & = & 0\\
	%\label{score4}   (\by-\bX \bbeta_{\hat{q}_i^{ELB}})^\prime \bV_i^{-1}\bI_n \bV_i^{-1}  (\by-\bX \bbeta_{\hat{q}_i^{ELB}})- tr \left [\bV_i^{-1}\bI_n \bV_i^{-1}\bI_n \right ] & = & 0.
	%\end{eqnarray}
	%\normalsize
	%The variance components can be estimates also via restricted maximum likelihood (REML) method or method of moments.
	\item[Step 5] Repeat Steps 2-4 until convergence to obtain the estimated vector $\hat{\bphi}_i$. Convergence is achieved when the difference between the estimated model parameters obtained from two successive iterations is less than a small pre-specified value.
	%\item [Step 8] At convergence, predicted random effects for small area are obtained by solving the following estimating equation with respect to $\bgamma_i$
	%\begin{equation}\label{rre}
	%\bZ^{\prime} \bSigma_{\epsilon i}^{-1}(\by-\bX\bbeta_{\hat{q}_i}-\bZ\bgamma)  -\bSigma_{\gamma i}^{-1} \bgamma=\bzero.
	%\end{equation}
\end{itemize}
\textcolor{black}{Like any other iterative algorithm, the proposed procedure requires initial values for the parameters. As a result, using well-defined starting values for the fixed and variance parameters is advisable for reducing the computational time. Here we suggest to employ a deterministic strategy for initialisation, based on considering the estimates of regression coefficients and variance components obtained by the standard linear mixed model as starting points for all areas. Obviously, this strategy can be substantially improved by adopting a multi-start random initialisation, as the one we have used in the analysis of real data examples (see Section \ref{emap:data}). However, this strategy may significantly increase the global computational burden and, for this reason, it is not employed in this large scale simulation study.
}

For large $m$, under appropriate regularity conditions on the model and known $\psi_i$ function,  consistency of the estimator  $\hat{\bphi}_i$ of $\bphi_i$ can be established in a straightforward way, using Theorem 3.6 of \citet[][pp. 100-101]{jiang2017} or Theorem 4.1 of \citet{Jiang2002}. Assuming that the joint distribution of the observations $(\by_l)_{1\leqslant l \leqslant m}$ (vector-valued) depends on $\bphi_i$, equations \eqref{estequbeta}, \eqref{score3} and \eqref{score3a} can be written as:
\begin{equation}
	F_i(\bphi_i)=\sum_{l=1}^{m}f_l(\bphi_i, \by_l)+h(\bphi_i),
\end{equation}
where $f_l(\bphi_i, \by_l)=(f_{l,k}(\bphi_i, \by_l))_{1\leqslant k \leqslant (p+3)}$ are vector-valued functions such that $E[f_l(\bphi_{0i}, \by_l)]=0$, $l=1,\dots, m$, $\bphi_{0i}$ is the true vector of parameters and $h(\bphi_i)$ is a vector-valued which may depend on the joint distribution of $(\by_l)_{1\leqslant l \leqslant m}$. As pointed out by \citet{Jiang2002} these equations maybe regarded as M-estimating equations and the estimators as M-estimators. Then we can state as it follows:
\begin{mydef}
	Suppose that regularity conditions (i)-(viii) \citep{Jiang2002, Cha14} given in Appendix hold \textcolor{black}{and $\psi_i$ functions are completely specified.}  Then the resulting vector of estimators $\hat\bphi_i$ is consistent.
	%and asymptotically normal.
\end{mydef}
\textbf{Proof} This result is a modified version of that obtained by \citet{Jiang2002} and \citet{jiang2017}.

Substituting $\hat{\bphi}_i$ for  $\bphi_i$ in equation (\ref{BP_prop}), we get empirical best predictor $\hat\theta_i^{EBP}\equiv \hat\theta_i(\hat{\bphi}_i)$ of $\theta_i$.  We now present asymptotic behavior of  the relative savings loss (RSL) of EBP over any direct estimator of $\theta_i$.  The concept of relative savings loss was introduced by Efron and Morris (1973).  In terms of RSL, the following result shows that $\hat\theta_i^{EBP}$ is closer to the optimal best predictor,  $\hat\theta_i^{BP}$, under  model (\ref{eq:mix:new}),  than any direct or synthetic estimator $\tilde{\theta}_i$.\\
\begin{mydef}Under the model (\ref{eq:mix:new}) and mild regularity conditions,
	$$RSL(\hat\theta_i^{EBP}, \tilde\theta_i)=\frac{MSE(\hat\theta_i^{EBP})-MSE(\hat\theta_i^{BP})} {MSE(\tilde\theta_i)-MSE(\hat\theta_i^{BP})}\longrightarrow 0,\;\mbox{as}\; m\rightarrow \infty, $$
	where MSE is the mean squared error under model (\ref{eq:mix:new}).
\end{mydef}
\textbf{Proof} An outline of the proof is as it follows. Using the definition of the best predictor, first note that  $$MSE(\hat\theta_i^{EBP})-MSE(\hat\theta_i^{BP})=E(\hat\theta_i^{EBP}-\hat\theta_i^{BP})^2.$$
	Under model (\ref{eq:mix:new}), regularity conditions and some algebra, we can show that
	$$\hat\theta_i^{EBP}-\hat\theta_i^{BP}\stackrel{P}\longrightarrow 0,\;\mbox{as}\; m\rightarrow \infty.$$
	Moreover, the uniform integrability of  $(\hat\theta_i^{EBP}-\hat\theta_i^{BP})^2$ in $m$ follows since
	$$\mbox{sup}_{m\ge 1}E(\hat\theta_i^{EBP}-\hat\theta_i^{BP})^{2+\delta}<\infty,$$
	for any $\delta>0,$ under model (\ref{eq:mix:new}), regularity conditions and considerable algebra.  Thus,
	$$MSE(\hat\theta_i^{EBP})-MSE(\hat\theta_i^{BP})\longrightarrow 0\;\mbox{as}\; m\rightarrow \infty.$$
	The result now follows because
	$$\mbox{inf}_{m\ge 1}\left [MSE(\tilde\theta_i)-MSE(\hat\theta_i^{BP})\right ]=\mbox{inf}_{m\ge 1}E(\hat\theta_i^{EBP}-\hat\theta_i^{BP})^2>0.$$
	The proof is technical and goes along the lines of \cite{GhoshLahiri87}.  The details are left out to save space, but are available from the authors upon request.

\vskip .2in
\noindent{\bf Remark 3:} Following \cite{Cha06}, one can obtain an MQ estimator of $\theta_i$ under model (\ref{eq:mix:new}) when $\tau_i$ is known.  We note that such an estimator would be a synthetic estimator of $\theta_i$ and so will be less efficient in terms of the above RSL  criterion.  Moreover, such an estimator of $\theta_i$ will not be design-consistent as the area specific sample size grows.  

\vskip .2in
\noindent{\bf Remark 4:} Following \cite{Tza10} \citep[see also equation (18) in][]{Fabrizi:BJ}, one can adjust the MQ estimator in Remark 3 for design-consistency.  Such an estimator would then be a direct estimator and so would be inferior to the EBP in terms of RSL.

\textcolor{black}{\subsection{Estimation of $\tau_i$}}\label{sectaui}
We now present a data-driven method to estimate $\tau_i$. 
For a fine grid $\tau \in \Omega,$ we fit a collection of regression models: 
$$y_{ij}=\beta_{0\tau}+\bx_{ij}'\bbeta_{\tau}+e_{ij},\;i=1,\dots,m;\;j=1,\dots,n_i,$$
using the standard quantile or M-quantile methods \citep{Koe05, Bre88}, 
where $\beta_{0\tau}$ and $\bbeta_\tau$ are fixed intercept and regression coefficients, respectively, and $e_{ij}$'s are standard random errors.

For each observation $y_{ij}$, we find the fitted line with minimum prediction error defined as the difference between $y_{ij}$ and the predicted value by the fitted regression at $\bx_{ij}$.  Let $\hat\tau_{ij}$ denote the value of $\tau$ in the grid for this best line. \textcolor{black}{\citet{Cha06} called $\hat\tau_{ij}$ estimated M-quantile coefficients. Their variability reflects variability at the unit level. If clustering exists, population units in the same cluster (or small area) will have similar M-quantile coefficients and these will be different from those of units that belong to other clusters (or areas). Provided that there are sample observations in area $i$, and a non-informative sampling method has been used to obtain them, an estimate of the area-$i$-specific M-quantile coefficient is the sample average of the estimated M-quantile coefficients for that area, $\bar{\hat\tau}_i=n_i^{-1}\sum_{j\in s_i}\hat\tau_{ij}$.} Since $n_i$ is typically small,  $\bar{\hat\tau}_i$ is likely to be unstable.  We propose the following empirical linear best (ELB) predictor of $\tau_i$ \textcolor{black}{, which improves on $\bar{\hat\tau}_i$ because it uses data from all areas using the following model:}
\begin{enumerate}
	\item[(a)] $E[\hat \tau_{ij}|\tau_i]=\tau_i$, $V[\hat \tau_{ij}|\tau_i]=\nu^2$, 
	\item[(b)] $E[\tau_i]=\mu$, $V[\tau_i]=\eta^2$.
\end{enumerate}
Assuming $E[\tau_i|\hat\btau]=\alpha+\beta \bar{\hat\tau}_i$, the linear best (LB) predictor of $\tau_i$ can be written as
\beq\label{LBq}
\hat\tau_i^{LB}=(1-B_i)\bar{\hat\tau}_i+B_i \mu,
\eeq
where $B_i=\frac{\nu^2/n_i}{\nu^2/n_i+\eta^2}$. Substituting $\mu$, $\eta^2$ and $\nu^2$ by their consistent estimators $\hat\mu$, $\hat\eta^2$ and $\hat\nu^2$, we obtain the following empirical linear best (ELB) predictor:
\beq\label{LBq1}
\hat \tau_i^{ELB}=(1-\hat{B}_i)\bar{\hat\tau}_i+\hat{B}_i \hat{\mu},
\eeq
where $\hat B_i=\frac{\hat\nu^2/n_i}{\hat\nu^2/n_i+\hat\eta^2}$. We refer to \cite{Gho97} for a detailed theory on empirical linear best estimator in the context of  finite population sampling. 

\textcolor{black}{The concept of the proposed data driven estimation of $\tau_i$ for the specific choice of $\psi$ function is grounded on the basis of increasing predictive power of our proposed method.  When $n_i$'s are small, resulting estimators of the model parameters $\bphi$ are not consistent, like the MLE.  In order to achieve consistency, we need large $n_i$ at the minimum.  For the regression context, \citet{bianchi:salvati:2015} considered consistency of similar tuning parameter estimators. In this paper, since $n_i$'s are assumed to be small, we evaluate the performance of our proposed methodology through extensive simulation and demonstrate the utility of our method over existing rival methods.} 

% For the set $E = \{i|n_i = 0\}$ of the out of sample areas, that is areas where $n_i= 0$, consistently with \citet{Cha06} we may define $\hat\theta_i^{EBP-MQ}=\hat\beta_{0}+\bar{\bX}_i \hat\bbeta_{0.5}$, that is a synthetic estimator based on the M-regression.

%Finally, the EBP \eqref{EBP_prop} becomes an estimator that is called hereafter M-quantile Best Predictor:
%\begin{equation}\label{EBP_prop_MQ}
%\tilde\theta_i^{MQBP}=\bar{\bX}_i \hat\bbeta_{\hat{q}_i}+(1-\hat{B}_i)(\bar{y}_i-\bar{\bx}_i ^\prime \hat\bbeta_{\hat{q}_i}),
%\end{equation}
%where $\hat{B}_i=\frac{\hat\sigma_{\eps i}^2/n_i}{\hat\sigma_{\eps i}^2/n_i+\hat\sigma_{\gamma i}^2}$.
%%%%%%%%%%%%%%%%%%%%%%%%%%%%%%%%%%%%%%%%%%%%%%%%%%%%%%%%%%%%%%%%%%%%%%%%%
%%%%%%%%%%%%%%%%%%%%%%%%%%%%%% McJack     %%%%%%%%%%%%%%%%%%%%%%%%%%%%%%
%%%%%%%%%%%%%%%%%%%%%%%%%%%%%%%%%%%%%%%%%%%%%%%%%%%%%%%%%%%%%%%%%%%%%%%%%

\section{Uncertainty measures}\label{sec:mcjk}
{\color{black}{For known $\tau_i$, our estimators of uncertainty measures (not necessarily MSE) tend to the true corresponding uncertainty measures in probability as $m$ tends to infinity, under certain regularity conditions, including the regularity conditions of Proposition 1.  
		For unknown $\tau_i$, such a convergence does not hold unless possibly within area sample sizes $n_i$ are large. We do not pursue this research here because our focus is on bounded $n_i$.  As stated in the concluding remarks, investigation of such asymptotic properties could be an interesting topic for future research.}}

In this section, we discuss the estimation of a general class of uncertainty measures for a small area estimator, say $\hat\theta_i$, not necessarily an empirical best predictor.  An uncertainty measure in this general class is denoted by $f(\mbox{E}[d(\hat \theta_i,\theta_i)])$, where $f(\cdot)$ denotes a known, possibly non-linear, function, $d(\hat \theta_i,\theta_i)$ is a known distance measure between $\hat\theta_i$ and $\theta_i$, and $E$ is the expectation with respect to the assumed model. Examples of such uncertainty measures include commonly used root mean squared error (RMSE) and  relative root mean squared error (RRMSE) defined as
\beq\label{MSE_mcjk}
\mbox{RMSE}[\hat\theta_i]=\sqrt{\mbox{MSE}[\hat\theta_i]},
\eeq
where $\mbox{MSE}[\hat\theta_i]=E[(\hat\theta_i-\theta_i)^2]$, and 
\beq\label{RRMSE_mcjk}
\mbox{RRMSE}[\hat\theta_i]=
\frac{\mbox{RMSE}[\hat\theta_i]}{E[\hat\theta_i]},
\eeq
respectively. As noted in the introduction, a customary estimator of $RMSE$ is a plug-in estimator $\sqrt{\widehat{\mbox{MSE}}}$, where $\widehat{\mbox{MSE}}$ is a second-order unbiased estimator of MSE.  But, a function of an second-order unbiased estimator of MSE does not necessarily yield a second-order unbiased estimator of the corresponding function of MSE. 

We propose a different approach that could potentially justify the use of a much simpler estimator of $f(\mbox{E}[d(\hat \theta_i,\theta_i)])$.  To this end, we propose a more flexible modeling that allows us to combine a large number of small areas so that it suffices to {\color{black}{use a simpler probabilistic convergence criterion.}}  For example, unlike the traditional nested error regression model, we assume that regression coefficients and variance components in the nested error regression model vary across small areas - model \eqref{eq:mix:new}.  This milder model assumption  allows us to include more small areas to be combined than the corresponding traditional nested error regression model and thereby making {\color{black}{the proposed probabilistic convergence criterion}} more reasonable.  

In some cases, it may be possible to apply an analytical method to produce {\color{black}{a reasonable estimator}} of $f(\mbox{E}[d(\hat \theta_i,\theta_i)])$. For example, {\color{black}{when $\tau_i$ is known, an estimator}} of RMSE for EBP, under the assumed model \eqref{eq:mix:new}, is given by $\sqrt{g_{1i}(\hat{\bphi}_i)}$, where
\beq\label{sqrt_g1}
g_{1i}(\bphi_i)=\frac{\sigma_{\gamma}^2\sigma_{\epsilon i}^2/n_i}{\sigma_{\gamma}^2+\sigma_{\epsilon i}^2/n_i},
\eeq 
and $\hat{\bphi}_i$ is a consistent estimator of $\bphi_i$, for large $m$.
Note that a new derivation for such an analytical estimator of $f(\mbox{E}[d(\hat \theta_i,\theta_i)])$ would be necessary as we change  the model, the estimator $\hat\theta_i$, the distance function $d$ or the $f$ function.  This makes such analytical method unattractive to analysts.

We now propose a simple parametric bootstrap method that can be applied to produce {\color{black}{a reasonable estimator}} of $f(\mbox{E}[d(\hat \theta_i,\theta_i)])$.  
First, note that the joint distribution $\{(\theta_i,y_{ij}),\;i=1,\dots,m;\;j=1,\dots,n_i\}$ is known except  possibly for the model parameters $\bphi_i=(\beta_0,\bbeta_{i},\sigma_{\gamma}^2, \sigma_{\eps i}^2)$. Thus, $f(\mbox{E}[d(\hat \theta_i,\theta_i)])\equiv a_i(\bphi_i)$ is a function of $\bphi_i$. 
%  Let
% \beq\label{MSE_mcjk1}
% a_i(\bphi_i)=RMSE[\tilde\theta_i^{EBP-MQ}].
% \eeq

The proposed parametric bootstrap  procedure steps are described below:
\begin{itemize}
	\item[Step 1] Given $\bphi_i$, generate $R$  parametric bootstrap replicates $\{y_{ij}^{(r)},i=1,\dots,m;j=1,\dots,n_i,r=1,\dots,R\}$ using the following model:   
	$$y_{ij}^{(r)}=\beta_0+\bx_{ij}^\prime\bbeta_{i}+\gamma_i^{(r)}+\epsilon_{ij}^{(r)},
	$$ 
	where $\gamma_i^{(r)}|\sigma_{\gamma}^2\sim N(0,h_i {\sigma}_{\gamma}^2)$ and $\epsilon_{ij}^{(r)}|\sigma_{\eps i}^2\sim N(0,k_{ij}\sigma_{\eps i}^2)$ are all independently distributed, $i=1,\dots,m;\;j=1,\dots,n_i$.
	\item[Step 2] For each replication $r$, compute the simulated parameter of interest: $\theta_i^{(r)}=\beta_0+\bar\bX'_i\bbeta_{i}+\gamma_{i}^{(r)},\;r=1,\dots,R$.
	\item[Step 3] For each replication $r$, compute $\hat{\bphi}_i^{(r)}$ using the estimation algorithm described in Section \ref{alg_pred}  and compute $\hat{\theta}_i^{(r)}$, which may depend on $\hat{\bphi}_i^{(r)}$  $\;r=1,\dots,R$.
	\item[Step 4] We propose the following parametric bootstrap estimator of $a_i(\bphi_i)$:  $a_{i;boot}\equiv a_{i;boot}(\bphi_i)= f\left (\mbox{E}_{*}[d(\hat \theta_i^{*},\theta_i^{*})]\right )$, where $\mbox{E}_{*}$ is the expectation with respect to the parametric bootstrap distribution.  In practice, we approximate $a_{i;boot}$ by 
	\beq\label{MC_step}
	a_{i;boot}\approx f\left (\frac{1}{R} \sum_{r=1}^{R}  d(\hat \theta_i^{(r)},\theta_i^{(r)})\right ).
	\eeq
\end{itemize}
{\color{black}{For known $\tau_i$,  $a_{i;boot}$ converges in probability to $a_i(\bphi_i)$ as $m$ tends to infinity.  This can be proved by first noting that $a_{i;boot}$ converges in probability to $a_i(\hat{\bphi}_i)$ as $m\rightarrow \infty$ }}and then applying the Taylor series expansion of $a_i(\hat{\bphi}_i)$ around $\bphi_i$ and consistency of $\hat{\bphi}_i$ as an estimator of $\bphi_i$.

{\color{black}{For known $\tau_i$,}} if $f$ is a smooth function well-defined in the real line (e.g., logarithmic function), it is possible to correct for the bias of $a_{i;boot}(\bphi_i)$ by a jackknife method and produce a second-order unbiased estimator of $f(\mbox{E}[d(\hat \theta_i,\theta_i)])$.    This is essentially a simple extension of the  Monte-Carlo jackknife (hereafter McJack) method, proposed by \citet{Jiang2018}, to a general class of uncertainty measures.

To elaborate the McJack method, suppose that $\hat{\bphi}_i$ is an estimator of $\bphi_i$ obtained using the procedure described in Section \ref{alg_pred}. Let $\hat\bphi_{i;-l}$ be the estimated parameters $\hat\bphi_i$ by deleting the $l$th area data set from the full data set. Then the McJack estimator of $f(\mbox{E}[d(\hat \theta_i,\theta_i)])$ is given by
\beq\label{mse_mcjk}
a_{i;mcjack}={a}_{i;boot}(\hat \bphi_i)-\frac{m-1}{m}\sum_{\ell=1}^{m}\{{a}_{i;boot}(\hat \bphi_{i;-l})-{a}_{i;boot}(\hat \bphi_i) \}.
\eeq
Note that {\color{black}{for known $\tau_i$,}} under appropriate regularity conditions,  $a_{i;mcjack}$ is a second-order unbiased estimator of $f(\mbox{E}[d(\hat \theta_i,\theta_i)])$ -- proof follows along the lines of \citet{Jiang2018} and is not included in this paper. 

\vskip .2in
\noindent{\bf Remark 5:} An efficient implementation of the estimators proposed is provided in the \texttt{saebpmq} \texttt{R} package, which includes a main function for model fitting, and a variety of auxiliary functions for prediction and for MSE estimation. The package is available from the authors upon request.

%Write the above procedure as a function, say, $\tilde{a}_i(\bphi_i)=mcjack(\bphi_i)$, that compute \eqref{MC_step} for every given $\bphi_i$. As for the jackknife estimator, we consider $q_i=\hat{q}_i^{ELB}$ as known and suppose that $\hat\bphi_i$ is an estimator of $\bphi_i$ by using the procedure proposed in Section \ref{alg_pred}. Let $\hat\bphi_{-\ell}$ be the estimated parameters $\hat\bphi$ by deleting the $\ell$th area data set from the full data set. The McJack estimator of \eqref{MSE_mcjk} is given by
%\beq\label{mse_mcjk}
% \widehat{a_i(\bphi_i)}=\tilde{a}_i(\hat \bphi_i)-\frac{m-1}{m}\sum_{\ell=1}^{m}\{\tilde{a}_i(\hat \bphi_{-\ell})-\tilde{a}_i(\hat \bphi_i) \}.
%\eeq
%\red{NICOLA: As for the jackknife method proposed in Section \ref{sec:jk}, if we need to consider the variability of $\hat{q}_i$, the $\bphi=(\bbeta_{q_i},\sigma_{\gamma i}^2, \sigma_{\eps i}^2,q_i)$ and the The McJack estimator becomes
%\beq\label{mse_mcjk1}
%\widehat{a_i(\bphi)}=\tilde{a}_i(\hat \bphi)-\frac{m-2}{m-1}\sum_{\ell=1,\ell \neq i}^{m}\{\tilde{a}_i(\hat \bphi_{-\ell})-\tilde{a}_i(\hat \bphi) \}.
%\eeq}
\section{Monte Carlo simulation studies}\label{sim:sec}

In this section, we first discuss findings from a model-based simulation to compare  different estimators/predictors of a small area mean and assess different measures of uncertainty of our proposed EBP.  In a model-based simulation, a synthetic population is repeatedly generated using a model and a sample is drawn from each generated population using a probability sampling design.  Model-based relative bias (RB) and relative root mean squared error (RRMSE) of an estimator/predictor of a small area mean are approximated using values of the estimator/predictor of the small area mean and the corresponding known  small area population mean from the replicated samples. Since different models can be used to generate such synthetic populations, model-based simulation has the flexibility for  evaluating different estimators/predictors under different simulation conditions. 

We also devise a design-based simulation experiment to understand the performance of different estimators/predictors of a small area mean in terms of design-based relative bias and relative root mean squared error criteria. In a design-based simulation, a synthetic population is first constructed using a real-life data. Repeated samples are then independently drawn from this synthetic population using a probability sampling design.  Relative bias and relative root mean squared error of an estimator/predictor of a small area mean are approximated using values of the estimator/predictor from different samples and the fixed known small area population mean. Such a design-based simulation is a fair way to compare different estimators/predictors because synthetic population is generated using a real-life data and not using a hypothetical model that may favor one model-based estimator/predictor over the others.

In both model-based and design-based simulations, we used a simple random sampling from each small area population and consider the following estimators of the small area mean: 
\textcolor{black}{\begin{description}
		\item[(A)] Direct estimator (sample mean),
		\item[(B)] Empirical best linear unbiased predictor (EBLUP) under a nested error regression model,
		\item[(C)] M-quantile estimator of \citet{Cha06} (MQ),
		\item[(D)] Empirical best linear unbiased predictor (EBLUP-RS) under a random regression coefficient model \citep{hobza2013},
		\item[(E)] Empirical best linear unbiased predictor (EBLUP-H) under a heteroscedastic nested error regression model \citep{Jiang2012, Kubokawa2016, sugasawa2017},
		\item[(F)] Observed best predictor (OBP) under a nested-error regression model \citep{Jiang2011, Jiang2015}, using area level covariates $\bar{\bX}_i$,
		\item[(G)] The proposed empirical best predictor (EBP) based on the proposed nested error regression model with high dimensional parameter. 
	\end{description}   
}
%the traditional M-quantile estimator with the small area quantiles are computed by using the Linear Best Predictor (MQ-LBP), 
%the REBLUP estimator of \citet{Sin09}, the robust synthetic estimator (SYNTH) and the proposed 
%the M-quantile Best Predictor where the small area quantiles are computed as average of individual quantiles (MQBP-Naive), 

\textcolor{black}{The nested error regression model and the random regression coefficient model are fitted using the REML option of the \texttt{lmer} function \citep{bates2015} in \texttt{R}. The M-quantile linear regression model is fitted using a modified version of the \texttt{rlm} function \citep[][Sect. 8.3]{Venables2002} in \texttt{R} and so uses iteratively re-weighted least squares \citep{Cha06}. An extended version of an \texttt{R} script that includes numerous functions, available from the authors, is used to fit the nested error regression model with high dimensional parameter.}
In this simulation experiment, M-quantile regression models are fitted by setting the value of the tuning constant in the Huber influence function to $c = 1.345$. This value gives $95\%$ efficiency in the normal case while protecting against outliers \citep{Hub81}. \textcolor{black}{We assume that the $\tau_i$ value is unknown to evaluate the performance and the properties of our proposal and of the corresponding MSE estimator when the tuning parameter is estimated.} The parameters of EBP and ${\tau}_i$ are estimated following the algorithm shown in Section \ref{alg_pred} using the Huber influence function $\psi_i$ with tuning constant equal to $1.345$. \textcolor{black}{Estimated model coefficients obtained from these fits are used to compute EBLUP,  EBLUP-RS, MQ and the proposed EBP. In the simulation the performance of EBP based on estimators obtained by maximising the log-likelihood function \eqref{loglikeML} is also investigated (EBP-MLE). %, as well as the EBP based on grouped areas as defined in Section \ref{sectaui}. 
	The small area estimates EBLUP-H are obtained using the \texttt{RHNERM} function of the package \texttt{rhnerm} in \texttt{R} performing the heteroscedastic nested error regression model of \citet{Kubokawa2016}. The OBP is computed using an \texttt{R} script developed following the procedure described in \citet{Jiang2015}.}

% \textcolor{blue}{NICOLA: Maybe here we need to write that for OBP we used $\bar{\bX}_i$ instead of $\bar{\bx}_i$ as Jiang et al. have written in JASA and SM paper. Moreover here we need to add that we explore also the EBP grouped, but I don't know where we can add it in the paper.}

\subsection{Model-based simulations}\label{sec:mb:sim}
% M-quantile methods are useful to address some of the issues.  The method implicitly assumes regression coefficients and variance components to vary across small area.  But since the method estimates the area specific regression quantiles using small data from that small area only the regression quantiles could be subject to high variability and hence  highly unstable. Moreover, the method is essentially  synthetic and do not produce the desired design-consistent small area estimation.  Thus, if the underlying model fails, the method could provide estimates that could deviate from the true values considerably even when the area specific sample size for the area is large.

% In this paper, we borrow strength from both empirical best prediction method and M-quantile method in an effort to address shortcomings of both methods.  First, we show how one can incorporate heteroscedasticity of variance components and regression coefficients without going through complex random effects models and estimating all these unknown parameters using data from all areas.  We show that these estimators are all consistent for known area specific regression quantiles.  Unlike the existing M-quantile methods, we then improve on the area specific regression quantiles by borrowing strength from all areas covered by the estimation procedure through an exchangeable model.  Our method is essentially EBP and hence design-consistent.

% We provide model-based simulation results illustrating the performances of the various small area predictors in comparison with the proposal of this paper. 

Population data are generated for \textcolor{black}{$m = 100$} small areas, with samples selected by simple random sampling without replacement within each area. The population and sample sizes are the same for all areas and are fixed at $N_i =100$ and \textcolor{black}{$n_i =4$}, respectively. Values for the auxiliary variable $x$ are generated independently from a common log-normal distribution with a mean of $1.0$ and a standard deviation of $0.5$ on a logarithmic scale that yields to an adjusted $R^2$ of about 0.7. Values for $y$ are generated using the following linear mixed model:
$$y_{ij} = 10+\beta_i x_{ij} + \gamma_i + \epsilon_{ij}~~i=1,\dots,100; j=1,\dots,4,$$
where the slope $\beta_i$, the random-area effects $\{\gamma_i\}$ and sampling errors $\{\epsilon_{ij}\}$ are independently generated according the following three different simulation conditions denoted by:
\begin{description}
	\item[(i) $(0,0)$:] $\beta_i=5$ for all the small areas, $\gamma_i\sim N(0,3)$ and $\epsilon_{ij} \sim N(0,6)$ -- this model is essentially the nested error regression model \eqref{eq:mix} under which EBLUP is developed;
	\item[(ii) $(\beta,0)$:] \textcolor{black}{$\beta_i=5$ for $i=1,\dots,50$ and $\beta_i=-5$ for $i=51,\dots,100$ and it is kept fixed over the simulations,} $\gamma_i\sim N(0,3)$ and $\epsilon_{ij}\sim N(0,6)$ -- this model violates assumptions of the nested error regression model \eqref{eq:mix} because slopes vary across small areas;
	%   \item[$(\beta,0)^a$] $\beta=5$ for areas 1-30 and $\beta=15$ for areas 31-40, $u\sim N(0,3)$ and $\epsilon \sim N(0,6)$;
	\item[(iii) $(\beta,~\sigma_\eps^2)$:] \textcolor{black}{$\beta_i=5$ for $i=1,\dots,50$ and $\beta_i=-5$ for $i=51,\dots,100$ and it is kept fixed over the simulations, $\gamma_i\sim N(0,3)$, $\epsilon_{ij}\sim N(0,\sigma_{\epsilon i}^2)$, $\sigma_{\epsilon i}^2\sim N(6,2)$ for $i=1,\dots,50$ and $\sigma_{\epsilon i}^2\sim N(12,2)$ for $i=51,\dots,100$} -- this model violates assumptions of nested error regression model \eqref{eq:mix} because both slopes and sampling variances vary across small areas.
	%   \item[$(\beta,\sigma^2)^a$] $\beta=5$ for areas 1-30 and $\beta=15$ for areas 31-40, $u\sim N(0,3)$, $\epsilon_{ij}\sim N(0,\sigma_{\epsilon_i}^2)$, $\sigma_{\epsilon_i}^2\sim N(6,2)$.
	%\item[(e,u,0,0)] $\beta=5$, outliers in both area and individual effects, $u \sim \xi_u  N(0,3)+(1-\xi_u)N(0,20)$ and $\epsilon \sim \xi_\epsilon  N(0,6)+(1-\xi_\epsilon)N(0,150)$, where $\xi_u$ and $\xi_\epsilon$ are independently generated Bernoulli random variable with  $Pr(\xi_u=1)= 0.90$ and $Pr(\xi_\epsilon=1)= 0.97$, respectively, i.e. the individual effects are independent draws from a mixture of two normal distributions, with $97\%$ on average drawn from a `well-behaved' $N(0, 6)$ distribution and $3\%$ on average drawn from an outlier $N(0, 150)$ distribution.
	%\item[(e,u,$\beta$,0)] $\beta\sim N(5,4)$ and $u$ and $\epsilon$ generated as in (e,u,0,0);
	%\item[(e,u,$\beta^a$,0)] outliers in both area and individual effects and in $\beta$. In particular $\beta \sim \xi_\beta  N(5,4)+(1-\xi_\beta)N(5,25)$, where $\xi_\beta$ is an independently generated Bernoulli random variable with  $Pr(\xi_\beta=1)= 0.90$.
\end{description}

% Each scenario is independently simulated 1000 times. %\red{or we can keep fixed the population}. 
% For each simulation the population values are generated according to the underlying scenario, a sample is selected in each area and the sample data are then used to compute estimates of each of the actual area means for $y$.

\textcolor{black}{Each scenario is independently simulated $T=1,000$ times.} The performance of the estimators (A)-(G), under the above three simulation conditions (i)-(iii), is assessed using the following three criteria:
\begin{description}
	\item[(a)] Median absolute relative bias (ARB), median being taken over all 100 small areas; for a given area, ARB of an estimator is defined as the ratio of the absolute value of the average difference between the estimate and the corresponding true simulated small area mean to the average true simulated small area mean, average being taken over simulations;
	\item[(b)] Relative root mean squared error (RRMSE) is defined as the ratio of the square root of the average  squared difference between the estimate and the corresponding true simulated small area mean to the average true simulated small area mean, average being taken over simulations;
	\item[(c)]  Efficiency (EFF) measured as the ratio of the RMSE of each estimator/predictor to the RMSE of the corresponding EBLUP estimator.
	%\item[(d)]  Relative savings loss (RSL) of EBP over any other estimator of $\theta_i$.
\end{description}

\textcolor{black}{Figure 2 compares performances of our proposed algorithm and ML in estimating regression coefficients and variance components of our proposed general nested error regression model with high dimensional parameter (\ref{eq:mix:new}). In estimating regression coefficients, neither method exhibits any clear sign of bias though MLEs of the slope tend to be more variable than our proposed method under  all three simulation conditions.  The two methods differ substantially in estimating  $\sigma_{\eps i}^2$. The box-plots of MLEs  exhibit downward bias (around $18\%-20\%$ in each scenario), suggesting possible inconsistency of MLEs, whereas the box-plots of the estimates obtained with the proposed method exhibit lower bias ($-1.1\%$, $-2.5\%$, $-6.2\%$ in scenarios $(0,0)$, $(\beta,0)$ and $(0,\sigma_\eps^2)$, respectively).  Moreover, MLEs are generally much more variable than the estimates of the proposed method based on GEE.} 

\begin{figure}[h]
	\centering   
	\includegraphics[scale = 0.20]{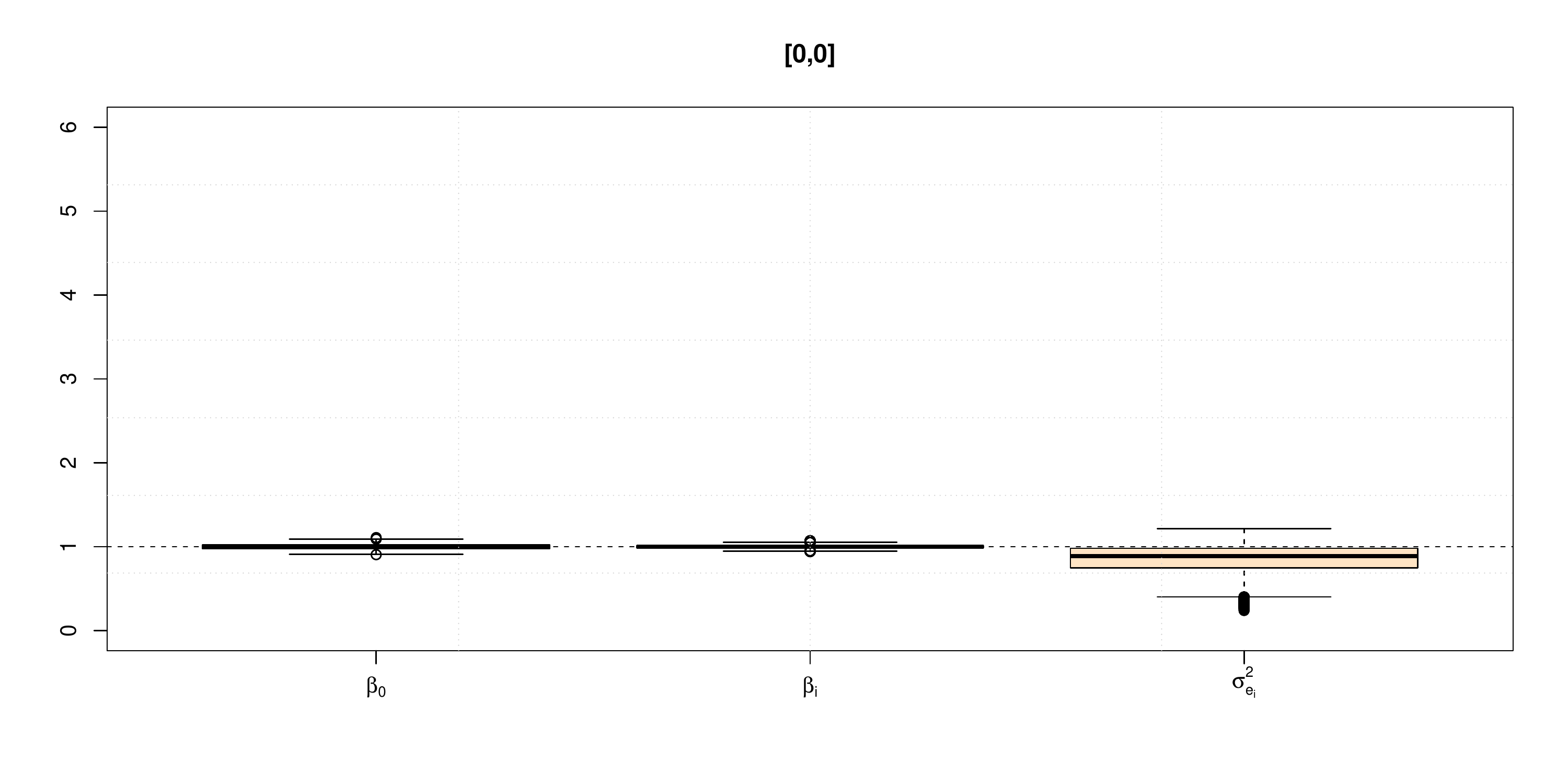} \includegraphics[scale = 0.20]{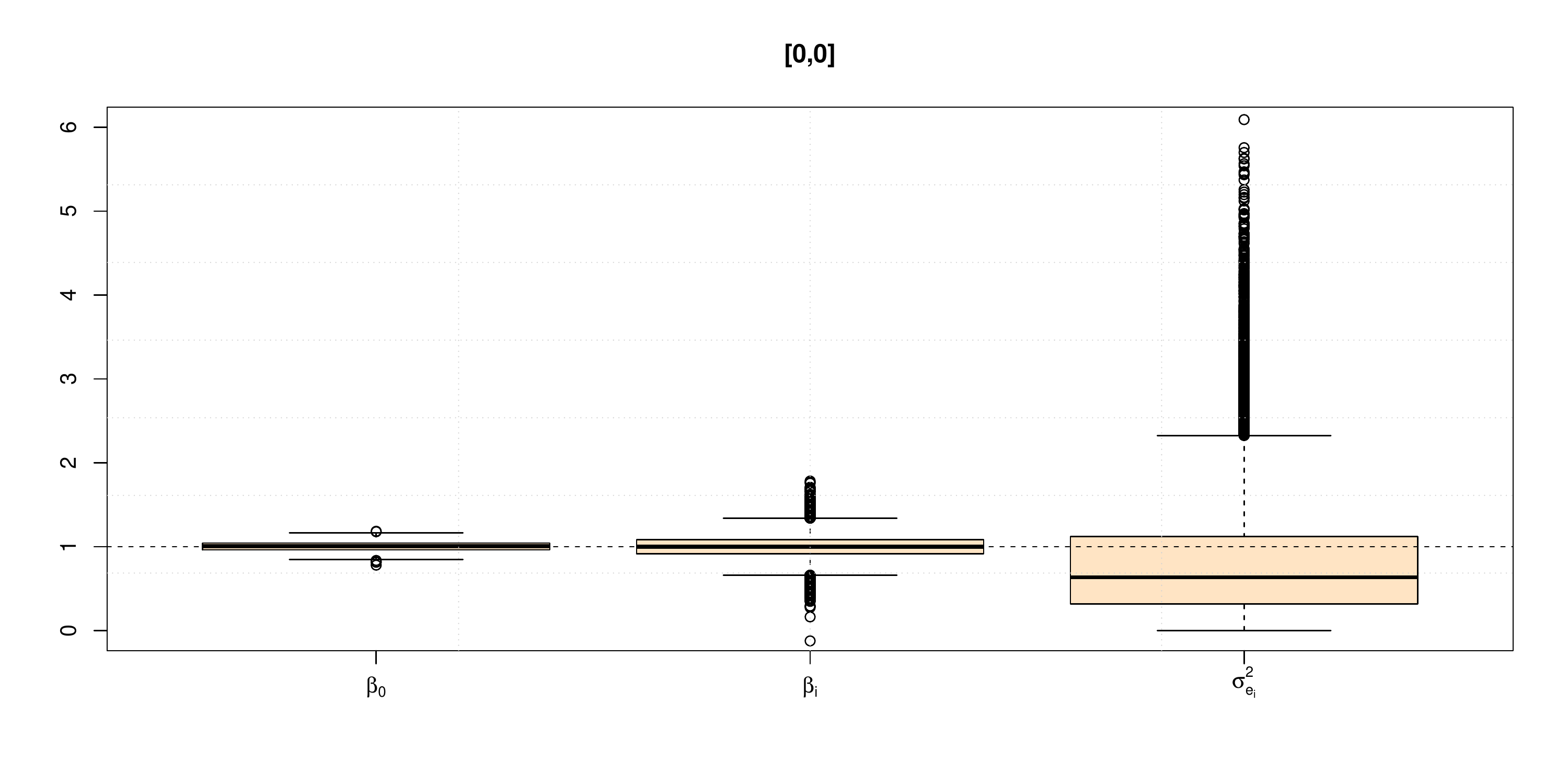} \\
	\includegraphics[scale = 0.20]{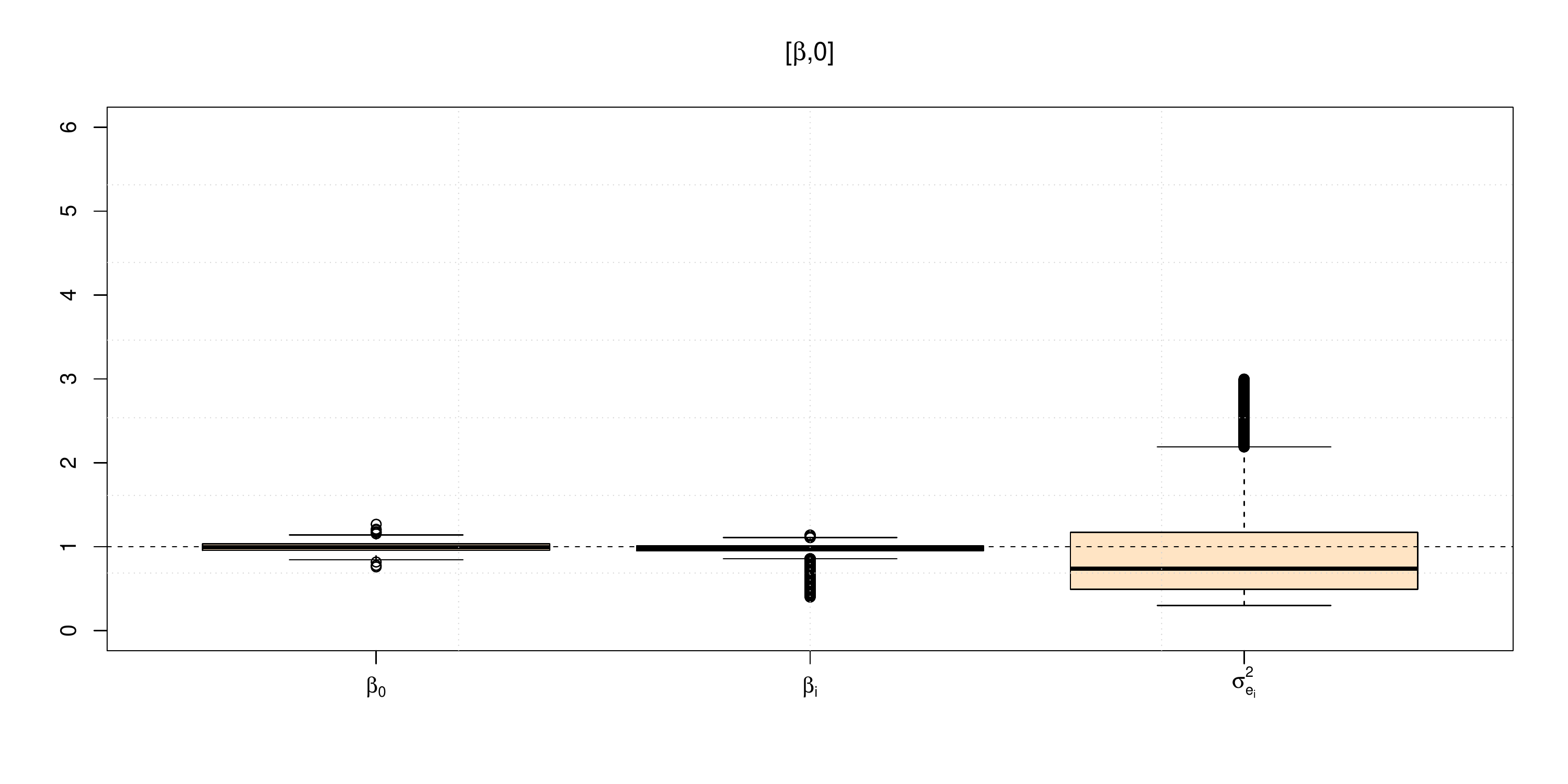}
	\includegraphics[scale = 0.20]{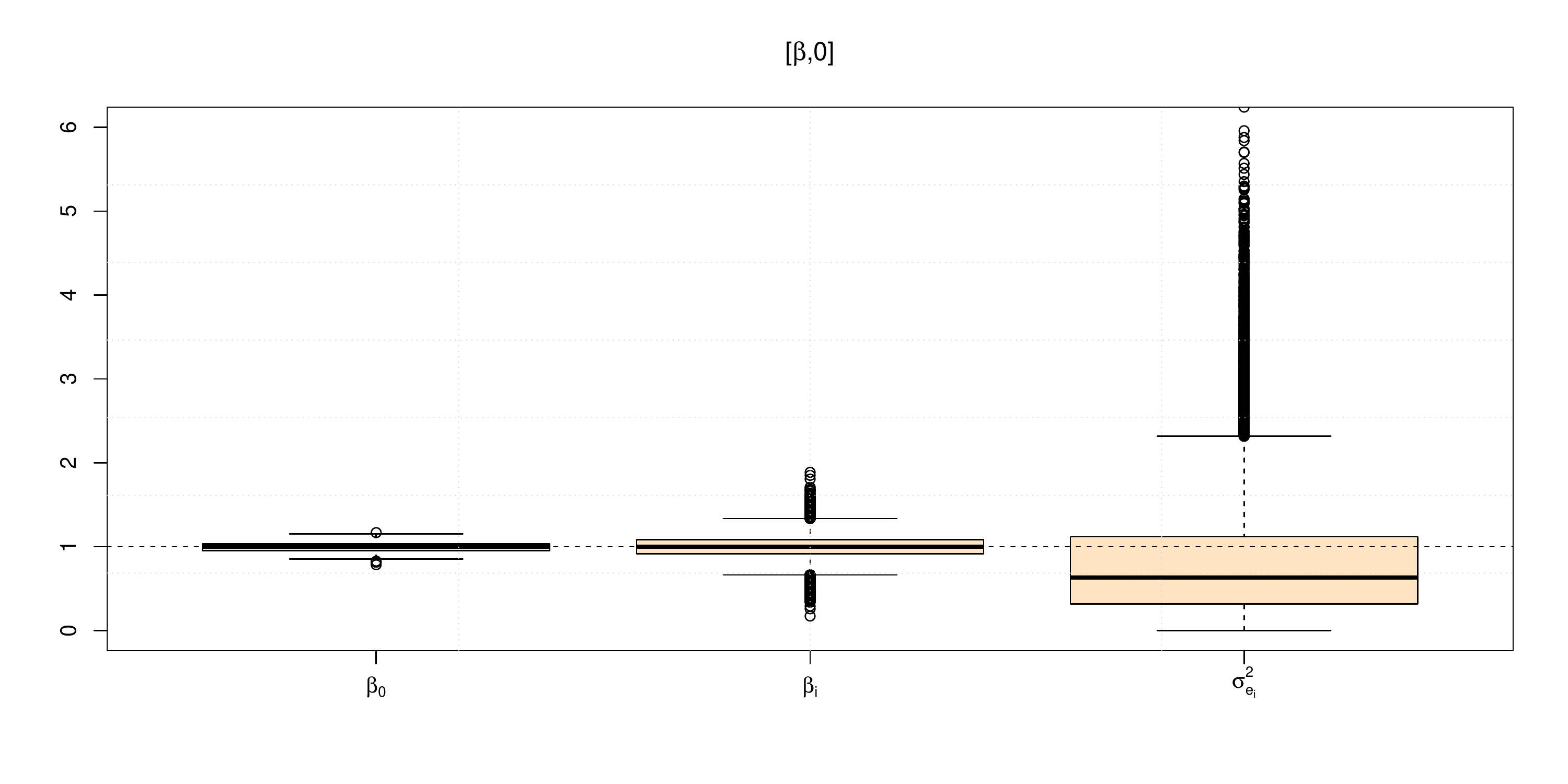} \\
	\includegraphics[scale = 0.20]{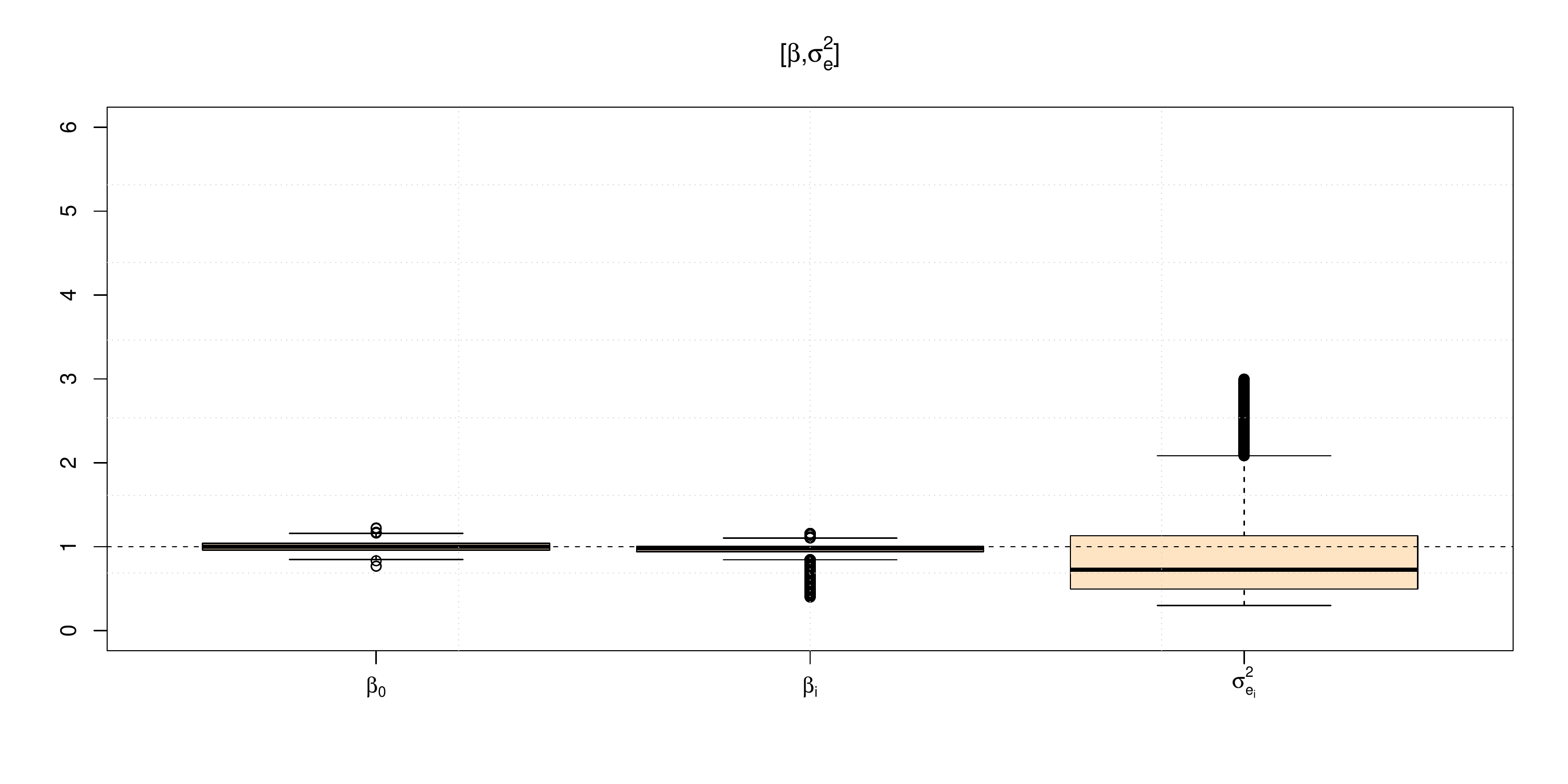}
	\includegraphics[scale = 0.20]{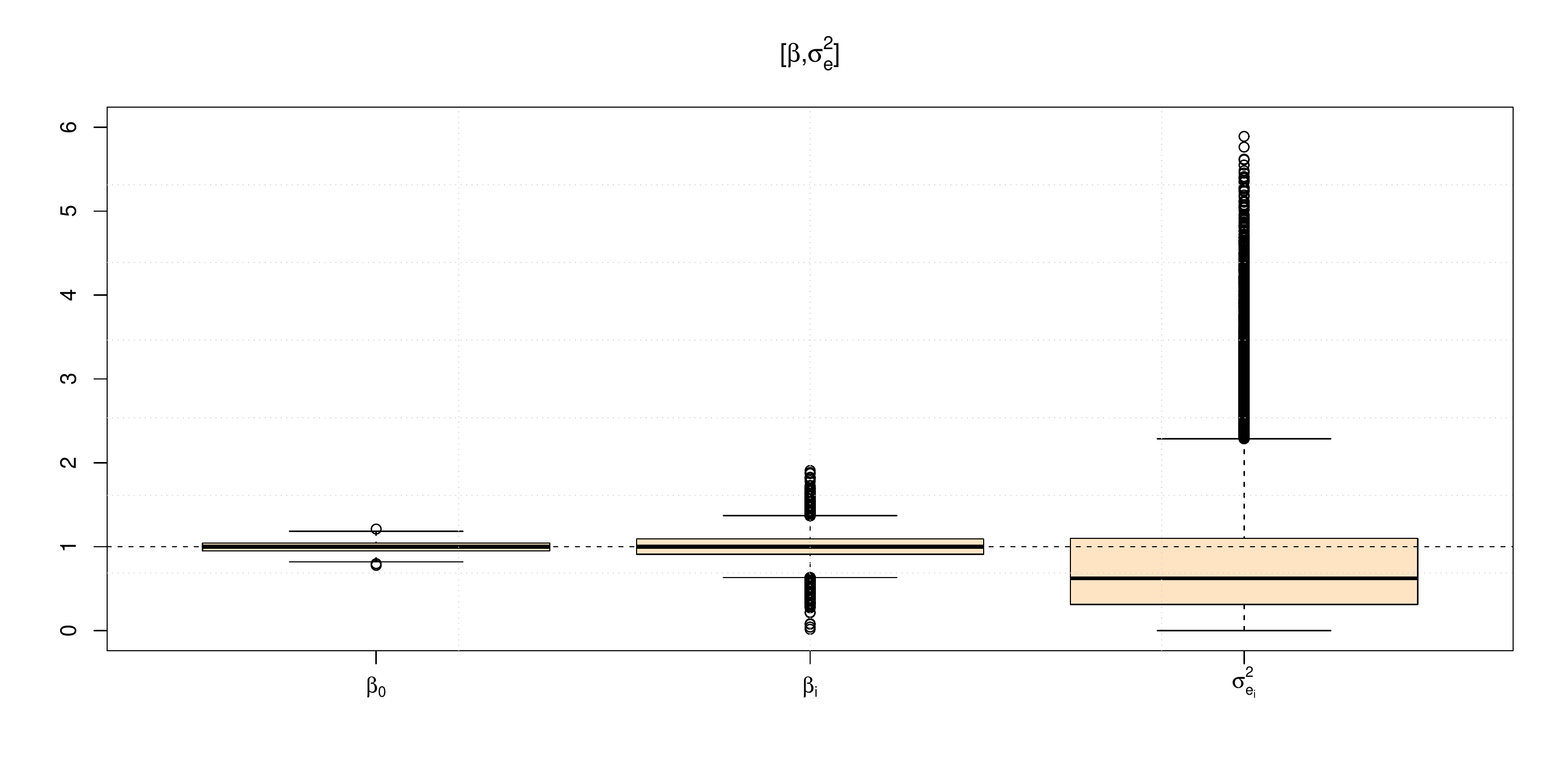} \\
	\caption{\label{Res_reg_coef_vc} Box-plots displaying ratios of estimates of regression coefficients and area specific sampling variances to their corresponding true values under repeated sampling in our model-based simulation experiment; left and right panels use the proposed algorithm and ML method, respectively; different rows of the panel graph correspond to different simulation scenarios. }
\end{figure}

\begin{figure}[h]
	\centering    
	\includegraphics[scale = 0.20]{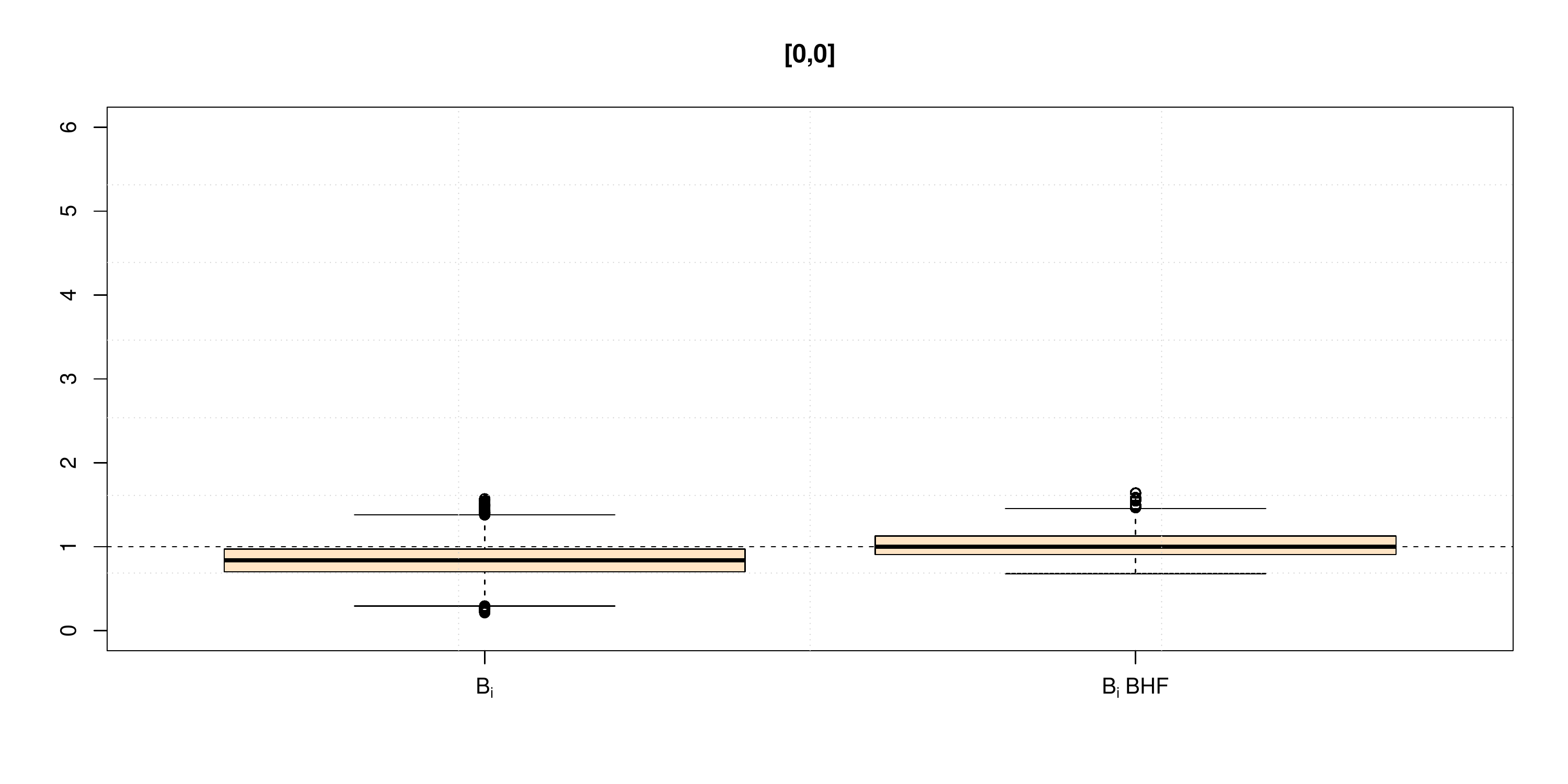} \includegraphics[scale = 0.20]{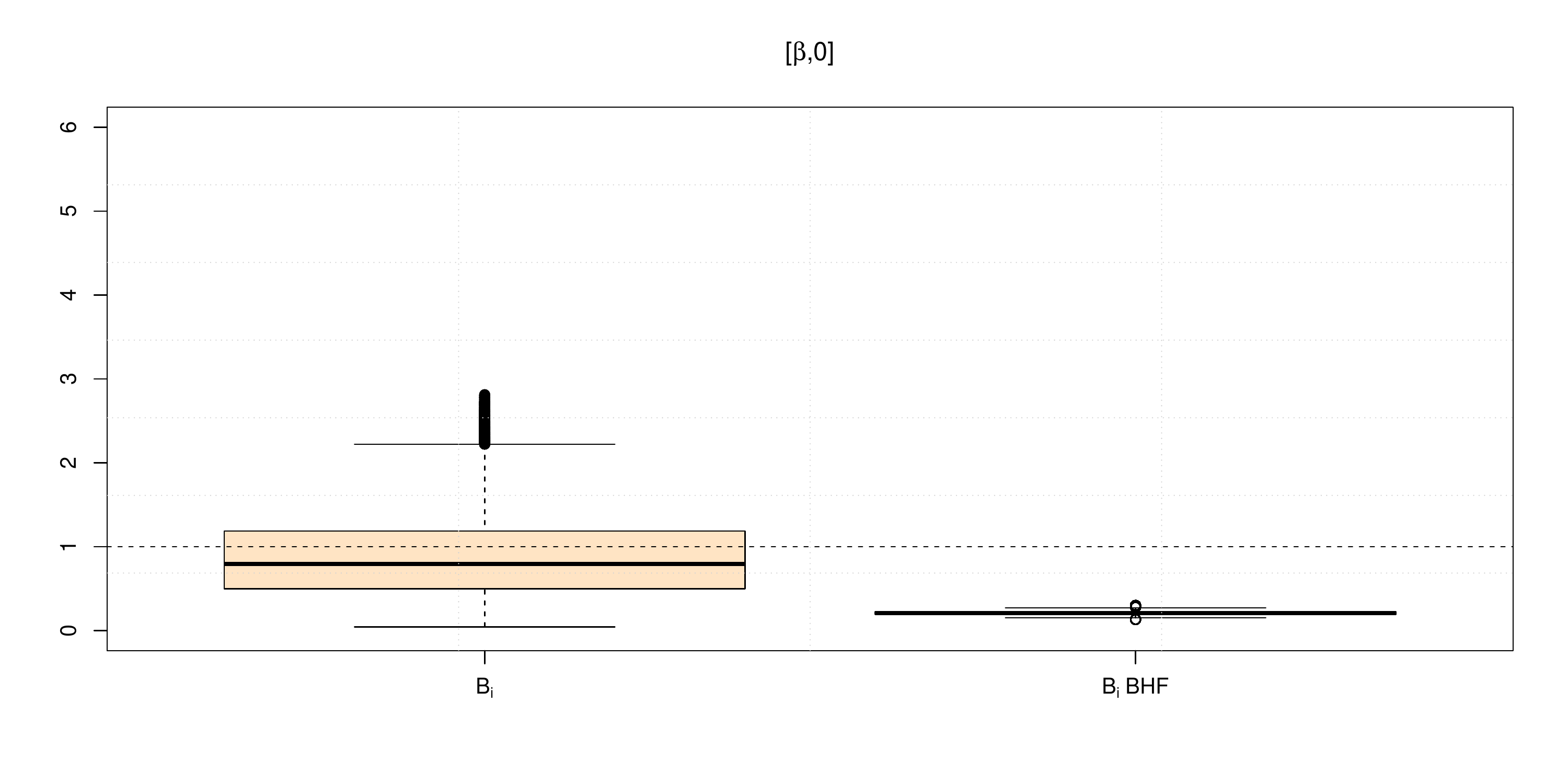} \\
	\includegraphics[scale = 0.20]{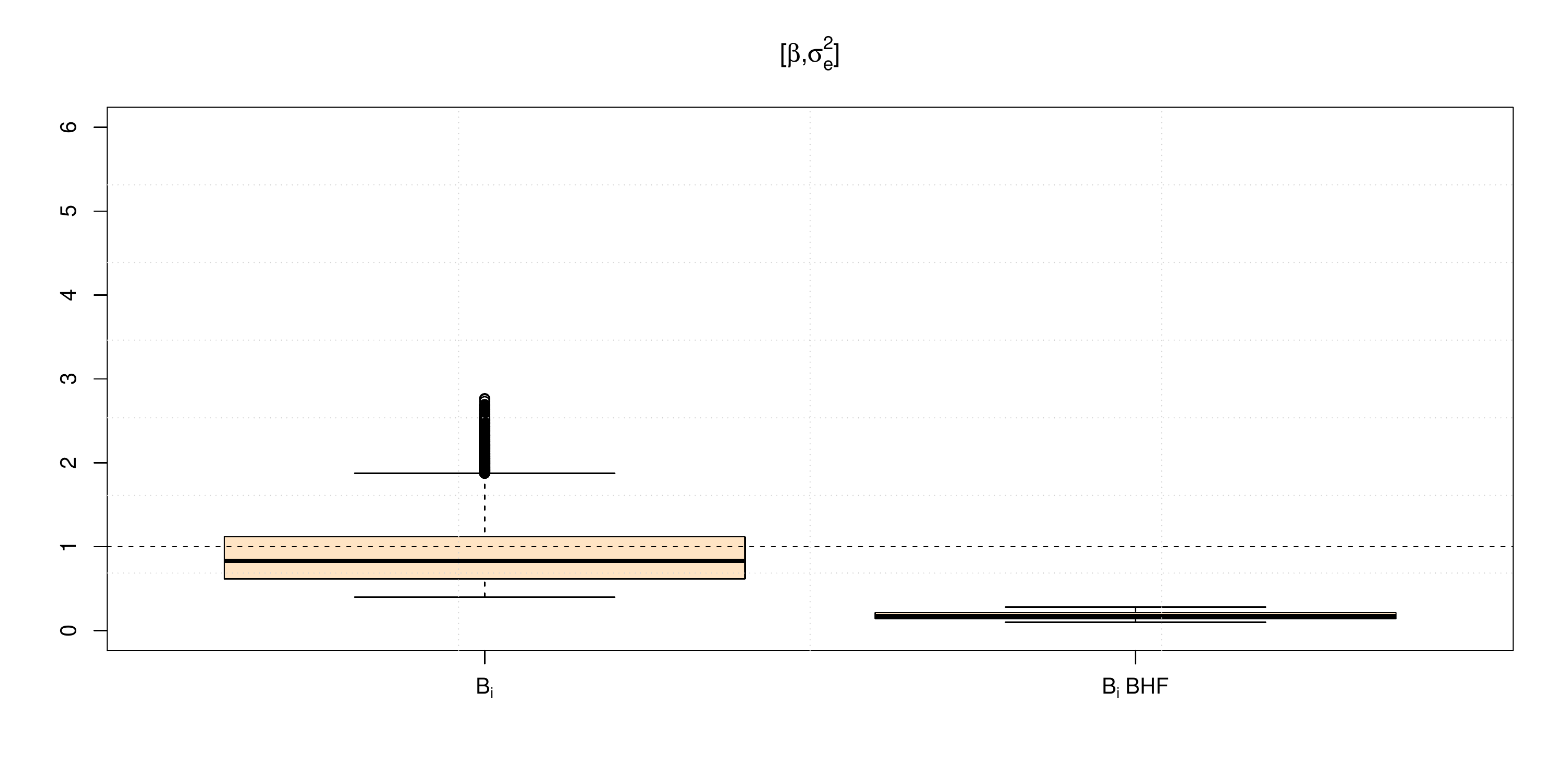} %\\ \includegraphics[scale = 0.20]{EBP_MQ_Bi_bsigma_Jiongo.pdf} 
	\caption{\label{Res_Bi} Boxplots displaying ratios of estimates of $B_i$ to their corresponding true values under repeated sampling in our  model-based simulation experiment; three graphs correspond to the three simulation 
		scenarios; in each graph, estimates of nested regression model with high dimensional parameter and nested error regression model of \citep{Bat88} are denoted by   $B_i$ and $B_i BHF$, respectively.}
\end{figure}

\textcolor{black}{We now study the effects of violation of the exchangeability  of the regression coefficients and sampling variance components across areas in the nested error regression model of \citet{Bat88} on the REML estimates of the shrinkage factors $B_i$.   For different simulation conditions, Figure \ref{Res_Bi} compares proposed estimates of $B_i$, under our nested error regression model with high dimensional parameter (\ref{eq:mix:new}), with the REML estimates of $B_i$ under the nested error regression model of \citep{Bat88}.
	% shows the behaviour of $\hat{B}_i$ for EBP and EBLUP under traditional mixed effects model (indicated as $B_i$ BHF in the plots). The figure shows the ratio of the estimated $B_i$ to the true values. The values of $\hat{B}_i$ for EBP have been computed using the estimation procedure of $\sigma_\gamma^2$ and $\sigma_{\eps i}^2$ proposed in the paper. 
	Under scenario  $(0,0)$, i.e., when the nested error model of \citet{Bat88} is indeed the correct model, the REML estimates of $B_i$, obtained under the correct model, perform slightly better than our proposed estimates under the nested error model with high dimensional parameter.} 
% shrinkage factor estimated under \citet{Bat88} is unbiased and is less variable of the $B_i$ estimated under the high dimensional model. 
\textcolor{black}{When the regression coefficients or/and the sampling variances vary across the areas, the estimates of $B_i$ under the nested error model with high dimensional parameter exhibit considerably less bias than the REML under the nested error model at the expense of more variability.} 

Table \ref{table_scen12} reports the median values of the ARB, RRMSE and EFF for the various simulation scenarios and estimators. \textcolor{black}{The proposed EBP exhibits the least ARB among all predictors for scenarios $(\beta,0)$ and $(\beta,\sigma_\eps^2)$. As expected, for simulation scenario $(0,0)$,{\color{black}{ i.e. for the nested error regression model}}, the EBLUP performs the best in terms of RRMSE. In contrast, when either slopes or both slopes and the sampling variances vary across small areas we note that the proposed EBP performs much better than other predictors in terms of RRMSE. {\color{black}{If we compare the EBP with EBP-MLE, the EBP shows the best performance in terms of RRMSE whereas the EBP-MLE exhibits lowest bias values. In the second and third scenarios, the bias and variability of the OBP  are similar to those of EBLUP under the nested error regression model.  In our simulation experiment, we observe poor performance of the OBP with unit level covariates, as given in \citet{Jiang2011} and \citet{Jiang2015},  unless the within area sample sizes are large.  However, for the second and third scenarios, the bias and variability of the OBP with area level covariates  are similar to those of EBLUP. Thus, in Table \ref{table_scen12}, we only report results of OBP with area level covariates.}}} 

The results presented in Table \ref{table_scen12} are supported by the results from two additional simulation experiments. First, we replicated the experiment for  $n_i=10$ and $m=40$. The results (reported in the Supplementary Material) are similar to those of  Table \ref{table_scen12}, i.e., the proposed EBP exhibits smaller bias and higher efficiency compared to rival small area predictors. As expected, RRMSEs and absolute relative biases for all predictors decrease with the increase of area specific sample sizes.
% \textcolor{cyan}{Table \ref{table_scen12} reports also the median values of the RLS of EBP over Direct, EBLUP and MQ estimator of $\theta_i$. The results show that $\hat\theta_i^{EBP}$ is closer to the optimal best predictor,  $\hat\theta_i^{BP}$, than any other estimator of $\theta_i$ when ....holds.}

% These additional results are presented in Tables \ref{table_scen12_ni5} and \ref{table_scen12_ni20}.

Moreover, in the second additional experiment, for assessing the performance of the proposed EBP in case of outlying values, we have mixed the scenarios of \cite{Cha14} with the ones proposed in this article. The EBP exhibits better performance in terms of both bias and variability compared to the other predictors. The results are reported in Supplementary Material for $n_i=5$ and $m=100$ and for $n_i=10$ and $m=40$.

\begin{table}[h!]
	\caption{\label{table_scen12} Model-based simulation results: performance of estimators/predictors of small area means; the number of small areas considered is $100$; population and sample sizes for each area are $100$ and $4$, respectively; median is over 100 small areas; numbers in parenthesis are the values of the efficiency over EBLUP in terms of RMSE. }
	\centering
	\fbox{
		\scalebox{1.0}{
			\begin{tabular}{lrrr}\hline 
				Predictor &\multicolumn{3}{c}{Results ($\%$) for the following scenarios} \\ 
				& $(0,0)$ & $(\beta,0)$ & $(\beta,\sigma_\eps^2)$ \\\hline
				&\multicolumn{3}{c}{Median absolute relative bias} \\ 
				Direct & 0.535 & 0.927  & 1.083  \\
				EBLUP & 0.132& 9.129  & 9.155  \\
				MQ & 0.127 & 5.070  & 7.315  \\ 
				EBLUP-RS & 0.120 & 0.672  & 0.783  \\ 
				EBLUP-H & 0.130 & 8.876  & 8.903  \\ 
				OBP & 0.222 & 8.691  & 9.074  \\ 
				%REBLUP & 0.087 &0.224  & 0.332 & 0.118  & 0.235 & 0.225\\
				%SYNTH & 0.254 & 0.700 & 0.927 & 0.230 & 1.214 & 1.259  \\
				EBP & 0.136 & 0.634 & 0.671    \\
				%MQBP-LBP$^{\psi} $& 0.102 & 0.164 & 0.238 & 0.144  & 0.237 & 0.225  \\
				%MQBP-LBP$^{\psi, Rob} $& 0.106 & 0.165 & 0.201 & 0.147  & 0.251 & 0.220  \\
				%MQBP-Naive & 0.123 & 0.170 &0.178 & 0.152 & 0.206 & 0.204 \\
				%MQ-LBP & 0.138 & 0.328& 0.315& 0.186 & 0.336 & 0.403 \\
				EBP-MLE & 0.183 & 0.232  & 0.271  \\ 
				%EBP-G & 0.128 & 0.509  & 0.643  \\ 
				&\multicolumn{3}{c}{Median  RRMSE} \\ 
				Direct  &16.640 (17.887) & 44.259 (1.074)  & 45.770 (1.083)  \\
				EBLUP   & 3.922 (1.000) & 43.119 (1.000) & 44.188 (1.000)\\
				MQ & 4.105 (1.103) & 14.774 (0.132) & 20.101 (0.186)   \\ 
				EBLUP-RS   & 3.931 (1.006) & 13.991 (0.103) & 18.283 (0.144)\\
				EBLUP-H   & 3.924 (1.003) & 50.139 (1.359) & 53.093 (1.291)\\
				OBP   & 6.969 (3.152) & 43.273 (1.001) & 44.287 (1.002)\\
				%REBLUP  &6.206 (1.034)  & 11.680 (1.065)  & 14.317 (1.023) & 6.677 (0.870) & 12.404 (0.990) & 14.362 (1.018) \\
				%SYNTH &11.252 (1.871) & 41.197 (3.736)  & 41.010 (2.924) & 14.023 (1.831) & 42.099 (3.368) & 51.205 ( 3.643) \\
				EBP &4.002 (1.047) & 12.065 (0.087)  &  15.596 (0.118)  \\
				%MQBP-LBP$^{\psi} $ &6.126 (1.018) & 8.388 (0.758)  & 10.947 (0.798) & 7.508 (0.979)  & 10.276 (0.829) & 11.071 (0.788)\\
				%MQBP-LBP$^{\psi, Rob} $ &6.111 (1.017) & 8.383 (0.757)  & 11.457 (0.811) & 7.692 (1.004)  & 10.328 (0.831) & 11.052 (0.787)\\
				%MQBP-Naive & 6.483 (1.073)& 8.260 (0.752)  & 10.881 (0.790) & 7.798 (1.016) & 10.229 (0.823) & 10.888 (0.774)\\
				%MQ-LBP &7.218 (1.198)& 14.311 (1.302) & 14.799 (1.062) & 8.916 (1.169) & 15.704 (1.250) & 20.126 (1.406)\\
				EBP-MLE & 5.279 (1.798) & 14.357 (0.108) & 18.685 (0.147)   \\ 
				%EBP-G & 4.092 (1.098) & 10.836 (0.066) & 13.585 (0.083)   \\ 
	\end{tabular}}}
\end{table}

We now examine the performance of our proposed bootstrap and McJack estimators of RMSE of EBP in comparison with that of the naive RMSE estimator that uses $g_{1i}(\hat{ \phi}_i)$ given by  \eqref{sqrt_g1}. \textcolor{black}{The bootstrap and the McJack procedures have been implemented by generating $100$ bootstrap samples in each Monte Carlo run. The data are generated according to  scenarios $(0,0)$, $(\beta,0)$, and $(\beta,\sigma_\eps^2)$. The proposed McJack procedure for the {\color{black}{nested error regression model with high dimensional parameter}} is very computational intensive because for each area $R$ bootstrap iterations are needed. For this reason, to evaluate the performance of the RMSE estimators, we have decided to decrease the number of small areas from $100$ to $40$.} The median values of area-specific relative biases (RB) and relative root mean squared errors (RRMSE) of different estimators of RMSE  are displayed in Table \ref{table_MSE}.  The table also reports the median values of empirical coverage rates (CR) for nominal $95\%$ confidence intervals of small area means, where confidence intervals are constructed using different RMSE estimators. In the case of naive and and McJack RMSE estimators, these confidence intervals are constructed using the small area estimate plus or minus twice the value of the of the corresponding RMSE estimates. The bootstrap confidence intervals are based on the $2.5$ and the $97.5$ percentiles of the corresponding bootstrap distributions. Compared to the naive RMSE estimator, the bootstrap and McJack RMSE estimators are more stable and less bias and provide confidence intervals with coverage rates. The McJack RMSE estimator provides slightly better performance than the bootstrap estimator, but McJack, as stated above, is time consuming.

\begin{table}[h!]
	\caption{\label{table_MSE} Model-based simulation results: performance of RMSE estimators for the proposed EBP of a small area mean;  the number of small areas considered is $40$; population and sample sizes for each area are $100$ and $10$, respectively; median is over 40 small areas.}
	\centering
	\fbox{
		\scalebox{1.00}{
			\begin{tabular}{lrrr}\hline 
				RMSE estimator &\multicolumn{3}{c}{Results ($\%$) for scenarios} \\ 
				& $(0,0)$ & $(\beta,0)$ & $(\beta,\sigma^2)$  \\\hline
				&\multicolumn{3}{c}{Median relative bias} \\ 
				Naive & -14.9 & -23.9  & -25.2 \\
				Bootstrap & -3.6& -8.1  &  -11.7  \\
				%REBLUP & 0.087 &0.224  & 0.332 & 0.118  & 0.235 & 0.225\\
				%SYNTH & 0.254 & 0.700 & 0.927 & 0.230 & 1.214 & 1.259  \\
				McJack & -4.9 &-4.8 & -4.9 \\
				%MQBP-LBP$^{\psi} $& 0.102 & 0.164 & 0.238 & 0.144  & 0.237 & 0.225  \\
				%MQBP-LBP$^{\psi, Rob} $& 0.106 & 0.165 & 0.201 & 0.147  & 0.251 & 0.220  \\
				%MQBP-Naive & 0.123 & 0.170 &0.178 & 0.152 & 0.206 & 0.204 \\
				%MQ-LBP & 0.138 & 0.328& 0.315& 0.186 & 0.336 & 0.403 \\
				&\multicolumn{3}{c}{Median RRMSE} \\ 
				Naive   &18.4 & 30.2  &  32.5  \\
				Bootstrap  &14.5 & 26.4  & 28.7 \\
				%REBLUP  &6.206 (1.034)  & 11.680 (1.065)  & 14.317 (1.023) & 6.677 (0.870) & 12.404 (0.990) & 14.362 (1.018) \\
				%SYNTH &11.252 (1.871) & 41.197 (3.736)  & 41.010 (2.924) & 14.023 (1.831) & 42.099 (3.368) & 51.205 ( 3.643) \\
				McJack  &14.1 & 27.4  &   26.8   \\
				&\multicolumn{3}{c}{Median coverage rate} \\ 
				Naive  &91 &  84   &   84  \\
				Bootstrap  & 94 & 93  & 92  \\
				%REBLUP  &6.206 (1.034)  & 11.680 (1.065)  & 14.317 (1.023) & 6.677 (0.870) & 12.404 (0.990) & 14.362 (1.018) \\
				%SYNTH &11.252 (1.871) & 41.197 (3.736)  & 41.010 (2.924) & 14.023 (1.831) & 42.099 (3.368) & 51.205 ( 3.643) \\
				McJack  &93 & 93 &  94  \\
	\end{tabular}}}
\end{table}

%\begin{figure}[h]
%    \centering    
%	\makebox{\includegraphics[scale = 0.50]{MQ_LB_MSE_MB.pdf} }
%	\caption{\label{MQ_LB_MSE_MB} Boxplots showing area-specific values of the RMSE ratios for the MSE estimators of MQBP-LBP in the model-based
%scenarios (the RMSE ratio is defined as the ratio of the average over repeated sampling of the RMSE estimator
%for a predictor to the actual RMSE of this predictor under repeated sampling).}
%\end{figure}
%
%\begin{figure}[h]
%    \centering    
%	\makebox{\includegraphics[scale = 0.50]{MQ_LB_MSE_MB_1.pdf} }
%	\caption{\label{MQ_LB_MSE_MB_1} Boxplots showing area-specific values of the RMSE ratios for the MSE estimators of MQBP-LBP$^{\psi} $ in the model-based scenarios (the RMSE ratio is defined as the ratio of the average over repeated sampling of the RMSE estimator
%for a predictor to the actual RMSE of this predictor under repeated sampling).}
%\end{figure}

%\begin{figure}[h]
%    \centering    
%	\makebox{\includegraphics[scale=0.40]{MQ_LB_MSE_MCJACK_BOOT_G1_MB.pdf} }
%	\caption{\label{MQ_LB_MSE_MB_2} Boxplots showing area-specific values of the RMSE ratios for the MSE estimators of EBP-MQ in the model-based scenarios (the RMSE ratio is defined as the ratio of the average over repeated sampling of the RMSE estimator
%for a predictor to the actual RMSE of this predictor under repeated sampling). The analytic MSE estimator \eqref{sqrt_g1} has been compared with the bootstrap and the McJack estimators proposed in Section \ref{sec:mcjk}. }
%\end{figure}

\subsection{Design-based simulation}\label{sec:db:sim}
We use data collected in the 1995-1996 Australian Agricultural Grazing Industry Survey (AAGIS), conducted by the Australian Bureau of Agricultural and Resource Economics. In the original sample there were 759 farms from 12 regions (small areas of interest), which make up the wheat-sheep zone for Australian broad-acre agriculture. We use this sample data to generate a synthetic population of size $N = 39,562$ farms by inflating the original AAGIS sample of $n = 759$ farms by farm's sample weight \citep{scc:2012}. In our simulation, we define the 12 regions as small areas of interest.  We know that the proposed \textcolor{black}{nested error regression model with high dimensional parameter} and the EBP can work well with a large number of small areas. In this simulation experiment we are also interested in assessing the performance of the EBP when the number of regions is small, but the traditional assumptions of linear mixed model do not hold (e.g., regression coefficients and the sampling variances could vary across the areas). The outcome variable of interest is the total cash costs (TCC) with the number of closing sheep stock as the auxiliary variable.

Using this design-based simulation, we (a) compare the performance of different predictors of mean TCC in each region under repeated sampling from a fixed population with the same characteristics as the AAGIS sample, and  (b) evaluate the design-consistency properties of the proposed EBP and MQ. In the design-based simulation experiment we also evaluated the performance and the design-consistency property of an alternative MQ estimator of the mean proposed by \citet{Tza10}.  The estimator is based on a smearing argument discussed in \citet{Cha86a}. \citet{Tza10} noted that the estimator is subject to severe bias under the linear M-quantile regression model. The MQ estimator based on a smearing argument (hereafter MQCD) may be written as
\begin{equation}\label{mqcd}
	\hat{\theta}_i^{MQCD}=\bar{y}_i+(\bar{\bX}_i-\bar{\bx}_i)^{\prime}\hat{\bbeta}_i,
\end{equation}
where the vector $\bbeta_i$ is estimated by a method proposed by \citet{Cha06}. It resembles a GREG estimator of the small area mean, which is consistent under the assumption of simple random sampling or some other self-weighting design. Starting from equation \eqref{BP_prop} the EBP can be written as:
\begin{equation}\label{hatBP1}
	\hat{\theta}_i^{EBP}=\bar y_i+(\bar{\bX}_i-\bar{\bx}_i)'\hat{\bbeta}_i-\hat{B}_i (\bar y_i-\hat{\beta}_0-{\bar{\bx}}_i ^\prime \hat{\bbeta}_i).
\end{equation}
We can point out that estimators \eqref{mqcd} and \eqref{hatBP1} are similar. The EBP may be written as MQCD plus a component, $-\hat{B}_i (\bar y_i-\hat{\beta}_0-{\bar{\bx}}_i ^\prime \hat{\bbeta}_i)$, representing a part of the estimated specific-area random effect. For this reason we are interested in this design-based simulation experiment to evaluate the performance of MQCD.

For this simulation, we consider $1,000$ independent stratified random samples, each with regional sample sample size of $n_i=5$. This results in a total sample
size of $60$ locations within the 12 AAGIS regions. The experiment has been replicated with $n_i=10, ~15, ~20,~ 30,~40$ and $50$ for evaluating the design consistency.

% Model diagnostics on the generated synthetic population indicate that the Gaussian assumptions of the mixed model are not met. In particular, the Kolmogorov-Smirnov normality test rejects the null hypothesis that the residuals follow a normal distribution. From the diagnostic there is also evidence of small $R^2$ (around $24\%$) and of a non-stationary process. Using a model that relaxes the traditional normality and stationarity assumptions, such as the varying regression coefficients and random error  variance  model \eqref{eq:mix:new}, therefore seems reasonable for these data.

\begin{figure}[h!]
	\centering    
	\includegraphics[scale = 0.40]{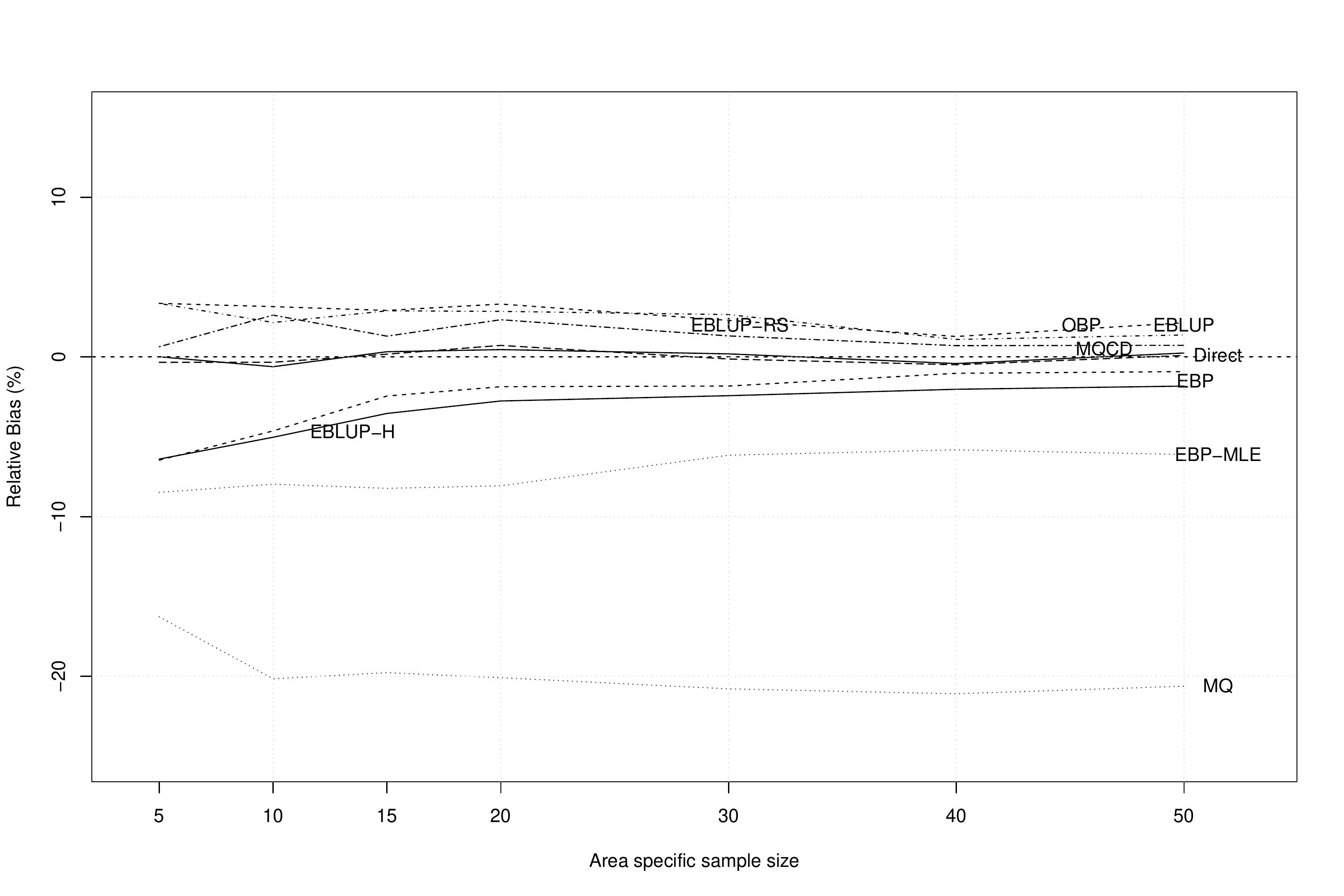}\\
	\includegraphics[scale = 0.40]{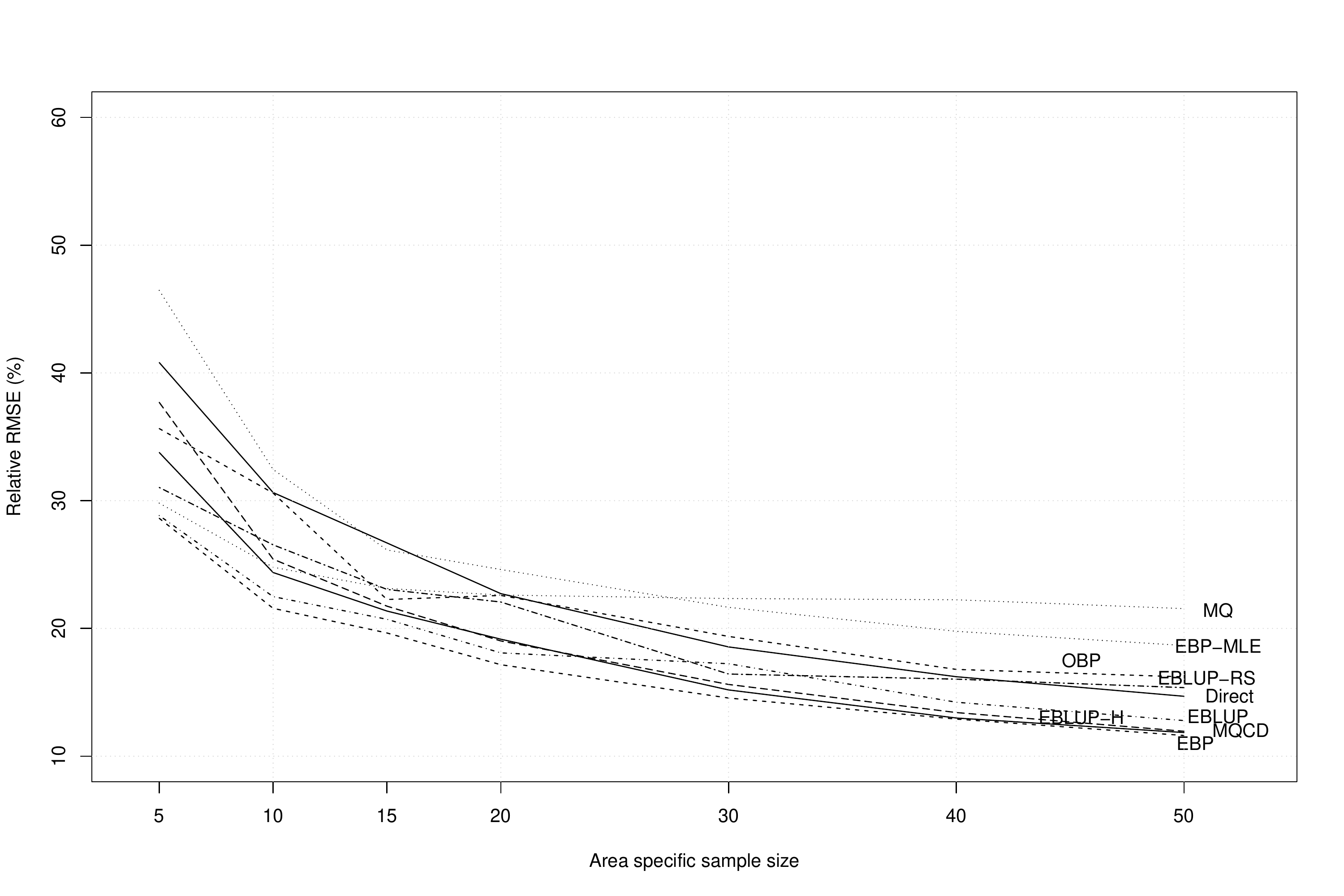}
	\caption{\label{Res_DB} Design-based simulation results: median of relative bias (top panel) and relative root mean squared error (bottom panel) of the small area predictors by area-specific sample size.}
\end{figure}

\begin{figure}[h!]
	\centering    
	\includegraphics[scale = 0.40]{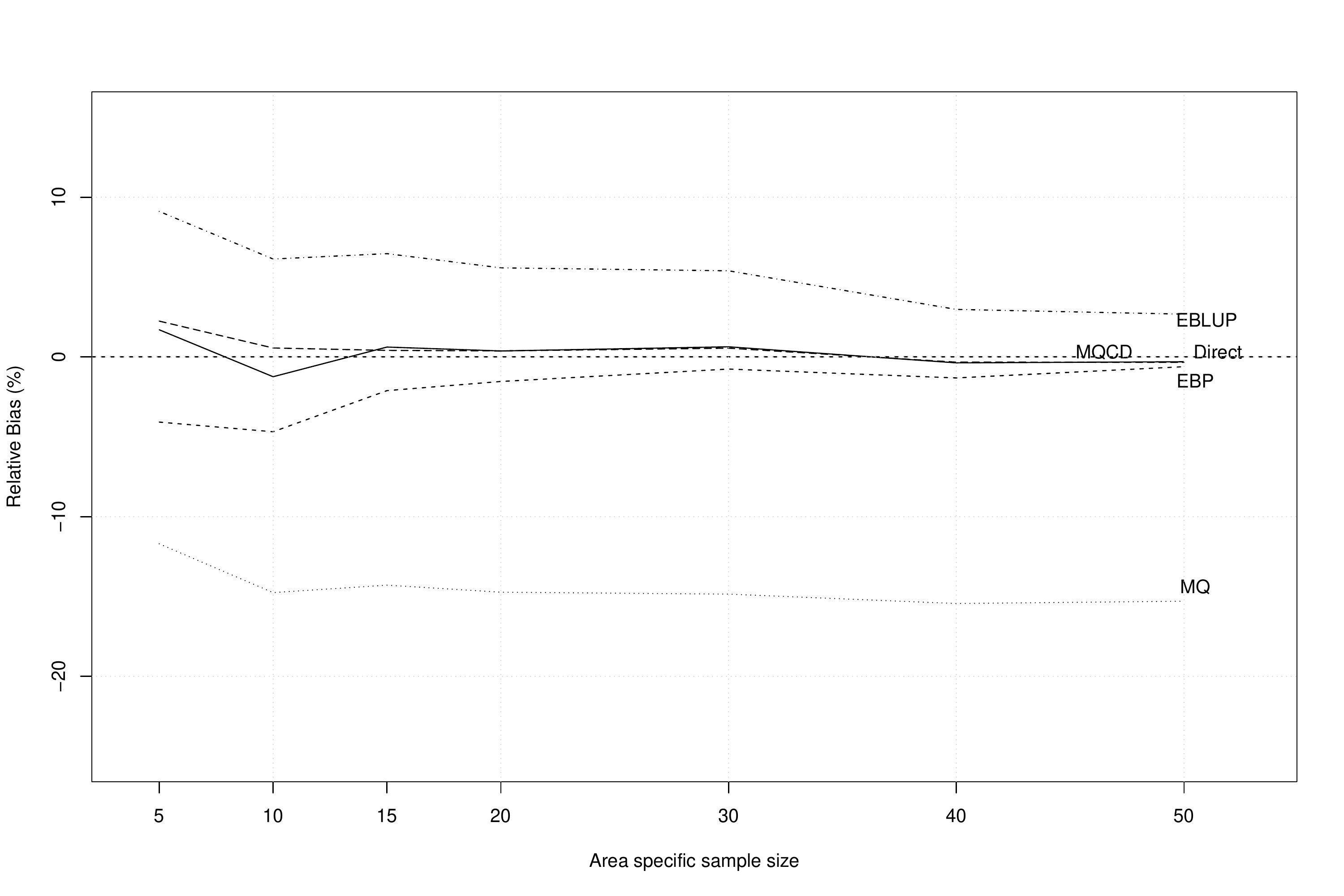}\\
	\includegraphics[scale = 0.40]{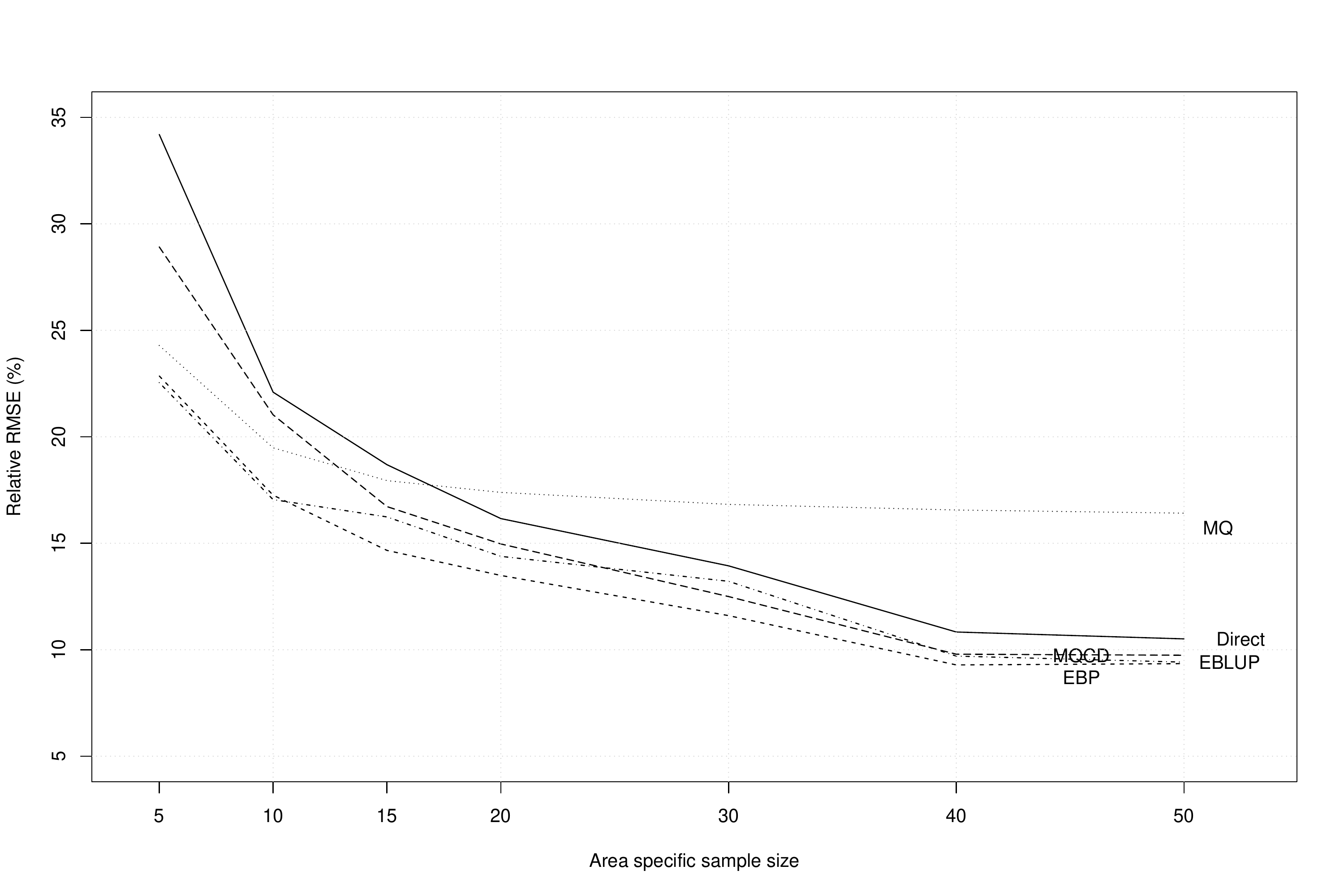}
	\caption{\label{Res_DB_area12} Design-based simulation results for area with smallest $N_i$: relative bias (top panel) and relative root mean squared error (bottom panel) of the small area predictors by area-specific sample size.}
\end{figure}

\textcolor{black}{We compute the relative bias (RB) and the relative root mean squared error (RRMSE) of each estimator/predictor of the mean value of TCC in each region.  Figure \ref{Res_DB} displays  median (over small areas) simulated RB and RRMSE of the estimators/predictors for different area specific sample sizes.  The median bias of EBP is lower than those of MQ, EBLUP, EBLUP-H, OBP, EBP-MLE.} In terms of median RRMSE criterion,  EBP performs the best among all estimators/predictors considered for all sample sizes. EBP outperforms the MQCD in terms of efficiency especially with small sample size. As expected, the difference between RRMSE of EBP and that of MQCD decreases as sample size increases.
% \textcolor{blue}{NICOLA: in my opinion we can leave the second plot as it was in the previous version. We don't need all the predictors. What do you think about it?}
The design consistency property of an estimator is better demonstrated if we focus on Figure \ref{Res_DB_area12}, which displays simulated RB and RRMSE by area-specific sample size for the area with smallest population size ($N_i=1450$).   We note that with the increase of area specific sample size, the simulated RB of EBP (or EBLUP) and MQCD approaches to zero  -- this is in line with their design consistency property.   This is, however, not the case with  the MQ -- even for large area specific sample size, the simulated RB of MQ is significant. For this specific small area, EBP is a winner with respect to RRMSE criterion.

% \begin{figure}[h]
%   \centering    
% 	\includegraphics[scale = 0.40]{AAGIS_estRMSE_g1_mcjakc_boot.pdf}
% 	\caption{\label{Res_DB_RMSE} Design-based simulation results: boxplot showing area-specific values of the RMSE ratios for the analytic $g_1$, MacJack and bootstrap estimators of EBP-MQ (the RMSE ratio is defined as the ratio of the average over repeated sampling of the RMSE estimator
% for a predictor to the actual RMSE of this predictor under repeated sampling). }
% \end{figure}

%%%%%%%%%%%%%%%%%%%%%%%%%%%%%%%%%%%%%%%%%%%%%%%%%%%%%%%%%%%%%%%%%%%%%%%%%
%%%%%%%%%%%%%%%%%% Illustrative application       %%%%%%%%%%%%%%%%%%%%%%%%%%%%%
%%%%%%%%%%%%%%%%%%%%%%%%%%%%%%%%%%%%%%%%%%%%%%%%%%%%%%%%%%%%%%%%%%%%%%%%%
\section{An application of the high dimensional parameter linear mixed  model: EBP estimates of the ecological condition of lakes in the northeastern USA}\label{emap:data}
%As an illustrative example we use the data collected from the US Environmental Protection Agency's Environmental Monitoring and Assessment Program (EMAP) Northeast lakes survey \citep{larsen2001, Ops08, salvati2012}. Between 1991 and 1995, researchers from the US Environmental Protection Agency conducted an environmental health study of the lakes in the north-eastern states of the USA. For this study, a sample of 334 lakes (or more accurately, lake locations) was selected from the population of 21,026 lakes in these states using a random systematic design. The lakes making up this population are grouped into 113 8-digit Hydrologic Unit Codes (HUCs), of which 27 did not contain observations. The HUCs are defined as the small areas of interest, with lakes grouped within HUCs and the variable of interest is the Acid Neutralising Capacity (ANC), an indicator of the acidification risk of water bodies. Factors affecting the ANC such as acid deposition and soil characteristics cut across HUCs, so overall spatial trends are also likely to be useful in predicting the ANC. The EMAP data set contains the elevation and geographical coordinates of the centroid of each lake in the target area. 

\textcolor{black}{We use as illustrative example the data collected from EMAP and presented in Section \ref{sec:motexample}.} Predicted values of average ANC for each HUC are calculated using the empirical version of \eqref{BP_prop} under {\color{black}{the nested error regression model with high dimensional parameter}} \eqref{eq:mix:new} with covariates equal to the elevation of each lake and location defined by the geographical coordinates of the centroid of each lake (in the UTM coordinate system). 

\begin{figure}[h!]
	\centering    
	\includegraphics[scale = 0.60]{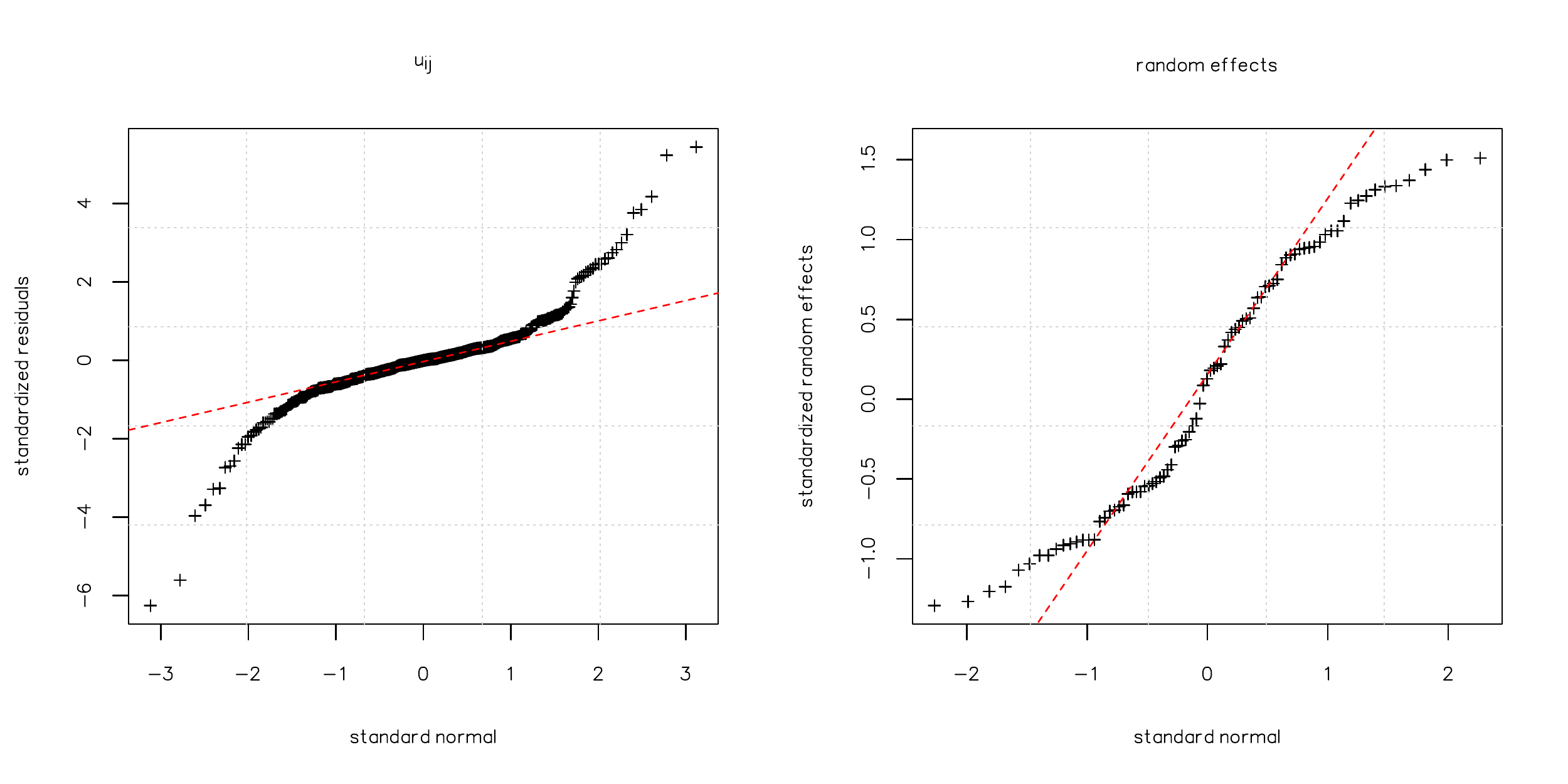} 
	\caption{\label{fig1:app} Normal probability plots of level 1 (left) and level 2 residuals (right) derived by fitting a two level linear mixed model to sample data.}
\end{figure}

Figure \ref{fig1:app} shows normal probability plots of level 1 transformed residuals \citep[$u_{ij}$,][]{Bat88} and level 2 standardized random effects \citep{lange:1989} obtained by fitting a two-level (level 1 is the lake and level 2 is the HUC) linear mixed model to the sample data. The normal probability plots indicate that the Gaussian assumptions of the linear mixed model are not met. This is confirmed by a Shapiro-Wilk normality test, which rejects the null hypothesis that the residuals follow a normal distribution ($p$-values: level 1 = 2.2e-16, level 2 = 0.0006247).

\begin{figure}[h!]
	\centering    
	\includegraphics[scale = 0.60]{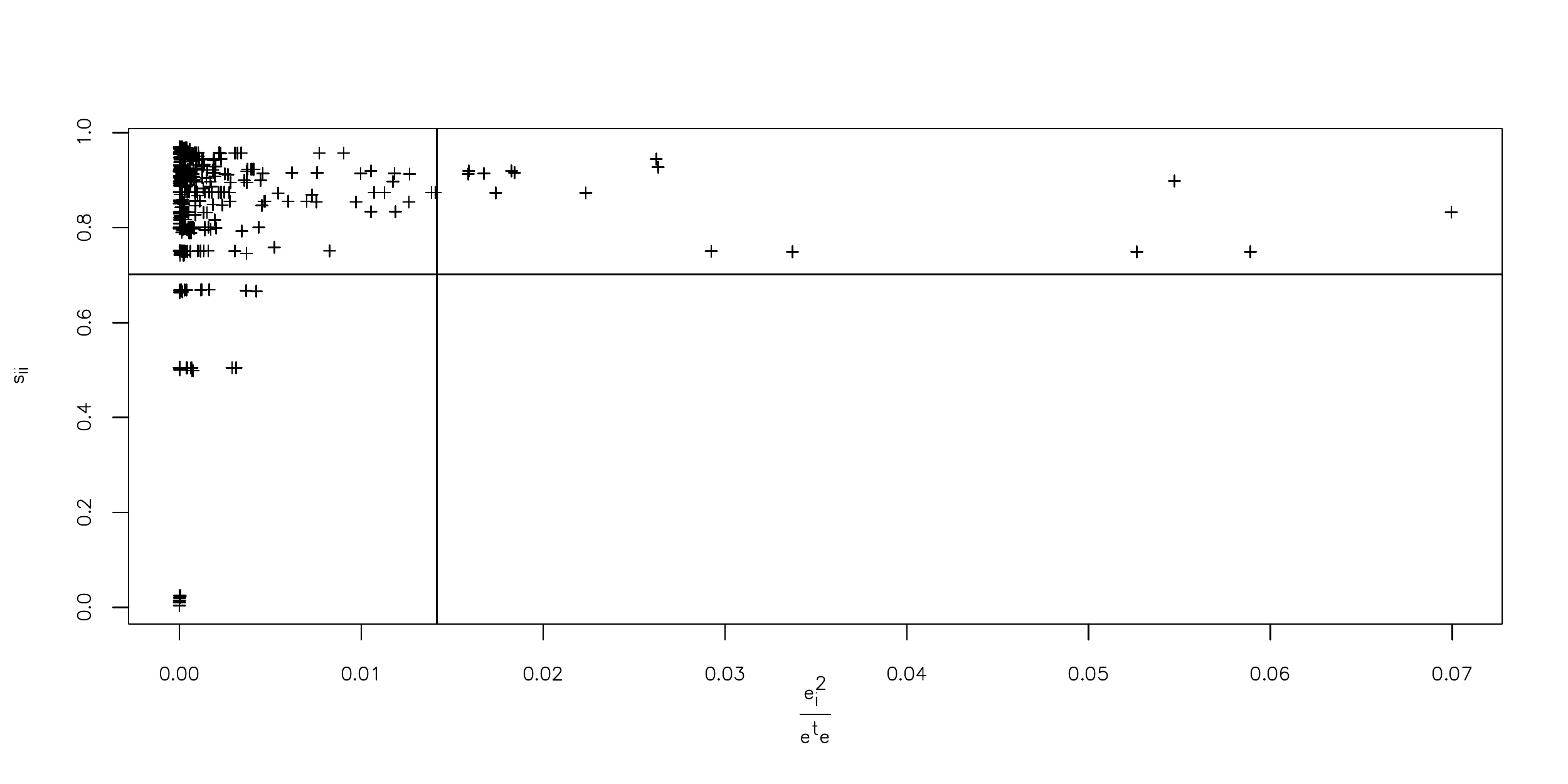}
	\caption{\label{fig2:app} Plot to detect both outliers and high leverage points: $s_{ii}$ versus $\hat{\eps}_i^2/\hat{\beps}^{\prime}\hat{\beps}$.}
\end{figure}

Following \citet{Zewotir2007} we study the detection of outliers and high leverage points in the sample data. The authors propose to use the diagonal elements $s_{ii}$ of the matrix $\bS=\sigma_{\eps}^{2}\bP$, with $\bP=\bV^{-1}-\bV^{-1}\bX(\bX^{\prime}\bV^{-1}\bX)^{-1}\bX^{\prime}\bV^{-1}$ (where $\bX$ is the design matrix and $\bV$ is the covariance matrix of the linear mixed model), to detect high leverage points. To detect both outliers and high leverage points, they propose to examine a plot of $s_{ii}$ versus $\hat{\eps}_i^2/\hat{\beps}^{\prime}\hat{\beps}$, where $\hat{\eps}_i$ is the EBLUP residual. Points are expected to concentrate around the upper-left corner of the plot. Points separated from the main cloud of points that fall in the lower-left corner (small $s_{ii}$) are regarded as high leverage points, and points that appear separated on the right side (large relative squared $\hat{\eps}_i^2/\hat{\beps}^{\prime}\hat{\beps}$) are regarded as outliers. Using in Figure \ref{fig2:app} the rough cutoff values proposed by \citet{Zewotir2007} we note the presence of both outliers and high leverage points in the EMAP data.

These diagnostics suggest the use of a robust model that relaxes the assumptions of normality of the linear mixed model. The model \eqref{eq:mix:new} is fitted on the EMAP data by setting the value of the tuning constant in the Huber influence function to $c = 1.345$. \textcolor{black}{We have studied {\color{black}{ the stability of our algorithm, i.e., we have ensured}} that the convergence of the algorithm does not depend on the starting values. For the data we used in this paper, when initialising the algorithm from different starting values, it always converged to the same point. {\color{black}{However, in some applications, convergence may not be stable  due to sparsity of the data.}}  Hence, users of the method should always test the convergence of the algorithm with their data sets to ensure that there are no convergence problems.} 

\begin{figure}[h!]
	\centering    
	\includegraphics[scale = 0.28]{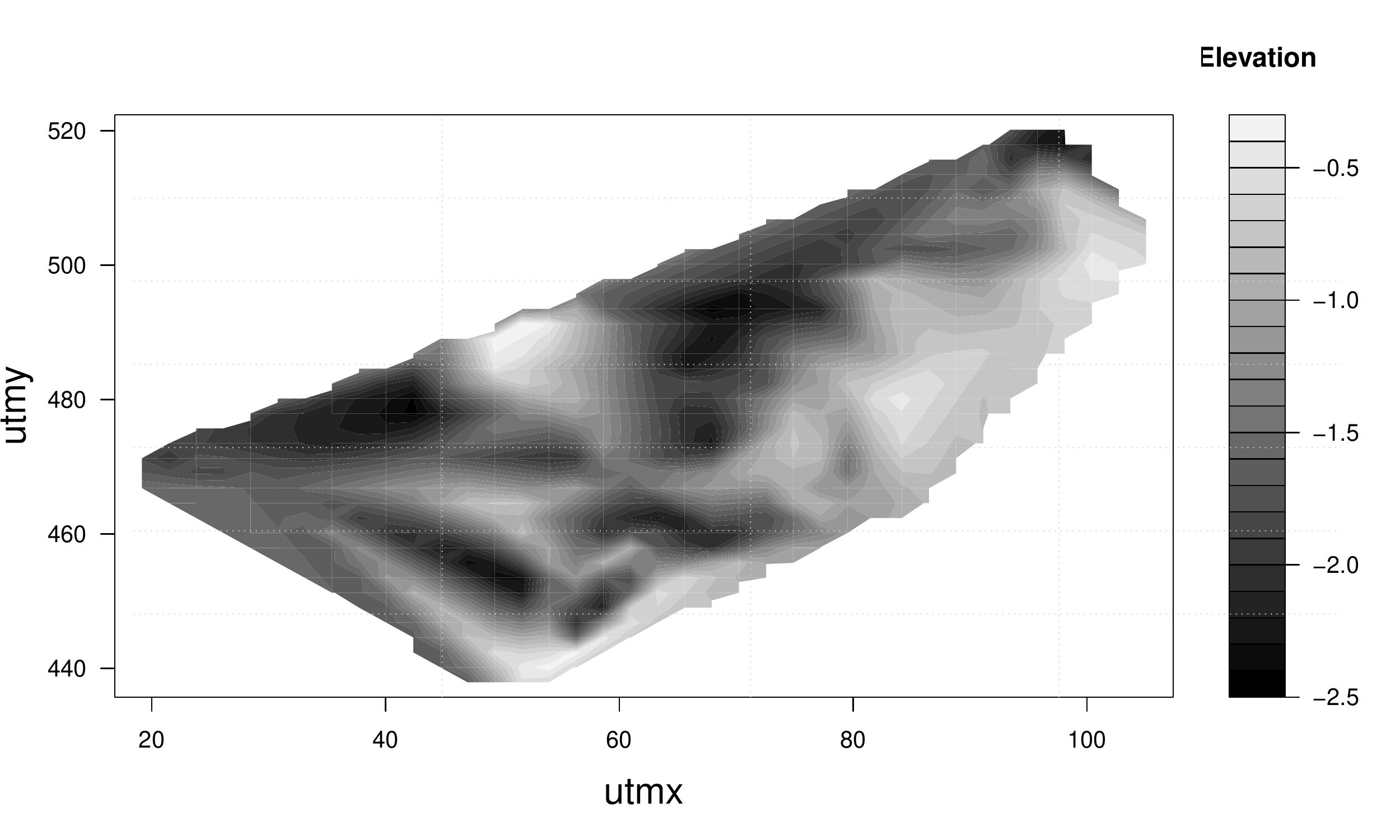} \includegraphics[scale = 0.28]{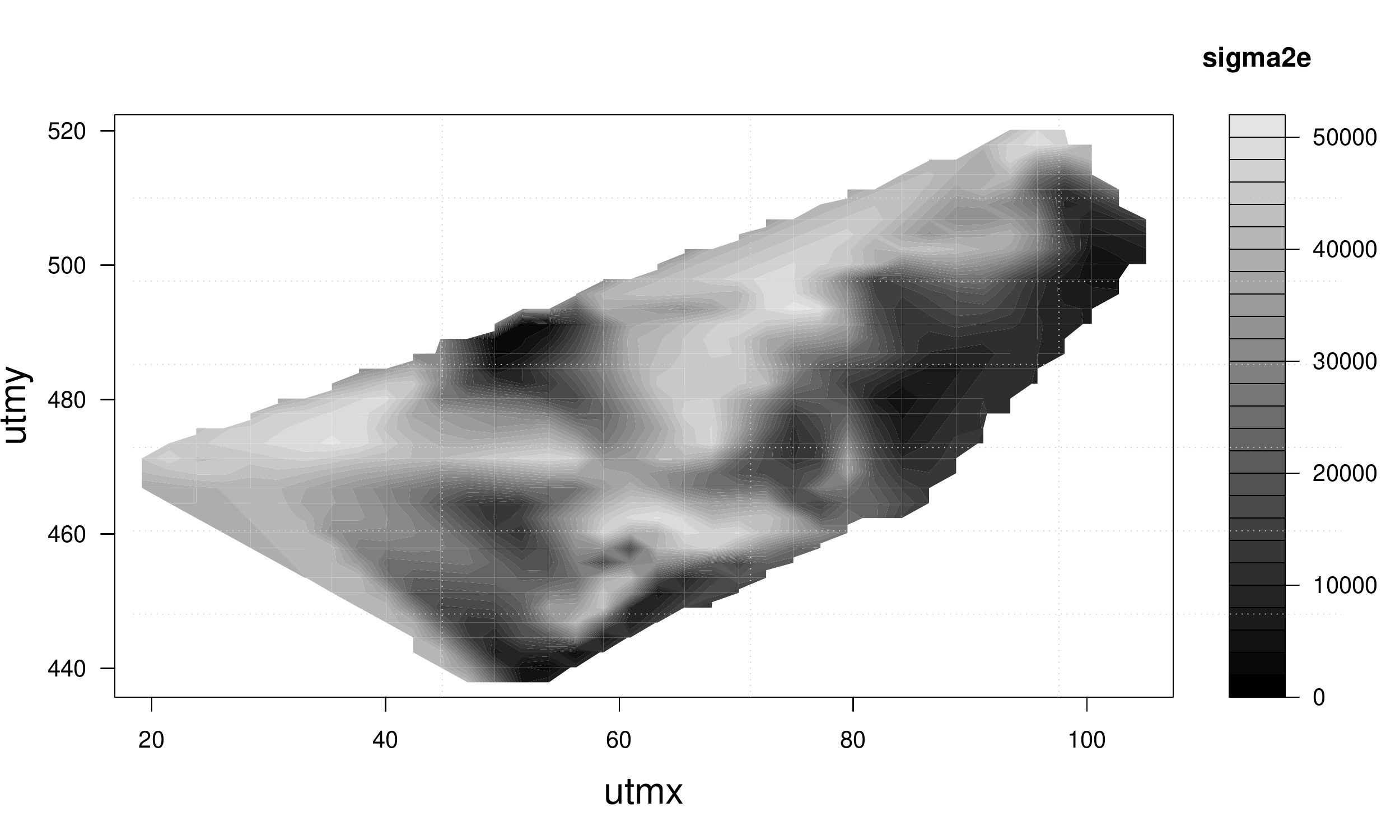} 
	\caption{\label{fig3:app} Maps showing the spatial variation in the HUC-specific area elevation slope coefficient (left) and sampling variance (right) estimates that are generated when the proposed nested error regression model with high dimensional parameter is fitted to the EMAP data.}
\end{figure}

From the estimates there is evidence of spatial variability of the regression coefficients and variance components. Figure \ref{fig3:app} shows contour maps of the estimated HUC-specific area elevation slope coefficient (left) and error variance model (right) from the fitted varying model. Examining the contours of the slope coefficients in Figure \ref{fig3:app} we see that the effect of elevation on ANC varies spatially, with these slope coefficients ranging from $-2.5$ to $-0.45$. The error variance component also shows spatial variation. In particular, the contour map of this component shows them ranging from a value close to $0$ (East) to $50,000$ (West).

These results are confirmed by the Hausman test \citep{Hausman1978, Bat88}, where the null hypothesis that slope parameters are the same within and among areas is rejected ($p$-value= 0.0021) and by the Raudenbush and Bryk's test \citep{Raudenbush2002}, where the null hypothesis of equality of the $\sigma_{\eps}^{2}$ between areas is rejected ($p$-value= 8e-04). 

To assess how EBP estimates are `close'  to the direct estimates we compute a goodness-of-fit diagnostic \citep{brown:2001}. This allows for evaluating if the model-based estimates are more precise than direct estimates. The goodness-of-fit diagnostic is computed as the value of the following Wald statistic:
$$W=\sum_{i}\frac{(\bar{y}_i-\hat{\theta}_i^{EBP})^2}{\widehat{var}(\bar{y}_i)+\widehat{MSE}(\hat{\theta}_i^{EBP})},$$
where $\widehat{var}(\bar{y}_i)$ is the estimated variance of the direct estimator and $\widehat{MSE}(\hat{\theta}_i^{EBP})$ is the estimated MSE of the EBP computed via bootstrap procedure.  The realized value of $W$ can then be compared against the $0.95$-quantile of a $\chi^2$-distribution with 86 degrees of freedom, i.e. $108.6479$. The values of the goodness-of-fit diagnostic is $19.21$ for EBP, i.e. the estimates are not statistically different from the direct estimates. The EBP estimates appear to be generally consistent with the direct estimates, with the correlation between the two sets of estimates being $0.98$.
To assess if EBP estimates are more precise than direct estimates, i.e. the potential gains in precision from using EBP instead of the direct estimates, we examine the distribution of the ratios of the estimated CVs of the direct and the EBP estimates for the EMAP data. A value greater than 1 for this ratio indicates that the estimated CV of the EBP estimate is smaller than that of the direct estimate. The average ratio across areas is $1.83$. It means a potential gain in precision from using EBP of about $83\%$.

Estimated values of average ANC for each HUC using EBP under \textcolor{black}{the nested regression model with high dimensional parameter} indicate that there are lower levels of average ANC (higher risk of water acidification) in the north-eastern part of the study region and they are consistent with the spatial distribution of ANC average values produced by previous non-parametric analyses of the EMAP data \citet{Ops08} and \citet{salvati2012}.

\begin{figure}[h]
	\centering    
	\includegraphics[scale = 0.60]{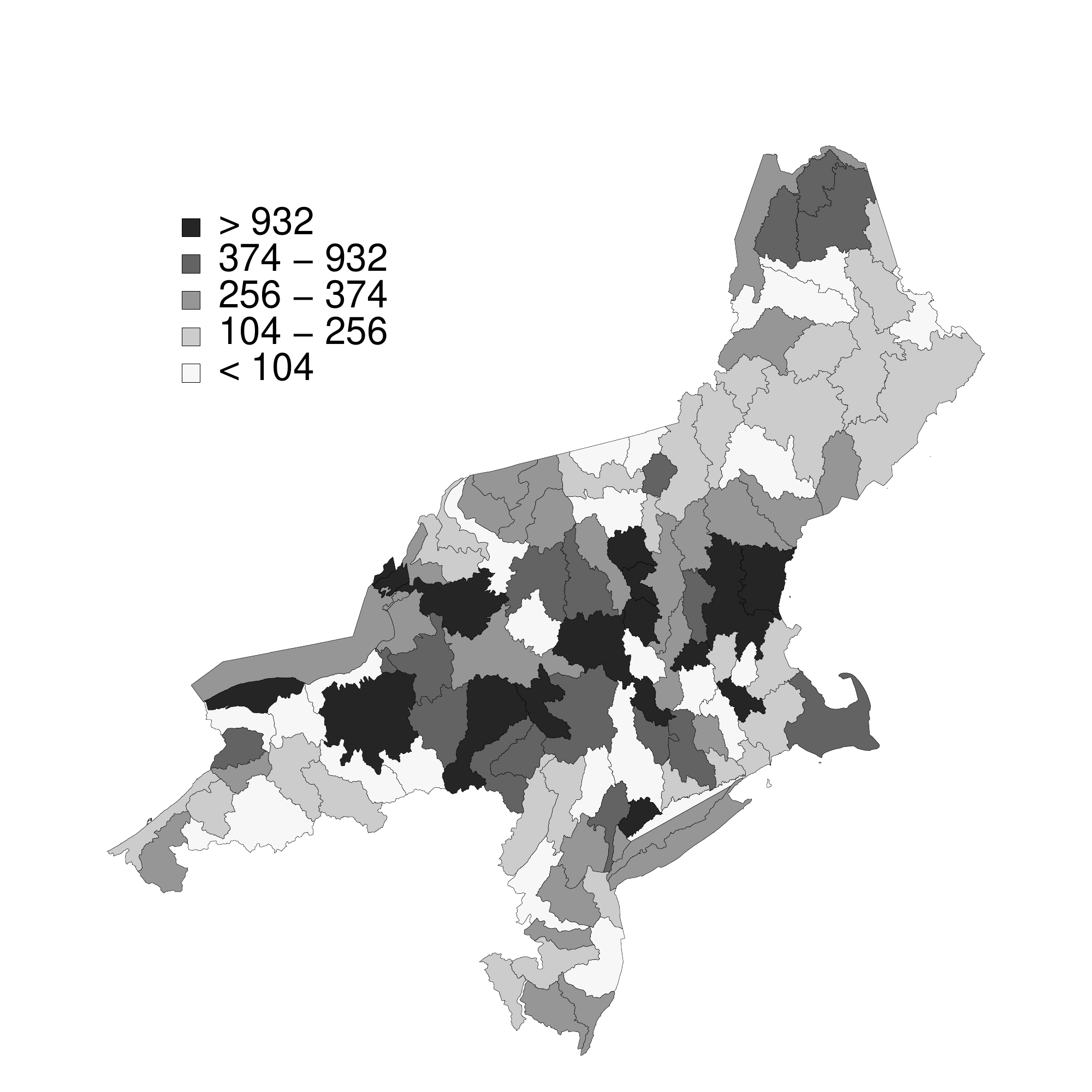}
	\caption{\label{fig4:app} Maps of estimated average ANC for HUCs using EBP under the nested error regression model with high dimensional parameter.}
\end{figure}

\textcolor{black}{In Figure \ref{fig4:app} we {\color{black}{display maps of estimates of average ANC for each HUC using EBP under the nested error model with high dimensional parameter.}} Estimates indicate that there are lower levels of average ANC (higher risk of water acidification) in the north-eastern part of the study region and they are consistent with the spatial distribution of ANC average values produced by previous non-parametric analyses of the EMAP data \citep{Ops08,salvati2012}.}

% \textcolor{blue}{Nicola: do we need to add a table with estimates and RMSE? The referee asked for: Data analysis: please provide some tables/figures showing the final EBP estimates.}

%%%%%%%%%%%%%%%%%%%%%%%%%%%%%%%%%%%%%%%%%%%%%%%%%%%%%%%%%%%%%%%%%%%%%%%%%
%%%%%%%%%%%%%%%%%% Conclusions             %%%%%%%%%%%%%%%%%%%%%%%%%%%%%%
%%%%%%%%%%%%%%%%%%%%%%%%%%%%%%%%%%%%%%%%%%%%%%%%%%%%%%%%%%%%%%%%%%%%%%%%%

\section{Conclusions}\label{sec:conclusion}
In this paper, we have demonstrated, through simulations and data analysis, unsuitability of the well-known nested error regression model for small area estimation when combining a large number of small areas.  As argued in the paper, {\color{black}{this could be due to}} the fact that the exchangeability assumption concerning the regression coefficients and fixed sampling variances does not hold for large number of small areas to be combined.  One potential solution to this problem is develop methodology for a joint random regression coefficients  and random sampling variances model. Such joint modeling strategy has not been tried in the small area literature and could be an excellent future research problem. Alternatively, one could consider allowing fixed area specific regression coefficients and sampling variances to circumvent problems associated with exchangeability of the regression coefficients and sampling variances across small areas.  However, this modeling strategy is likely to lead to an inefficient estimation of the area specific regression coefficients and sampling variances if traditional estimation methods are used because of small area specific samples.  In this paper, we develop a robust area specific estimating equations approach where different estimating equations are used for different small areas. Random effect is assumed on a  single area specific {\color{black}{tuning parameter of the estimation equation}}  (e.g., $\tau_i$, area specific M-quantile coefficient) to reduce dimensionality, which in turn, improves on estimation efficiency.

{\color{black}{Performances of the estimators of area specific regression coefficients and sampling variances $\phi_i$ and the associated EBP depend very much on the area specific tuning parameters $\tau_i$ of the estimating equations.  The case when $\tau_i$'s are known (e.g., from the census data) does not create any problem. Like in many other papers in small area estimation, our emphasize here is bounded $n_i$ and our evaluation for unknown $\tau_i$ case is exclusively by extensive simulation studies, which actually allow us to consider a variety of simulation conditions.  In most SAE asymptotics, $m$ tends to infinity but $n_i$'s are bounded. In such a setting, the estimators of $\tau_i$ considered in this paper are not consistent. As pointed out by Jiang and Lahiri (2006), the alternative asymptotic setting in which both number of areas $m$ and area specific sample sizes $n_i$ tend to    (possibly at differential rates) is indeed an important problem. This alternative asymptotic framework has received relatively less attention in SAE research.  Lye and Welsh (2021) recently put forward an asymptotic approach where both $m$ and   $n_i$ are allowed to tend to infinity.  It would be interesting to study the asymptotic properties of our estimators of our model parameters and predictors  under this alternative asymptotic setting. This is a topic for a future research.}}

In this paper, we also question the utility of the well-known second-order unbiasedness criterion of the mean squared error estimators. Besides the difficulty in establishing a rigorous theory \citep[see][]{Jiang2018} and increasing the computational burden, this property does not necessarily ensure similar second-order unbiasedness properties of other commonly used uncertainty measures such as coefficient of variation, relative root mean squared error, and others. In this paper, we downplay the second-order unbiasedness criterion and introduces a general parametric bootstrap method and a jackknife procedure for estimating commonly used uncertainly measures such as relative root mean square error and coefficient of variance.  Proposed methods perform well in our Monte Carlo simulations and real life data analysis.  In developing the methodology we ignored certain complex situations such as correlated data within the same PSU.  Also, the method is developed for continuous data. These present additional avenues for further research. \textcolor{black}{Finally, as alternative to the proposed nested error regression model with high dimensional parameter, an extension of linear models with both random coefficients and random dispersion values in Bayesian approach following \citet[][Chapter 8]{Hoff:2009} may be doable and it could be an objective of a future research.}

\vspace{20px}

\textsc{Acknowledgments:} The word of P. Lahiri was partially supported by the U.S. National Science Foundation grant SES-
1758808. The work of N. Salvati has been developed under the support of the Progetto di Ricerca di Ateneo From survey-based to register-based statistics: a paradigm
345 shift using latent variable models' (grant PRA2018-9).
The work of Salvati was carried out with the support of
the project InGRID-2 Integrating Research Infrastructure for European expertise on Inclusive Growth from data to policy (Grant Agreement N. 730998, EU)

\section*{Appendix: Regularity conditions for consistency of the estimators of $\bphi_i$}
\begin{itemize}
	\item[(i)] The influence function $\psi_i$ is a bounded continuous function with a derivative which, except for a finite number of points, is defined everywhere and it is also bounded.
	\item[(ii)] $|\bX_{l(p+1)} |$, $1\leqslant l \leqslant m$, is bounded as $m \rightarrow \infty$.
	\item[(iii)] The true parameter vector $\bphi_i \in \bPhi_i^{0}$, the interior of the parameter space for $\bphi_i$.
	\item[(iv)] For any compact set $\bB \in \bPhi_i^0$, the $sup_{\bphi_i \in \bB} || \cdot||$ of up to fourth derivatives of $ \bV_{i}$, $1\leqslant i \leqslant m$, are bounded, and $sup_{\bphi_i \in \bB} || \bV_{i} ||$, $1\leqslant i \leqslant m$, are bounded.
	\item[(v)] $\lambda_{min} \{\bX_{l(p+1)}^{\prime}\bU_{l
		;i}^{-1/2} \bD_{l;i}\bU_{l;i}^{1/2} \bV_{l;i}^{-1}\bX_{l(p+1)}\}$, $\lambda_{min}\left \{   tr \left [\bV_{l;i}^{-1}\bZ_l\bZ_l^{\prime} \bV_{l;i}^{-1}\bZ_l \bZ_l^{\prime}  \right ]     \right\}$ and \linebreak $\lambda_{min} \left\{  tr \left [\bV_{l;i}^{-1} \frac{\partial \bV_{l;i}}{\partial \sigma_{\epsilon i}^2}|_{\sigma_\gamma^2=\sigma_\gamma^{2} } \bV_{l;i}^{-1} \frac{\partial \bV_{l;i}}{\partial \sigma_{\epsilon i}^2}|_{\sigma_\gamma^2=\sigma_\gamma^{2} }  \right ]  \right\}$ are bounded away from zero, where $\lambda_{min} $ represents the smallest eigenvalue; here $\bD_{l;i}$ is a diagonal matrix with its $j$th diagonal element is the derivative of $\psi_i$ respect the $j$th residual;
	\item[(vi)] There are constants $\zeta>0$ and $L<  \infty$ such that, if $\br_{l;i}$, $1\leqslant l \leqslant m$,  then $E |\psi_{i}(\br_{l;i})|^{4+\zeta}$, $E || \partial \psi_{i}(\br_{l;i})||$ are all bounded by $L$.
	\item[(vii)] $E|\by_l |^{8+\varphi}$, $1\leqslant l \leqslant m$, are bounded for some $\varphi>0$.
	\item[(viii)] $|S|=O(m^\kappa)$ for some $0\leqslant \kappa \leqslant 3/(6+\varphi)$.
\end{itemize}

\bibliographystyle{apalike}
\bibliography{Datenbank}
\end{document}

\maketitle

%\begin{abstract}
%In this paper we propose a flexible {\color{black}{nested error regression small area model}} \textcolor{black}{with high dimensional parameter}  that incorporates heterogeneity in regression coefficients and variance components. 
%% Such modeling allows pooling information from a large number of areas and thus makes the notion of consistency more relevant.  
%We develop a new robust small area specific estimating equations method 
%% for producing consistent estimators of regression coefficients and variance components.  Moreover, since our theoretical framework
%that allows appropriate pooling of a large number of areas in estimating small area specific model parameters.
%% , the use of the first-order unbiasedness of the mean squared prediction error estimators is adequate and thus avoiding the cumbersome task of bias correction.  
%We propose a parametric bootstrap and jackknife method to estimate not only the mean squared  errors but also other commonly used uncertainty measures such as standard errors and coefficients of variation. We conduct both model-based and design-based simulation experiments and real-life data analysis to evaluate the proposed methodology. 
%
%\vspace{0.3cm}
%
%\noindent \textit{Keywords:} design consistency; M-estimation; root mean squared error estimation; AAGIS data; EMAP data.
%\end{abstract}

\section{Results of model-based simulation experiments}
In this section we report the results of the simulation experiments according the scenarios $(0,0)$, $(\beta,0)$ and $(\beta,\sigma_\eps)$ presented in the manuscript with lower number of small areas, $m=40$, and sample size in each small area equal to $n_i=10$. The results in Table \ref{table_scen1} confirm the ranking of the predictors in terms of performance presented in the manuscript with low levels of bias and variability. 

Moreover, we present the results of another scenario with $m=40, ~n_i=10$ and $m=100, ~n_i=4$. In this additional scenario values for $x_{ij}$ are generated as in the manuscript, as independently and identically distributed from a log-normal distribution with a mean of $1.0$ and a standard deviation of $0.5$ on the log-scale. Values for $y_{ij}$ are generated as $y_{ij} = 10 + \beta_i x_{ij}+ \gamma_i + \eps_{ij}$, where the fixed effects and the random area effects and sampling errors are independently generated as it follows: $\beta_i = 5$ for $i = 1,\dots,50$ and $\beta_i = -5$ for $i = 51,\dots,100$ (if $m=100$) and it is kept fixed over the simulations, $\gamma_i \sim N (0, 3)$ and $\eps_{ij}\sim \delta_{ij} N(0,6)+(1-\delta_{ij})N(20,150)$,where $\delta_{ij}$
is an independently generated Bernoulli random variable with $Pr(\delta_{ij} = 1) = 0.97$, i.e. the individual effects are independent draws from a mixture of two normal distributions, with $97\%$ on average drawn from a `well-behaved' $N(0, 6)$ distribution and $3\%$ on average drawn from an outlier $N(20, 150)$ distribution; this model violates assumptions of the nested error regression model (1) because slopes vary across small areas and the sampling errors have some outlying values. The results in Table \ref{table_scen2} show the superior outlier robustness of EBP and MQ compared with the other predictors certainly hold true, especially in terms of RRMSE. The EBP methods are less biased than the competitors.
\begin{table}[h!]
	\caption{\label{table_scen1} Model-based simulation results: performance of estimators/predictors of small area means; the number of small areas considered is $40$; population and sample sizes for each area are $100$ and $10$, respectively; median is over 40 small areas; numbers in parenthesis are the values of the efficiency over EBLUP in terms of RMSE. }
	\centering
	\fbox{
		\scalebox{1.0}{
			\begin{tabular}{lrrr}\hline 
				Predictor &\multicolumn{3}{c}{Results ($\%$) for the following scenarios} \\ 
				& $(0,0)$ & $(\beta,0)$ & $(\beta,\sigma_\eps^2)$ \\\hline
				&\multicolumn{3}{c}{Median absolute relative bias} \\ 
				Direct & 0.319 & 0.406  & 0.250  \\
				EBLUP & 0.055& 3.721  & 4.197  \\
				MQ & 0.054 & 5.690  & 8.805  \\ 
				EBLUP-RS & 0.053 & 0.177  & 0.289  \\ 
				EBLUP-H & 0.054 & 6.048  & 7.528  \\ 
				OBP & 0.145 & 3.531  & 4.050  \\ 
				%REBLUP & 0.087 &0.224  & 0.332 & 0.118  & 0.235 & 0.225\\
				%SYNTH & 0.254 & 0.700 & 0.927 & 0.230 & 1.214 & 1.259  \\
				EBP & 0.049 & 0.141 & 0.435    \\
				%MQBP-LBP$^{\psi} $& 0.102 & 0.164 & 0.238 & 0.144  & 0.237 & 0.225  \\
				%MQBP-LBP$^{\psi, Rob} $& 0.106 & 0.165 & 0.201 & 0.147  & 0.251 & 0.220  \\
				%MQBP-Naive & 0.123 & 0.170 &0.178 & 0.152 & 0.206 & 0.204 \\
				%MQ-LBP & 0.138 & 0.328& 0.315& 0.186 & 0.336 & 0.403 \\
				EBP-MLE & 0.069 & 0.131  & 0.095  \\ 
				%EBP-G & 0.052 & 0.101  & 0.702  \\ 
				&\multicolumn{3}{c}{Median  RRMSE} \\ 
				Direct  &10.175 (14.492) & 27.323 (1.023)  & 28.647 (1.025)  \\
				EBLUP   & 2.683 (1.000) & 26.930 (1.000) & 28.291 (1.000)\\
				MQ & 2.945 (1.230) & 10.877 (0.160) & 14.680 (0.229)   \\ 
				EBLUP-RS   & 2.689 (1.008) & 8.180 (0.97) & 10.700 (0.120)\\
				EBLUP-H   & 2.690 (1.005) & 35.067 (1.667) & 38.271 (1.556)\\
				OBP   & 6.206 (5.404) & 27.081 (0.999) & 28.321 (1.000)\\
				%REBLUP  &6.206 (1.034)  & 11.680 (1.065)  & 14.317 (1.023) & 6.677 (0.870) & 12.404 (0.990) & 14.362 (1.018) \\
				%SYNTH &11.252 (1.871) & 41.197 (3.736)  & 41.010 (2.924) & 14.023 (1.831) & 42.099 (3.368) & 51.205 ( 3.643) \\
				EBP &2.688 (1.011) & 7.984 (0.084)  &  10.230 (0.112)  \\
				%MQBP-LBP$^{\psi} $ &6.126 (1.018) & 8.388 (0.758)  & 10.947 (0.798) & 7.508 (0.979)  & 10.276 (0.829) & 11.071 (0.788)\\
				%MQBP-LBP$^{\psi, Rob} $ &6.111 (1.017) & 8.383 (0.757)  & 11.457 (0.811) & 7.692 (1.004)  & 10.328 (0.831) & 11.052 (0.787)\\
				%MQBP-Naive & 6.483 (1.073)& 8.260 (0.752)  & 10.881 (0.790) & 7.798 (1.016) & 10.229 (0.823) & 10.888 (0.774)\\
				%MQ-LBP &7.218 (1.198)& 14.311 (1.302) & 14.799 (1.062) & 8.916 (1.169) & 15.704 (1.250) & 20.126 (1.406)\\
				EBP-MLE & 3.130 (1.373) & 8.650 (0.106) & 11.105 (0.131)   \\ 
				%EBP-G & 2.684 (1.005) & 7.685 (0.078) & 9.838 (0.102)   \\ 
	\end{tabular}}}
\end{table}

%%%%%%%%%%%%%%%%%%%%%%%%%%%%%%%%New table
\begin{table}[h!]
	\caption{\label{table_scen2} Model-based simulation results for additional scenario based on individual outlying values: performance of estimators/predictors of small area means; the number of small areas considered is $40$ and $100$; population and sample sizes for each area are $100$ and $10$ (for $m=40$) and $4$ (for $m=100$), respectively; median is over $m$ small areas; numbers in parenthesis are the values of the efficiency over EBLUP in terms of RMSE. }
	\centering
	\fbox{
		\scalebox{1.0}{
			\begin{tabular}{lrr}\hline 
				Predictor &\multicolumn{2}{c}{Results ($\%$) for the following scenarios} \\ 
				&\multicolumn{2}{c}{Median absolute relative bias} \\ 
				& $m=40,~n_i=10$ & $m=100,~n_i=4$ \\\hline
				Direct & 0.475 & 0.960 \\
				EBLUP & 5.174& 10.654 \\
				MQ & 6.847 & 5.703  \\ 
				EBLUP-RS & 0.639 &2.345  \\ 
				EBLUP-H & 6.755 & 7.943 \\ 
				OBP & 5.034 & 10.295 \\ 
				%REBLUP & 0.087 &0.224  & 0.332 & 0.118  & 0.235 & 0.225\\
				%SYNTH & 0.254 & 0.700 & 0.927 & 0.230 & 1.214 & 1.259  \\
				EBP & 1.184 & 1.965    \\
				%MQBP-LBP$^{\psi} $& 0.102 & 0.164 & 0.238 & 0.144  & 0.237 & 0.225  \\
				%MQBP-LBP$^{\psi, Rob} $& 0.106 & 0.165 & 0.201 & 0.147  & 0.251 & 0.220  \\
				%MQBP-Naive & 0.123 & 0.170 &0.178 & 0.152 & 0.206 & 0.204 \\
				%MQ-LBP & 0.138 & 0.328& 0.315& 0.186 & 0.336 & 0.403 \\
				EBP-MLE & 0.437 & 0.623  \\ 
				%EBP-G & 1.221 & 1.784 \\ 
				&\multicolumn{2}{c}{Median  RRMSE} \\ 
				Direct  &33.621 (1.025) & 53.297 (1.087)  \\
				EBLUP   & 33.367 (1.000) &  52.183 (1.000)\\
				MQ & 16.753 (0.263) & 22.494 (0.204)  \\ 
				EBLUP-RS   & 16.113 (0.239) & 26.355 (0.275)\\
				EBLUP-H   & 41.368 (1.533) & 61.637 (1.346)\\
				OBP   & 33.179 (0.999) & 51.816 (1.003)\\
				%REBLUP  &6.206 (1.034)  & 11.680 (1.065)  & 14.317 (1.023) & 6.677 (0.870) & 12.404 (0.990) & 14.362 (1.018) \\
				%SYNTH &11.252 (1.871) & 41.197 (3.736)  & 41.010 (2.924) & 14.023 (1.831) & 42.099 (3.368) & 51.205 ( 3.643) \\
				EBP &14.513 (0.185) & 19.924 (0.160) \\
				%MQBP-LBP$^{\psi} $ &6.126 (1.018) & 8.388 (0.758)  & 10.947 (0.798) & 7.508 (0.979)  & 10.276 (0.829) & 11.071 (0.788)\\
				%MQBP-LBP$^{\psi, Rob} $ &6.111 (1.017) & 8.383 (0.757)  & 11.457 (0.811) & 7.692 (1.004)  & 10.328 (0.831) & 11.052 (0.787)\\
				%MQBP-Naive & 6.483 (1.073)& 8.260 (0.752)  & 10.881 (0.790) & 7.798 (1.016) & 10.229 (0.823) & 10.888 (0.774)\\
				%MQ-LBP &7.218 (1.198)& 14.311 (1.302) & 14.799 (1.062) & 8.916 (1.169) & 15.704 (1.250) & 20.126 (1.406)\\
				EBP-MLE &16.210 (0.247) & 26.260 (0.281)   \\ 
				%EBP-G & 13.783 (0.176) & 18.352 (0.121)   \\ 
	\end{tabular}}}
\end{table}

\section{Performance of the estimators of $\beta_i$ and $\sigma_{\eps i}^2$}
\begin{figure}[h!]
	\centering  
	$(0,0$)\\
	\includegraphics[scale = 0.40]{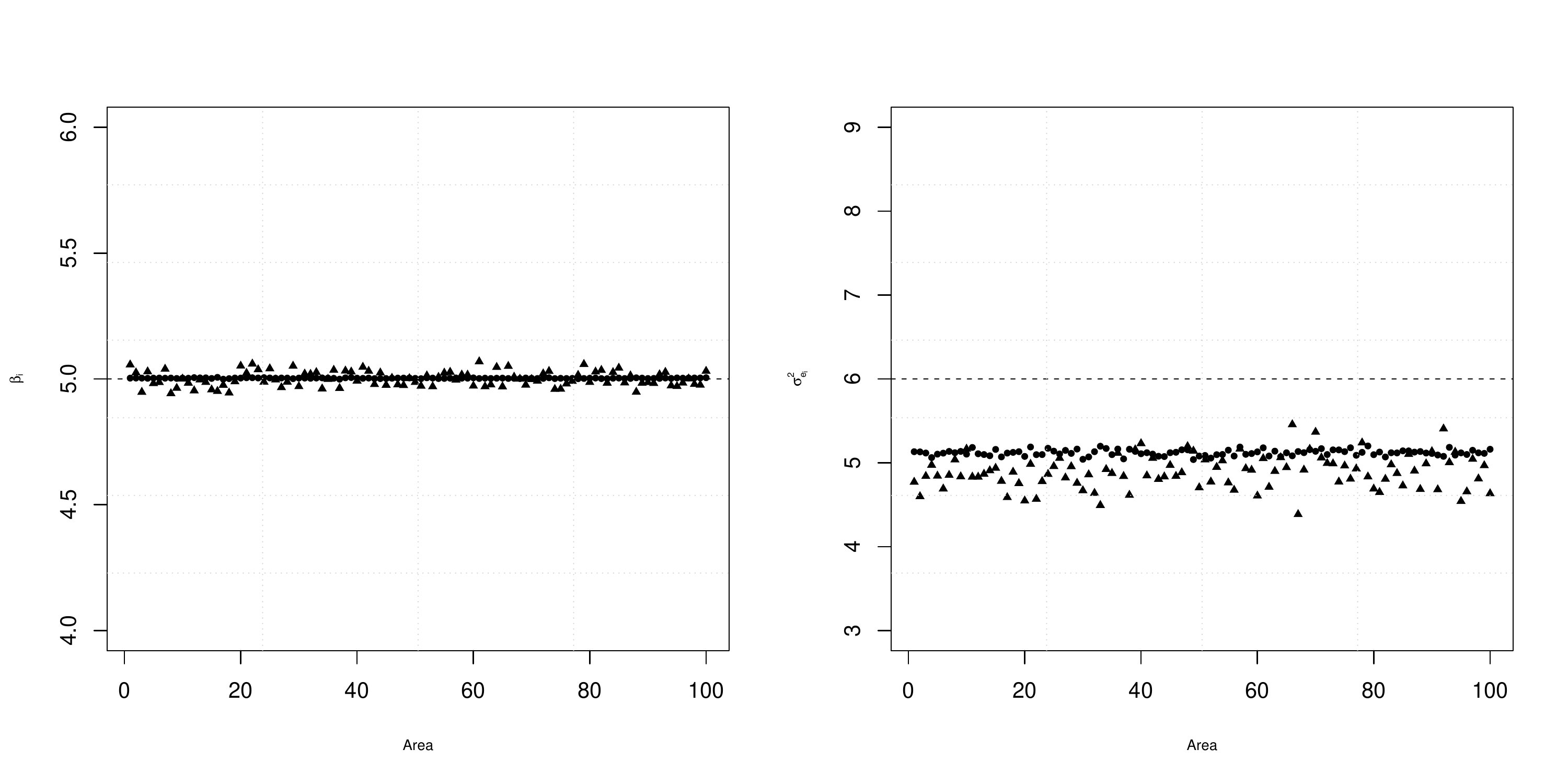}\\
	$(\beta,0)$\\
	\includegraphics[scale = 0.40]{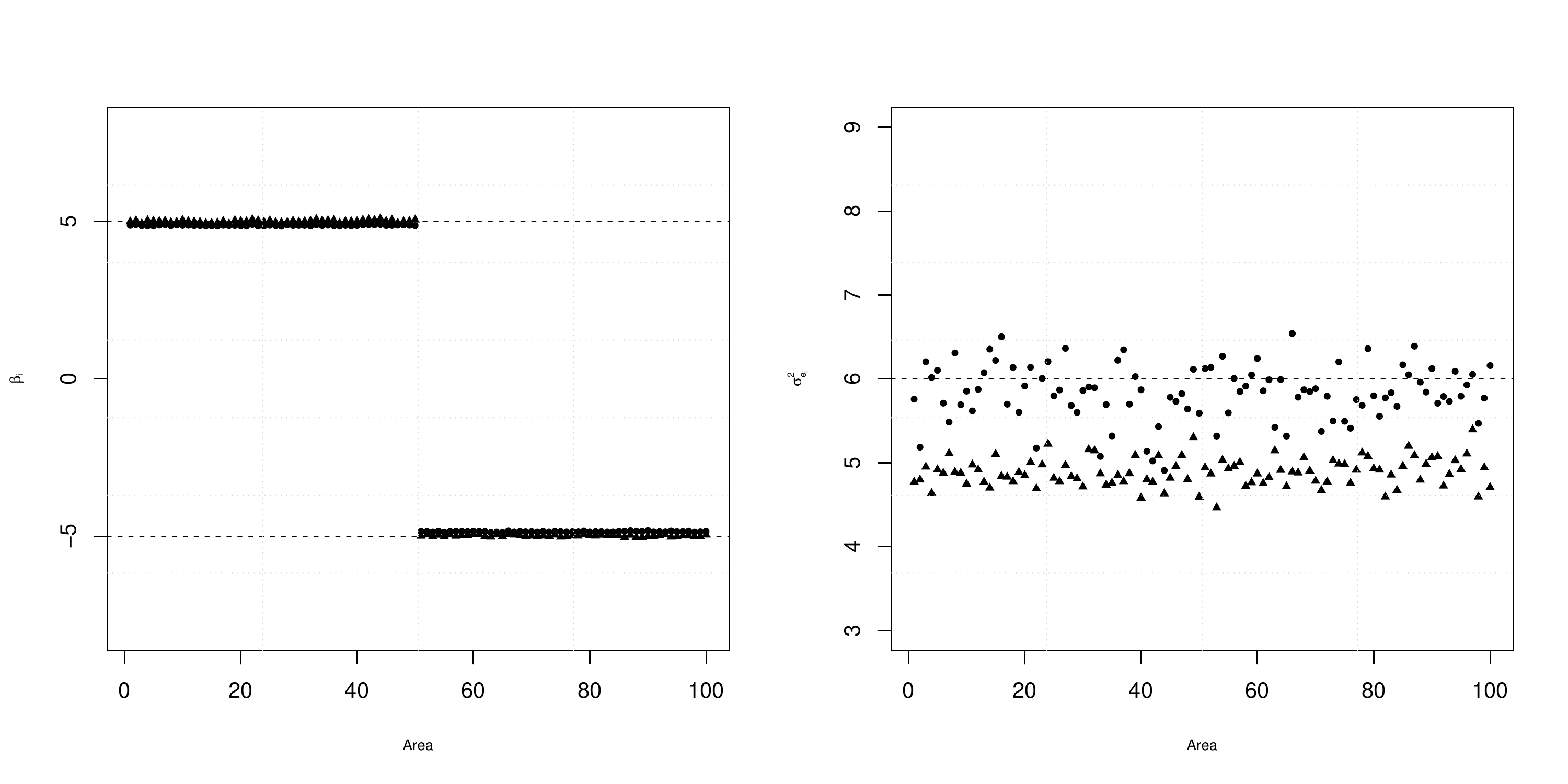}\\ 
	$(\beta,\sigma_\eps^2)$\\
	\includegraphics[scale = 0.40]{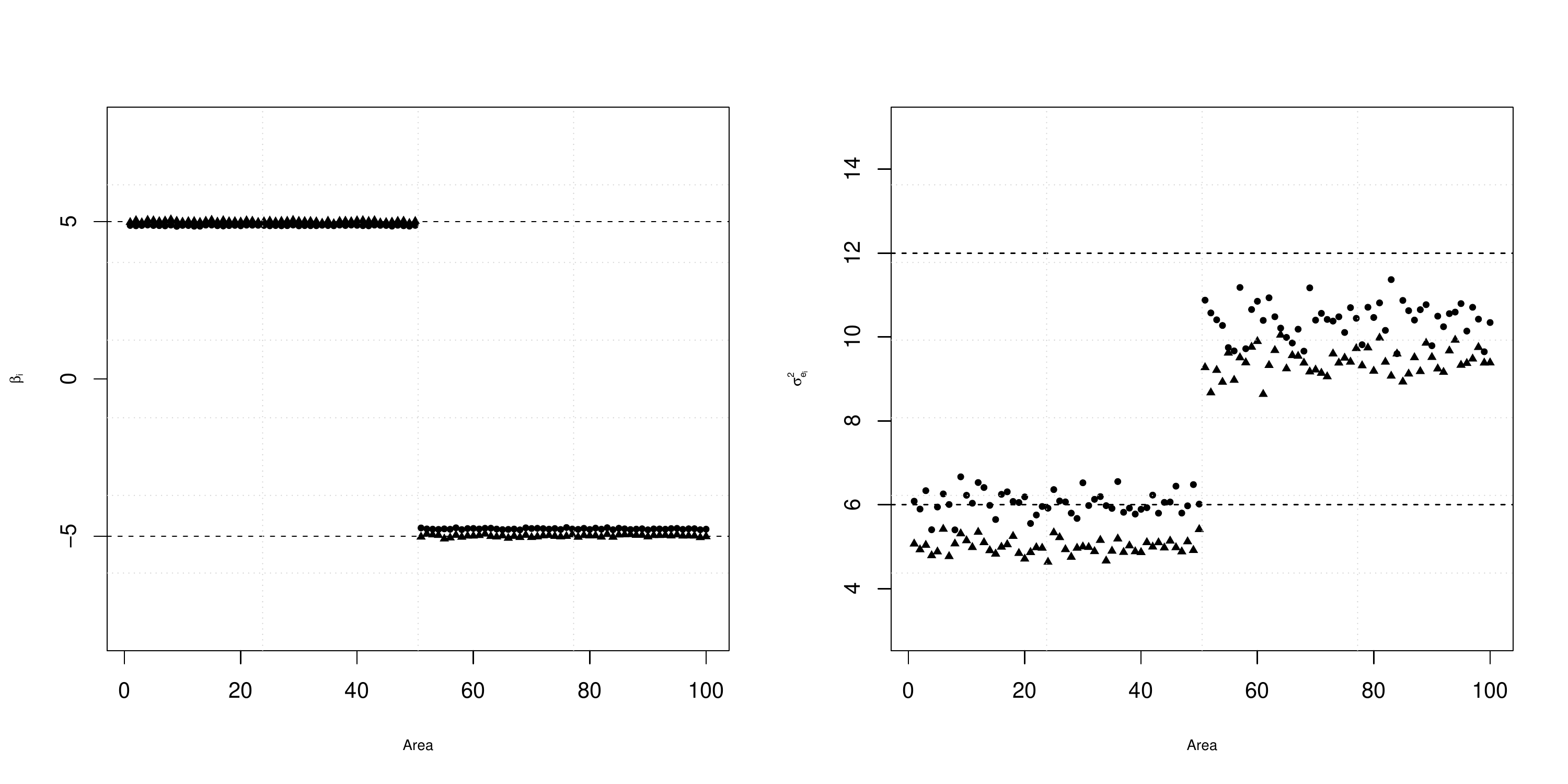} \\
	\caption{\label{Scatterplot} Scatterplot of three scenarios presented in the manuscript. The plots show the relationship between the true values of the parameters $\beta_i$ and $\sigma_{\eps i}^2$, on the y-axis, and the average estimated values of the parameters, over Monte Carlo runs, for each small area, on the x-axis. The estimates of the parameters of the nested regression model with high dimensional parameter based on GEE method are in filled circles, whereas those obtained by ML approach are represented by filled triangle.}
\end{figure}

In this section we present some plots which complete the information obtained in Figure 1 of the manuscript in order to evaluate the performance of the the estimators of $\beta_i$ and $\sigma_{\eps i}^2$ using the GEE and the ML methods. Figure \ref{Scatterplot} shows the relationship between the true values of the parameters $\beta_i$ and $\sigma_{\eps i}^2$, on the y-axis, and the average estimated values of the parameters, over Monte Carlo runs, for each small area, on the x-axis. The left panels present the performance of the estimators of $\beta_i$; the right panels indicate the behaviour of the estimators of $\sigma_{\eps i}^2$. The estimates of the parameters of the nested regression model with high dimensional parameter based on GEE method are in filled circles, whereas those obtained by ML approach are represented by filled triangle. In the three scenarios the parameter $\beta_i$ is estimated well both with GEE and ML approach. The latter shows estimated values with higher variability. Regarding the variance component $\sigma_{\eps i}^2$ we can note that the ML estimator is biased-low in all the scenarios. The estimator based on the GEE method perform better, even if it is slightly biased-low in scenario ($0,0$) and in scenario ($\beta_i, \sigma_{\eps}^2$) for these small areas where the expected value of the variance component is $12$.

